\documentclass[twoside,fleqn,letterpaper]{ActaStyle}
\usepackage[dvips]{graphicx}
\usepackage{times,cite}
       \def\authorheading{M. Creutz}
       \def\shorttitle{Confinement, chiral symmetry, and the lattice}

\newcommand{\Section}[1]{\section{#1} \setcounter{equation}{0}}

\long \def \blockcomment #1\endcomment{}

\def\slashchar#1{\setbox0=\hbox{$#1$}           
   \dimen0=\wd0                                 
   \setbox1=\hbox{/} \dimen1=\wd1               
   \ifdim\dimen0>\dimen1                        
      \rlap{\hbox to \dimen0{\hfil/\hfil}}      
      #1                                        
   \else                                        
      \rlap{\hbox to \dimen1{\hfil$#1$\hfil}}   
      /                                         
   \fi}                                         %

\begin{document}
       \pagerange{1}{126}

\title{Confinement, chiral symmetry, and the lattice}
\author{Michael Creutz
\email{creutz@bnl.gov}}
{Brookhaven National Laboratory\\
Upton, NY 11973
}

\abstract{Two crucial properties of QCD, confinement and chiral
  symmetry breaking, cannot be understood within the context of
  conventional Feynman perturbation theory.  Non-perturbative
  phenomena enter the theory in a fundamental way at both the
  classical and quantum levels.  Over the years a coherent qualitative
  picture of the interplay between chiral symmetry, quantum mechanical
  anomalies, and the lattice has emerged and is reviewed here.
}


\tableofcontents
\newpage

\Section{QCD}

Quarks interacting with non-Abelian gauge fields are now widely
accepted as the basis of the strong nuclear force.  This quantum field
theory is known under the somewhat whimsical name of Quantum
Chromodynamics, or QCD\footnote{If you prefer not to confuse this with
the 4000 Angstroms typical of color, you could regard this as an
acronym for Quark Confining Dynamics.}.  This system is remarkable in
its paucity of parameters.  Once the overall scale is set, perhaps by
working in units where the proton mass is unity, the only remaining
parameters are the quark masses. The quarks represent a new level of
substructure within hadronic particles such as the proton.

The viability of this picture relies on some rather unusual features.
These include confinement, the inability to isolate a single quark,
and the spontaneous breaking of chiral symmetry, needed to explain the
lightness of the pions relative to other hadrons.  The study of these
phenomena requires the development of techniques that go beyond
traditional Feynman perturbation theory.  Here we concentrate on the
interplay of two of these, lattice gauge theory and effective chiral
models.  

The presentation is meant to be introductory.  The aim is to provide a
qualitative picture of how the symmetries of this theory work together
rather than to present detailed methods for calculation.  In this
first section we briefly review why this theory is so compelling.

\subsection{Why quarks}

Although an isolated quark has not been seen, we have a variety of
reasons to believe in the reality of quarks as the basis for this next
layer of matter.  First, quarks provide a rather elegant explanation
of certain regularities in low energy hadronic spectroscopy.  Indeed,
it was the successes of the eightfold way
\cite{GellMann:1964nj} which originally motivated the quark model.  
Two ``flavors' of low mass quarks lie at the heart of isospin symmetry
in nuclear physics.  Adding the somewhat heavier ``strange'' quark
gives the celebrated multiplet structure described by representations
of the group $SU(3)$.

Second, the large cross sections observed in deeply inelastic
lepton-hadron scattering point to structure within the proton at
distance scales of less than $10^{-16}$ centimeters, whereas the
overall proton electromagnetic radius is of order $10^{-13}$
centimeters \cite{Mishra:1989jc}.  Furthermore, the angular
dependences observed in these experiments indicate that any underlying
charged constituents carry half-integer spin.

Yet a further piece of evidence for compositeness lies in the excitations of
the low-lying hadrons.  Particles differing in angular momentum fall
neatly into place along the famous ``Regge trajectories''
\cite{Collins:1977jy}.  Families of states group together as orbital
excitations of an underlying extended system.  The sustained rising of
these trajectories with increasing angular momentum points toward
strong long-range forces between the constituents.

Finally, the idea of quarks became incontrovertible with the discovery
of heavier quark species beyond the first three.  The intricate
spectroscopy of the charmonium and upsilon families is admirably
explained via potential models for non-relativistic bound states.
These systems represent what are sometimes thought of as the
``hydrogen atoms'' of elementary particle physics.  The fine details
of their structure provides a major testing ground for quantitative
predictions from lattice techniques.

\subsection{Gluons and confinement}

Despite its successes, the quark picture raises a variety of puzzles.
For the model to work so well, the constituents should not interact so
strongly that they loose their identity.  Indeed, the question arises
whether it is possible to have objects display point-like behavior in
a strongly interacting theory.  The phenomenon of asymptotic freedom,
discussed in more detail later, turns out to be crucial to realizing
this picture.

Perhaps the most peculiar aspect of the theory relates to the fact
that an isolated quark has never been observed.  These basic
constituents of matter do not copiously appear as free particles
emerging from high energy collisions.  This is in marked contrast to
the empirical observation in hadronic physics that anything which can
be created will be.  Only phenomena prevented by known symmetries are
prevented.  The difficulty in producing quarks has led to the concept
of a principle of exact confinement.  Indeed, it may be simpler to
have a constituent which can never be produced than an approximate
imprisonment relying on an unnaturally small suppression factor.  This
is particularly true in a theory like the strong interactions, which
is devoid of any large dimensionless parameters.

But how can one ascribe any reality to an object which cannot be
produced?  Is this just some sort of mathematical trick?  Remarkably,
gauge theories potentially possess a simple physical mechanism for
giving constituents infinite energy when in isolation.  In this
picture a quark-antiquark pair experiences an attractive force which
remains non-vanishing even for asymptotically large separations.  This
linearly rising long distance potential energy forms the basis of
essentially all models of quark confinement.

\begin{figure}[t]
\begin{center}
\includegraphics[width=3in]{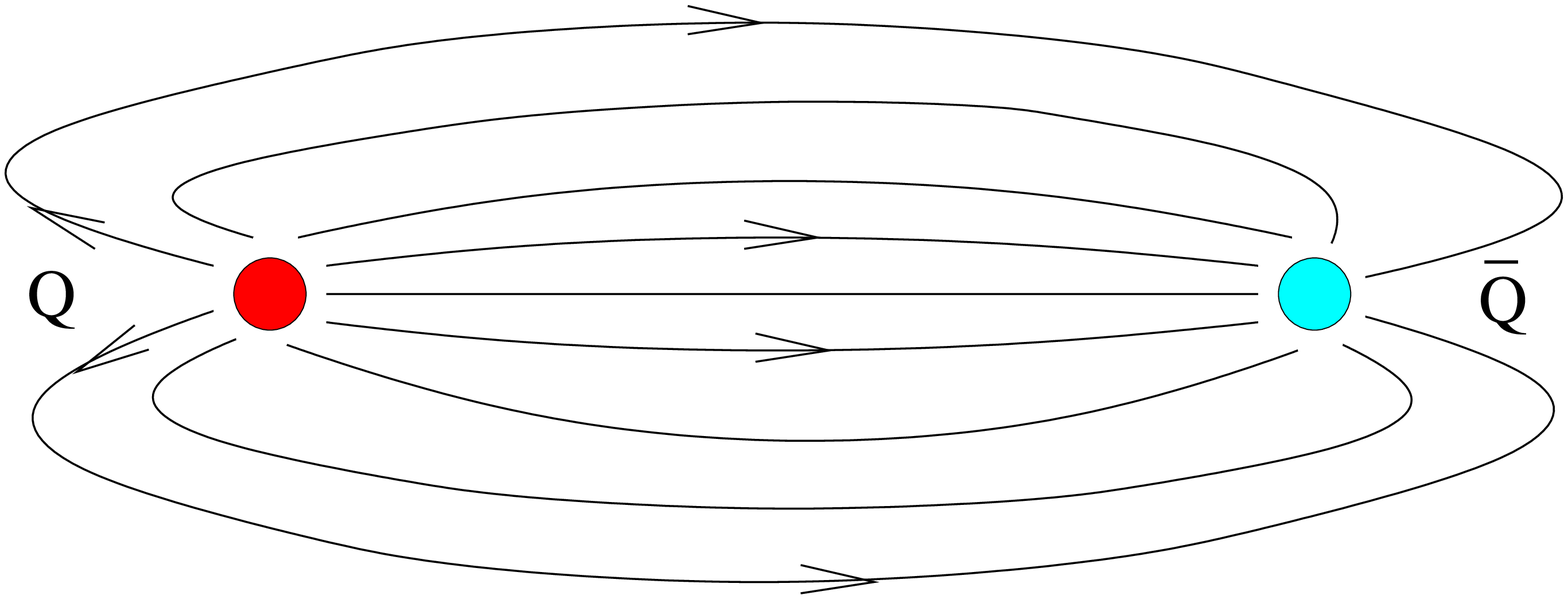}
\caption {A tube of gluonic flux connects quarks and anti-quarks.  The
strength of this string is 14 tons.}
\label{fig:flux}
\end{center}
\end{figure}

For a qualitative description of the mechanism, consider coupling the
quarks to a conserved ``gluo-electric'' flux.  In usual
electromagnetism the electric field lines thus produced spread and
give rise to the inverse square law Coulombic field.  If one can
somehow eliminate massless fields, then a Coulombic spreading will no
longer be a solution to the field equations.  If in removing the
massless fields we do not destroy the Gauss law constraint that the
quarks are the sources of electric fields, the electric lines must
form into tubes of conserved flux, schematically illustrated in
Fig.~\ref{fig:flux}.  These tubes begin and end on the quarks and
their antiparticles.  The flux tube is meant to be a real physical
object carrying a finite energy per unit length.  This is the storage
medium for the linearly rising inter-quark potential.  In some sense
the reason we cannot have an isolated quark is the same as the reason
that we cannot have a piece of string with only one end.  In this
picture a baryon would require a string with three ends.  It lies in
the group theory of non-Abelian gauge fields that this peculiar state
of affairs is allowed.

Of course a length of real string can break into two, but then each
piece has itself two ends.  In the QCD case a similar phenomenon
occurs when there is sufficient energy in the flux tube to create a
quark-antiquark pair from the vacuum.  This is qualitatively what
happens when a rho meson decays into two pions.

One model for this phenomenon is a type II superconductor containing
magnetic monopole impurities.  Because of the Meissner
effect \cite{meissner}, a superconductor does not admit magnetic
fields.  However, if we force a hypothetical magnetic monopole into
the system, its lines of magnetic flux must go somewhere.  Here the
role of the ``gluo-electric'' flux is played by the magnetic field,
which will bore a tube of normal material through the superconductor
until it either ends on an anti-monopole or it leaves the boundary of
the system \cite{Abrikosov:1956sx}.  Such flux tubes have been
experimentally observed in real superconductors \cite{Hess:1989zz}.

Another example of this mechanism occurs in the bag model
\cite{Chodos:1974je}.  Here the gluonic fields are unrestricted in the
bag-like interior of a hadron, but are forbidden by {\it ad hoc}
boundary conditions from extending outside.  In attempting to extract
a single quark from a proton, one would draw out a long skinny bag
carrying the gluo-electric flux of the quark back to the remaining
constituents.

The above models may be interesting phenomenologically, but they are
too arbitrary to be considered as the basis for a fundamental theory.
In their search for a more elegant approach, theorists have been drawn
to non-Abelian gauge fields \cite{Yang:1954ek}.  This dynamical system
of coupled gluons begins in analogy with electrodynamics with a set of
massless gauge fields interacting with the quarks.  Using the freedom
of an internal symmetry, the action also includes self-couplings of
the gluons.  The bare massless fields are all charged with respect to
each other.  The confinement conjecture is that this input theory of
massless charged particles is unstable to a condensation of the vacuum
into a state in which only massive excitations can propagate.  In such
a medium the gluonic flux around the quarks should form into the flux
tubes needed for linear confinement.  While this has never been proven
analytically, strong evidence from lattice gauge calculations
indicates that this is indeed a property of these theories.

The confinement phenomenon makes the theory of the strong interactions
qualitatively rather different from the theories of the
electromagnetic and weak forces.  The fundamental fields of the
Lagrangean do not manifest themselves in the free particle spectrum.
Physical particles are all gauge singlet bound states of the
underlying constituents.  In particular, an expansion about the free
field limit is inherently crippled at the outset.  This is perhaps the
prime motivation for the lattice approach.

In the quark picture, baryons are bound states of three quarks.  Thus
the gauge group should permit singlets to be formed from three objects
in the fundamental representation.  This motivates the use of $SU(3)$
as the underlying group of the strong interactions.  This internal
symmetry must not be confused with the broken $SU(3)$ represented in
the multiplets of the eightfold way.  Ironically, one of the original
motivations for quarks has now become an accidental symmetry, arising
only because three of the quarks are fairly light.  The gauge symmetry
of importance to us now is hidden behind the confinement mechanism,
which only permits observation of singlet states.

The presentation here assumes, perhaps too naively, that the nuclear
interactions can be considered in isolation from the much weaker
effects of electromagnetism, weak interactions, and gravitation.  This
does not preclude the possible application of the techniques to the
other interactions.  Indeed, unification may be crucial for a
consistent theory of the world.  To describe physics at normal
laboratory energies, however, it is only for the strong interactions
that we are forced to go beyond well-established perturbative methods.
Thus we frame the discussion around quarks and gluons.

\subsection{Perturbation theory is not enough}

The best evidence we have for confinement in a non-Abelian gauge
theory comes by way of Wilson's
\cite{Wilson:1974sk,Wilson:1975id} formulation on a space time
lattice.  At first this prescription seems a little peculiar because
the vacuum is not a crystal.  Indeed, experimentalists work daily with
highly relativistic particles and see no deviations from the
continuous symmetries of the Lorentz group.  Why, then, have theorists
spent so much time describing field theory on the scaffolding of a
space-time lattice?

The lattice should be thought of as a mathematical trick.  It provides
a cutoff removing the ultraviolet infinities so rampant in quantum
field theory.  On a lattice it makes no sense to consider momenta with
wavelengths shorter than the lattice spacing.  As with any regulator,
it must be removed via a renormalization procedure.  Physics can only
be extracted in the continuum limit, where the lattice spacing is
taken to zero.  As this limit is taken, the various bare parameters of
the theory are adjusted while keeping a few physical quantities fixed
at their continuum values.

But infinities and the resulting need for renormalization have been
with us since the beginnings of relativistic quantum mechanics.  The
program for electrodynamics has had immense success without recourse
to discrete space.  Why reject the time-honored perturbative
renormalization procedures in favor of a new cutoff scheme?

Perturbation theory has long been known to have shortcomings in
quantum field theory.  In a classic paper, Dyson \cite{Dyson:1952tj}
showed that electrodynamics could not be analytic in the coupling
around vanishing electric charge.  If it were, then one could smoothly
continue to a theory where like charges attract rather than repel.
This would allow creating large separated regions of charge to which
additional charges would bind with more energy than their rest masses.
This would mean there is no lowest energy state; creating
matter-antimatter pairs and separating them into these regions would
provide an inexhaustible source of free energy.

The mathematical problems with perturbation theory appear already in
the trivial case of zero dimensions.  Consider the toy path integral
\begin{equation}
 Z(m,g)=\int d\phi\ \exp(-m^2\phi^2-g\phi^4).
\end{equation}
Formally expanding and naively exchanging the integral with the sum
gives
\begin{equation}
Z(m,g)=\sum c_i g^i
\end{equation}
with
\begin{equation}
c_i={(-1)^i\over i!}
\int d\phi\ e^{-m^2\phi^2} \phi^{4i}
={(-1)^i (4i)! \over m^{2i+1} i!}.
\end{equation}
A simple application of Sterling's approximation shows that at large
order these coefficients grow faster than any power.  Given any value
for $g$, there will always be an order in the series where the terms
grow out of control.  Note that by scaling the integrand we can write
\begin{equation}
Z(m,g)=g^{-1/4} \int d\phi\ \exp(-m^2\phi^2/g^{-1/2}-\phi^4).
\end{equation}
This explicitly exposes a branch cut at the origin, yet another way of
seeing the non analyticity at vanishing coupling.

Thinking non-perturbatively sometimes reveals somewhat surprising
results.  For example, the $\phi^3$ theory of massive scalar bosons
coupled with a cubic interaction seems to have a sensible perturbative
expansion after renormalization.  However this theory almost certainly
doesn't exist as a quantum system.  This is because when the field
becomes large the cubic term in the interaction dominates and the
theory has no minimum energy state.  The Euclidean path integral is
divergent from the outset since the action is unbounded both above and
below.

Perhaps even more surprising, it is widely accepted, although not
proven rigorously, that a $\phi^4$ theory of bosons interacting with a
quartic interaction also does not have a non-trivial continuum limit.
The expectation here is that with a cutoff in place, the renormalized
coupling will display an upper bound as the bare coupling varies from
zero to infinity.  If this upper bound then decreases to zero as the
cutoff is removed, then the renormalized coupling is driven to zero
and we have a free theory.

This issue is sometimes discussed in terms of what is known as the
``Landau pole''
\cite{Landau:1955ip}.  In non-asymptotically free theories, such as
$\phi^4$ and quantum electrodynamics, there is a tendency for the
effective coupling to rise with energy.  A simple analysis suggests
the possibility of the coupling diverging at a finite energy.  Not
allowing this would force the coupling at smaller energies to zero.

The importance of non-perturbative effects is well understood in a
class of two dimensional models that can be solved via a technique
known as ``bosonization'' \cite{Mandelstam:1975hb,Coleman:1975qj}.
This includes massless two dimensional electrodynamics, i.e. the
Schwinger model \cite{Schwinger:1962tp}, the sine-Gordon
model \cite{Coleman:1974bu}, and the Thirring
model \cite{Thirring:1958in}.  These solutions exploit a remarkable
mapping between fermionic and bosonic fields in two dimensions.  This
mapping is also closely related to the solution to the two dimensional
Ising model \cite{Itzykson:1989sx}.  The Schwinger model in particular
has several features in common with QCD.  First of all it confines,
i.e. the physical ``mesons'' are bound states of the fundamental
fermions.  With multiple ``flavors'' the theory has a natural current
algebra \cite{Creutz:2005ra} and the spectrum in the presense of a
small fermion mass has both multiple light ``pions'' and a heavier
eta-prime meson.  Finally, the massive theory naturally admits the
introduction of a CP violating parameter.

Returning to the main problem, QCD, we are driven to the lattice by
the necessary prevalence of non-perturbative phenomena in the strong
interactions.  Most predominant of these is confinement, but issues
related to chiral symmetry and quantum mechanical anomalies, to be
discussed in later sections, are also highly non-perturbative.  The
theory at vanishing coupling constant has free quarks and gluons and
bears no resemblance to the observed physical world of hadrons.
Renormalization group arguments explicitly demonstrate essential
singularities when hadronic properties are regarded as functions of
the gauge coupling.  To go beyond the diagrammatic approach, one needs
a non-perturbative cutoff.  Herein lies the main virtue of the
lattice, which directly eliminates all wavelengths less than the
lattice spacing.  This occurs before any expansions or approximations
are begun.

This situation contrasts sharply with the great successes of quantum
electrodynamics, where perturbation theory is central.  Most
conventional regularization schemes are based on the Feynman
expansion; some process is calculated diagrammatically until a
divergence is met, and the offending diagram is regulated.  Since the
basic coupling is so small, only a few terms give good agreement with
experiment. While non-perturbative effects are expected, their
magnitude is exponentially suppressed in the inverse of the coupling.

On a lattice, a field theory becomes mathematically well-defined and
can be studied in various ways.  Conventional perturbation theory,
although somewhat awkward in the lattice framework, should recover all
conventional results of other regularization schemes.  Discrete
space-time, however, allows several alternative approaches.  One of
these, the strong coupling expansion, is straightforward to implement.
Remarkably, confinement is automatic in the strong coupling limit
because the theory reduces to one of quarks on the end of strings with
finite energy per unit length.  While this realization of the flux
tube picture provides insight into how confinement can work,
unfortunately this limit is not the continuum limit.  The latter, as
we will see later, involves the weak coupling limit.  To study this
one can turn to numerical simulations, made possible by the lattice
reduction of the path integral to a conventional but large
many-dimensional integral.

Non-perturbative effects in QCD introduce certain interesting aspects
that are invisible to perturbation theory.  Most famous of these is
the possibility of having an explicit CP violating term in the theory.
In the classical theory this involves adding a total derivative term
to the action.  This can be rotated away in the perturbative limit.
However, as we will discuss extensively later, in the quantum theory
there are dramatic physical consequences.

Non-perturbative effects also raise subtle questions on the meaning of
quark masses.  Ordinarily the mass of a particle is correlated with
how it propagates over long distances.  This approach fails due to
confinement and the fact that a single quark cannot be isolated.  With
multiple quarks, we will also see that there is a complicated
dependence of the theory on the number of quark species.  As much of
our understanding of quantum field theory is based on perturbation
theory, several of these effects remain controversial.

This picture has evolved over many years.  One unusual result is that,
depending on the parameters of the theory, QCD can spontaneously break
CP symmetry.  This is tied to what is known as Dashen's
phenomenon \cite{Dashen:1970et}, first noted even before the days of
QCD.  In the mid 1970's, 't Hooft \cite{'tHooft:1976fv} elucidated the
underlying connection between the chiral anomaly and the topology of
gauge fields.  This connection revealed the possible explicit CP
violating term, usually called $\Theta$, the dependence on which does
not apper in perturbative expansions.  Later
Witten \cite{Witten:1980sp} used large gauge group ideas to discuss
the behavior on $\Theta$ in terms of effective Lagrangeans.
Refs.~\cite{Rosenzweig:1979ay, Arnowitt:1980ne,Nath:1980nf,
Kawarabayashi:1980dp,Kawarabayashi:1980uh,Ohta:1981ai} represent a few
of the many early studies of the effects of $\Theta$ on effective
Lagrangeans via a mixing between quark and gluonic operators.  The
topic continues to appear in various contexts; for example,
Ref.~\cite{Boer:2008ct} contains a different approach to understanding
the behavior of QCD at $\Theta=\pi$ via the framework of the
two-flavor Nambu Jona-Lasinio model.

All these issues are crucial to understanding certain subtleties with
formulating chiral symmetry on the lattice.  Much of the picture
presented here is implicit in the discussion of
Ref.~\cite{Creutz:1995wf}.  Since then the topic has raised
some controversial  issues, including 
the realization that the ambiguities
in defining quark masses precludes a vanishing up quark mass as a
solution to the strong CP problem
\cite{Creutz:2003xc}.  The non-analytic behavior in the number of
quark species reveals an inconsistency with one of the popular
algorithms in lattice gauge theory \cite{Creutz:2008nk}.  These
conclusions directly follow from the intricate interplay of the
anomaly with chiral symmetry.  The fact that some of these issues
remain disputed is much of the motivation for this review.

The discussion here is based on a few reasonably uncontroversial
assumptions.  First, QCD with $N_f$ light quarks should exist as a
field theory and exhibit confinement in the usual way.  Then we assume
the validity of the standard picture of chiral symmetry breaking
involving a quark condensate $\langle\overline\psi\psi\rangle\ne 0$.
The conventional chiral perturbation theory based on expanding in
masses and momenta around the chiral limit should make sense.  We
assume the usual result that the anomaly generates a mass for the
$\eta^\prime$ particle and this mass survives the chiral limit.
Throughout we consider $N_f$ small enough to avoid any potential
conformal phase of QCD \cite{Banks:1981nn}.

\newpage
\Section{Path integrals and statistical mechanics}
\label{pathintegrals}
Throughout this review we will be primarily focussed on the Euclidean
path integral formulation of QCD.  This approach to quantum mechanics
reveals deep connections with classical statistical mechanics.  Here
we will explore this relationship for the simple case of a
non-relativistic particle in a potential.  Starting with a partition
function representing a path integral on an imaginary time lattice, we
will see how a transfer matrix formalism reduces the problem to the
diagonalization of an operator in the usual quantum mechanical Hilbert
space of square integrable functions \cite{Creutz:1976ch}.  In the
continuum limit of the time lattice, we obtain the canonical
Hamiltonian.  Except for our use of imaginary time, this treatment is
identical to that in Feynman's early work \cite{Feynman:1948ur}.

\subsection{Discretizing time}

We begin with the Lagrangean for a free
particle of mass $m$ moving in potential $V(x)$
\begin{equation}
L(x,\dot x)=K(\dot x) + V(x)
\label{nrl}
\end{equation}
where $K(\dot x)={1\over 2} m\dot x^2$ and $\dot x$ is the time
derivative of the coordinate $x$.  Note the unconventional relative
positive sign between the two terms in Eq.~(\ref{nrl}).  This is
because we formulate the path integral directly in imaginary time.
This improves mathematical convergence, yet we are left with the usual
Hamiltonian for diagonalization.

\begin{figure}
\centering
\includegraphics[width=3in]{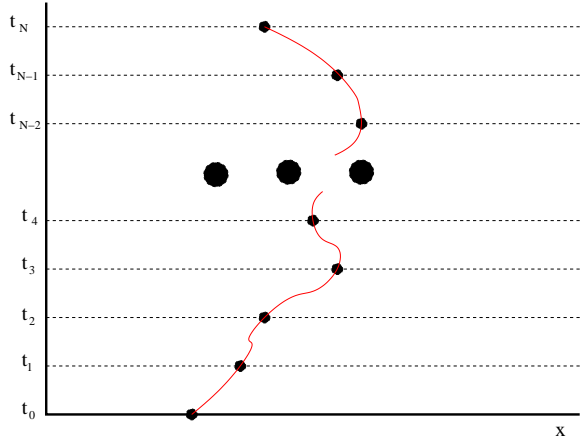}
\caption {Dividing time into a lattice of $N$ slices of timestep $a$.}
\label{slicing}
\end{figure}

For a particle traversing a trajectory $x(t)$, we have the action
\begin{equation}
S=\int dt\ L(\dot x(t),x(t)).
\end{equation}
This appears in the path integral
\begin{equation}
Z=\int(dx) e^{-S}.
\label{pathintegral}
\end{equation}
Here the integral is over all possible trajectories $x(t)$.  As it
stands, Eq.~(\ref{pathintegral}) is rather poorly defined.  To
characterize the possible trajectories we introduce a cutoff in the
form of a time lattice.  Putting our system into a temporal box of
total length $T$, we divide this interval into $N={T\over a}$ discrete
time slices, where $a$ is the timelike lattice spacing.  Associated
with the $i$'th such slice is a coordinate $x_i$.  This construction
is sketched in Figure \ref{slicing}.  Replacing the time derivative of
$x$ with a nearest-neighbor difference, we reduce the action to a sum
\begin{equation}
S=a\sum_i\left[{1\over 2}m\left({x_{i+1}-x_i\over a}\right)^2+V(x_i)\right].
\end{equation}
The integral in Eq.~(\ref{pathintegral}) is now defined as an ordinary
integral over all the coordinates
\begin{equation} 
Z=\int\left(\prod_i dx_i\right)\ e^{-S}.
\end{equation}

This form for the path integral is precisely in the form of a
partition function for a statistical system.  We have a one
dimensional polymer of coordinates $x_i$.  The action represents the
inverse temperature times the Hamiltonian of the thermal analog.  This
is a special case of a deep result, a $D$ space-dimensional quantum
field theory is equivalent to the classical thermodynamics of a $D+1$
dimensional system.  In this example, we have one degree of freedom
and $D$ is zero; for the lattice gauge theory of quarks and gluons,
$D$ is three and we work with the classical statistical mechanics of a
four dimensional system.

We will now show that the evaluation of this partition function is
equivalent to diagonalizing a quantum mechanical Hamiltonian obtained
from the action via canonical methods.  This is done with the use of
the transfer matrix.

\subsection{The transfer matrix}

The key to the transfer-matrix analysis is to note that the local
nature of the action permits us to write the partition function as a
matrix product
\begin{equation}
Z=\int\prod_i dx_i\ T_{x_{i+1},x_i}
\end{equation}
where the transfer-matrix elements are
\begin{equation}
T_{x^\prime,x}=\exp\left[-{m\over 2a}(x^\prime-x)
-{a\over 2}(V(x^\prime)+V(x))\right].
\label{transfer}
\end{equation}
The transfer matrix itself is an operator in the Hilbert space of
square integrable functions with the standard inner product
\begin{equation}
\langle\psi^\prime\vert\psi\rangle=\int dx {\psi^\prime}^*(x)\psi(x).
\end{equation}
We introduce the non-normalizable basis states $\vert x \rangle$ such
that
\begin{eqnarray}
&&\vert\psi\rangle=\int  dx\ \psi(x)\ \vert x \rangle\\
&&\langle x^\prime \vert x \rangle=\delta(x^\prime-x)\\
&&1=\int dx\ \vert x\rangle\langle x\vert.
\end{eqnarray}
Acting on the Hilbert space are the canonically conjugate operators
$\hat p$ and $\hat x$ that satisfy
\begin{eqnarray}
&&\hat x\vert x\rangle=x\vert x\rangle\cr
&&[{\hat p,\hat x}]=-i\cr
&&e^{-i\hat py}\vert x\rangle=\vert x+y\rangle.
\end{eqnarray}
The operator $T$ is defined via its matrix elements
\begin{equation}
\langle x^\prime\vert T\vert x\rangle
=T_{x^\prime,x}
\end{equation}
where $T_{x^\prime,x}$ is given in Eq.~(\ref{transfer}).  With periodic
boundary conditions on our lattice of $N$ sites, the path integral is
compactly expressed as as a trace over the Hilbert space
\begin{equation}
Z={\rm Tr}\ T^N.
\end{equation}

Expressing $T$ in terms of the basic operators $\hat p,\hat x$ gives
\begin{equation} 
T=\int dy\ e^{-y^2/(2a)}\ e^{-aV(\hat x)/2}\ e^{-i\hat py}
\ e^{-aV(\hat x)/2}.
\end{equation}
To prove this, check that the right hand side has the appropriate
matrix elements.  The integral over $y$ is Gaussian and gives 
\begin{equation}
T=\left({2\pi a\over m}\right)^{1/2} \ e^{-aV(\hat x)/2}
e^{-a\hat p^2/(2m)}
e^{-aV(\hat x)/2}.
\end{equation}
The connection with the usual quantum mechanical Hamiltonian appears
in the small lattice spacing limit.  When $a$ is small, the exponents
in the above equation combine to give
\begin{equation}
T=\left({2\pi a\over m}\right)^{1/2}\ e^{-aH+O(a^2)}
\end{equation}
with 
\begin{equation}
H={\hat p^2\over 2m}+V(\hat x).
\end{equation}
This is just the canonical Hamiltonian operator following from our
starting Lagrangean.

The procedure for going from a path-integral to a Hilbert-space
formulation of quantum mechanics consists of three steps.  First
define the path integral with a discrete time lattice.  Then construct
the transfer matrix and the Hilbert space on which it operates.
Finally, take the logarithm of the transfer matrix and identify the
negative of the coefficient of the linear term in the lattice spacing
as the Hamiltonian.  Physically, the transfer matrix propagates the
system from one time slice to the next.  Such time translations are
generated by the Hamiltonian.

The eigenvalues of the transfer matrix are related to the energy
levels of the quantum system.  Denoting the $i$'th eigenvalue of $T$
by $\lambda_i$, the path integral or partition function
becomes 
\begin{equation} 
Z=\sum_i\lambda_i^N.  
\end{equation} 
As the number of time slices goes to infinity, this expression is
dominated by the largest eigenvalue 
$\lambda_0$ \begin{equation}
  Z=\lambda_0^N\times[1+O(\exp[-N\log(\lambda_0/\lambda_1)])].  
\end{equation}
In statistical mechanics the thermodynamic properties of a system
follow from this largest eigenvalue.  In ordinary quantum mechanics
the corresponding eigenvector is the lowest eigenstate of the
Hamiltonian.  This is the ground state or, in field theory, the vacuum
$\vert 0\rangle$.  Note that in this discussion, the connection
between imaginary and real time is trivial.  Whether the generator of
time translations is $H$ or $iH$, we have the same operator to
diagonalize.

In statistical mechanics one is often interested in correlation
functions between the statistical variables at different points.  This
corresponds to a study of the Green's functions of the corresponding
field theory.  These are obtained upon insertion of polynomials of the
fundamental variables into the path integral.  

An important feature of the path integral is
that a typical path is
non-differentiable \cite{feynmanhibbs,Creutz:1980gp}.  Consider the
discretization of the time derivative
\begin{equation}
\dot x \sim {x_{i+1}- x_i \over a}.
\end{equation}
The kinetic term in the path integral controls how close the fields
are on adjacent sites.  Since this appears as simple Gaussian factor
$\exp(-(x_{i+1}- x_i)^2m/a)$ we see that
\begin{equation}
{1\over 2} m \langle \dot x^2\rangle=O(1/ma).
\end{equation} 
This diverges as the lattice spacing goes to zero.

One can obtain the average kinetic energy in other ways, for example
through the use of the virial theorem or by point splitting.  However,
the fact that the typical path is not differentiable means that one
should be cautious about generalizing properties of classical fields
to typical configurations in a numerical simulation.  We will see that
such questions naturally arise when considering the topological
properties of gauge fields.

\newpage
\Section{Quark fields and Grassmann integration}

Of course since we are dealing with a theory of quarks, we need
additional fields to represent them.  There are subtle complications
in defining their action on a lattice; we will go into these in some
detail later.  For now we just assume the quark fields $\psi$ and
$\overline\psi$ are associated with the sites of the lattice and carry
suppressed spinor, flavor, and color indices.  Being generic, we take
an action which is a quadratic form in these fields $\overline \psi
(D+m) \psi$.  Here we formally separate the kinetic and mass
contributions.  For the path integral, we are to integrate over $\psi$
and $\overline\psi$ as independent Grassmann variables.  Thus $\psi$
and $\overline\psi$ on any site anti-commutes with $\psi$ and
$\overline\psi$ on any other site.

Grassmann integration is defined formally as a linear function
satisfying a shift symmetry.  Consider a single Grassmann variable
$\psi$.  Given any function $f$ of $\psi$, we impose
\begin{equation}
\label{gint}
\int d\psi \ f(\psi)=\int d\psi \ f(\psi+\chi)
\end{equation}
where $\chi$ is another fixed Grassmann variable.  Since the square of
any Grassmann variable vanishes, we can expand $f$ in just two terms
\begin{equation}
f(\psi)=\psi a + b.
\end{equation}
Assuming linearity on inserting this into Eq.~(\ref{gint}) gives
\begin{equation}
\left(\int d\psi \ \psi\right) a+ \left(\int d\psi \ 1\right) b
=\left(\int d\psi \ \psi\right) a + \left(\int d\psi \ 1\right)(a \chi + b).
\end{equation}
This immediately tells us $\int d\psi \ 1$ must vanish.  The
normalization of $\int d\psi \ \psi$ is still undetermined; the
convention is to take this to be unity.  Thus the basic Grassmann
integral of a single variable is completely determined by
\begin{eqnarray}
\int d\psi \ \psi = 1\\
\int d\psi \ 1 = 0.
\end{eqnarray}

Note that the rule for Grassmann integration seems quite similar to
what one would want for differentiation.  Indeed, it is natural to
define derivatives as anticommuting objects that satisfy
\begin{eqnarray}
{d\over d\psi} \ \psi = 1\\
{d\over d\psi} \ 1 = 0.
\end{eqnarray}
This is exactly the same rule as for integration.  For Grassmann
variables, integration and differentiation are the same thing.  It is
a convention what we call it.  For the path integral it is natural to
keep the analogy with bosonic fields and refer to integration.  On the
other hand, for both fermions and bosons we refer to differentiation
when using sources in the path integral as a route to correlation
functions.

We can make changes of variables in a Grassmann integration in a
similar way to ordinary integrals.  For example, if we want to change
from $\psi$ to $\chi=a\psi$, the above integration rules imply
\begin{equation}
\int d(\psi) f(a\psi)=a\int d(\chi) f(\chi)
\end{equation}
or simply $d(a\psi)= d\chi={1\over a} d\psi$.  We see that the primary
difference from ordinary integration is that the Jacobean is
inverted.  If we consider a multiple integral and take
$\chi=M\psi$ with $M$ a matrix, the transformation generalizes to
\begin{equation}
d\chi=d(M\psi)={1\over \det (M)}\ d\psi.
\end{equation}
A particularly important consequence is that we can formally evaluate
the Gaussian integrals that appear in the path integral as
\begin{equation}
\label{fermiint}
\int d\psi d\overline\psi\ \exp\left(\overline\psi(D+m)\psi\right)
={1\over \det (D+m)}=\det\left((D+m)^{-1}\right).
\end{equation}
The normalization is fixed by the earlier conventions.  Note that in
the path integral formulation $\psi$ and $\overline\psi$ represent
independent Grassmann fields; in the next subsection we will discuss
the connection between these and the canonical anti-commutation
relations for fermion creation and annihilation operators in a quantum
mechanical Hilbert space..

In practice Eq.~(\ref{fermiint}) allows one to replace fermionic
integrals with ordinary commuting fields $\phi$ and $\overline\phi$ as
\begin{equation}
\int d\psi d\overline\psi\ \exp\left(\overline\psi(D+m)\psi\right)
\propto \int d\phi d\overline\phi\ 
\exp\left(\overline\phi(D+m)^{-1}\phi\right).
\end{equation}
This forms the basis for most Monte Carlo algorithms, although the
intrinsic need to invert the large matrix $D+m$ makes such simulations
extremely computationally intensive.  This approach is, however,
still much less demanding than any known way to deal directly with the
Grassmann integration in path integrals \cite{Creutz:1998ee}.

\subsection{Fermionic transfer matrices}

The concept of continuity is lost with Grassman variables.  There is
no meaning to saying that fermion fields at nearby sites are near
each other.  This is closely tied to the doubling issues that we will
discuss later.  But is also raises interesting complications in
relating Hamiltonian quantum mechanics with the Euclidian formulation
involving path integrals.  Here we will go into how this connection is
made with an extremely simple zero space-dimensional model.

Anti-commutation is at the heart of fermionic behavior.  This is true
in both the Hamiltonian operator formalism and the Lagrangean path
integral, but in rather complementary ways.  Starting with a
Hamiltonian approach, if an operator $a^\dagger$ creates a fermion in
some normalized state on the lattice or the continuum, it satisfies
the basic relation
\begin{equation}
[a,a^\dagger]_+\equiv a a^\dagger + a^\dagger a =1.
\end{equation}
This contrasts sharply with the fields in a path integral, which all
anti-commute 
\begin{equation}
[\chi,\chi^\dagger]_+=0.
\end{equation}
The connection between the Hilbert space approach and the path
integral appears through the transfer matrix formalism.  For bosonic
fields this is straightforward \cite{Creutz:1976ch}, but for fermions
certain subtleties arise related to the doubling issue \cite
{Creutz:1999zy}.

To be more precise, consider a single fermion state created by the
operator $a^\dagger$, and an antiparticle state created by another
operator $b^\dagger$.  For an extremely simple model, consider the
Hamiltonian
\begin{equation}
H=m (a^\dagger a+b^\dagger b).
\end{equation}
Here $m$ can be thought of as a ``mass'' for the particle.
What we want is an exact path
integral expression for the partition function
\begin{equation}
Z={\rm Tr} e^{-\beta H}.
\end{equation}
Of course, since the Hilbert space generated by $a$ and $b$ has only
four states, this is trivial to work out: $
Z=1+2e^{-\beta m}+e^{-2\beta m} $.  However, we want this in a form
that easily generalizes to many variables.

The path integral for fermions uses Grassmann variables.  We introduce
a pair of such, $\chi$ and $\chi^\dagger$, which will be connected to
the operator pair $a$ and $a^\dagger$, and another pair, $\xi$ and
$\xi^\dagger$, for $b$, $b^\dagger$.  All the Grassmann variables
anti-commute.  Integration over any of them is determined by the
simple formulas mentioned earlier
\begin{equation}
\int d\chi \ 1 = 0\ ; \qquad \int d\chi\ \chi = 1.
\end{equation}
For notational simplicity combine the individual Grassmann variables
into spinors
\begin{equation}\matrix{
\psi = \pmatrix{\chi\cr\xi^\dagger\cr};&
\psi^\dagger = \pmatrix{\chi^\dagger&\xi\cr}.\cr
}
\end{equation}
To make things appear still more familiar, introduce a ``Dirac
matrix''
\begin{equation}
\gamma_0=\pmatrix{1&0\cr 0 & -1\cr}
\end{equation}
and the usual
\begin{equation}
\overline\psi=\psi^\dagger\gamma_0.
\end{equation}
Then we have
\begin{equation}
\overline \psi \psi = \chi^\dagger \chi + \xi^\dagger \xi.
\end{equation}
where the minus sign from using $\xi^\dagger$ rather than $\xi$ in
defining $\psi$ is removed by the $\gamma_0$ factor.  The temporal
projection operators
\begin{equation}
P_\pm={1\over 2}(1\pm\gamma_0)
\end{equation}
arise when one considers the fields at
two different locations
\begin{equation}
\chi_i^\dagger  \chi_j +\xi_i^\dagger  \xi_j =
\overline \psi_i P_+\psi_j+\overline \psi_j P_-\psi_i.
\end{equation}
The indices $i$ and $j$ will soon label the ends of a temporal hopping
term; this formula is the basic transfer matrix justification for the 
Wilson projection operator formalism that we will return to in later
sections.

\subsection{Normal ordering and path integrals}

For a moment ignore the antiparticles and consider some general
operator $f(a,a^\dagger)$ in the Hilbert space.  How is this
related to an integration in Grassmann space?  To proceed we need a
convention for ordering the operators in $f$.  We adopt the usual
normal ordering definition with the notation $:f(a,a^\dagger):$
meaning that creation operators are placed to the left of destruction
operators, with a minus sign inserted for each exchange.  In this case
a rather simple formula gives the trace of the operator as a Grassmann
integration
\begin{equation}
\label{dchi}
{\rm Tr}\ :f(a,a^\dagger):\ = \int d\chi d\chi^\dagger 
e^{2\chi^\dagger\chi} f(\chi,\chi^\dagger).
\end{equation}
To verify, just check that all elements of the complete set of
operators $\{1,a,a^\dagger,a^\dagger a\}$ work.  However, this formula
is actually much more general; given a set of many Grassmann variables
with one pair associated with each of several fermion states, this
immediately generalizes to the trace of any normal ordered operator
acting in a many fermion Hilbert space.

What about a product of several normal ordered operators?  This leads
to the introduction of multiple sets of Grassmann variables and the
general formula
\begin{eqnarray}
&{\rm Tr}\ & 
\left(:f_1(a^\dagger,a):\ :f_2(a^\dagger,a): \ldots :f_n(a^\dagger,a):
\right)\cr
&=&\int d\chi_1\ d\chi_1^*\ldots d\chi_n\ d\chi_n^*
\ e^{\chi_1^*(\chi_1+\chi_n)} e^{\chi_2^*(\chi_2-\chi_1)} 
 \ldots e^{\chi_n^*(\chi_n-\chi_{n-1})} \cr 
&&\qquad \times f_1(\chi_1^*,\chi_1)f_2(\chi_2^*,\chi_2)
 \ldots f_n(\chi_n^*,\chi_n).
\end{eqnarray}
The positive sign on $\chi_n$ in the first exponential factor
indicates the natural occurrence of anti-periodic boundary conditions;
{\i.e.} we can define $x_0=-x_n$.  With just one factor, this formula
reduces to Eq.~(\ref{dchi}).  Note how the ``time derivative'' terms
are ``one sided;'' this is how doubling is eluded.

This exact relationship provides the starting place for converting our
partition function into a path integral.  The simplicity of our
example Hamiltonian allows this to be done exactly at every stage.
First we break ``time'' into a number $N$ of ``slices''
\begin{equation}
Z={\rm Tr} \left( e^{-\beta H/N}\right)^N. 
\end{equation}  
Now we need normal ordered factors for the above formula.  For this we
use
\begin{equation}
e^{\alpha a^\dagger a} = 1+(e^\alpha-1) a^\dagger a
=\ :e^{(e^\alpha-1) a^\dagger a }:\ ,
\end{equation}
which is true for arbitrary $\alpha$.\footnote{The definition of
normal ordering gives
$:(a^\dagger a)^2:=0.$}
This is all the machinery we need to write
\begin{equation}
Z=\int (d\psi d\overline\psi) e^{S}
\end{equation}
where
\begin{equation}
S=\sum_{i=1}^n 
\overline\psi_n 
(e^{-\beta m/N}-1)
\psi_n
+\overline\psi_n P_+ \psi_{n-1}
+\overline\psi_{n-1} P_ - \psi_{n}.
\end{equation}
Note how the projection factors of $P_\pm$ automatically appear for
handling the reverse convention of $\chi$ versus $\xi$ in our field
$\psi$.  Expanding the first term gives the $-\beta m/N$ factor
appearing in the Hamiltonian form for the partition function.

It is important to realize that if we consider the action as a
generalized matrix connecting fermionic variables
\begin{equation}
S=\overline\psi M \psi,
\end{equation}
the matrix $M$ is not symmetric.  The upper components propagate
forward in time, and the lower components backward.  Even though our
Hamiltonian was Hermitean, the matrix appearing in the corresponding
action is not.  With further interactions, such as gauge field
effects, the intermediate fermion contributions to a general path
integral may not be positive, or even real.  Of course the final
partition function, being a trace of a positive definite operator, is
positive.  Keeping the symmetry between particles and antiparticles
results in a real fermion determinant, which in turn is positive for
an even number of flavors.  We will later see that some rather
interesting things can happen with an odd number of flavors.

For our simple Hamiltonian, this discussion has been exact.  The
discretization of time adds no approximations since we could do the
normal ordering by hand.  In general with spatial hopping or more
complex interactions, the normal ordering can produce extra terms
going as $O(1/N^2)$.  In this case exact results require a limit of a
large number of time slices, but this is a limit we need anyway to
reach continuum physics.

\newpage
\Section{Lattice gauge theory}

Lattice gauge theory is currently the dominant path to understanding
non-perturbative effects.  As formulated by Wilson, the lattice cutoff
is quite remarkable in preserving many of the basic ideas of a gauge
theory.  But just what is a gauge theory anyway?  Indeed, there are
many ways to think of what is meant by this concept.

At the most simplistic level, a Yang-Mills \cite{Yang:1954ek} theory is
nothing but an embellishment of electrodynamics with isospin symmetry.
Being formulated directly in terms of the underlying gauge group, this
is inherent in lattice gauge theory from the start.

At a deeper level, a gauge theory is a theory of phases acquired by a
particle as it passes through space time.  In electrodynamics the
interaction of a charged particle with the electromagnetic field is
elegantly described by the wave function acquiring a phase from the
gauge potential.  For a particle at rest, this phase is an addition to
its energy proportional to the scalar potential.  The use of group
elements on lattice links directly gives this connection; the phase
associated with some world-line is the product of these elements along
the path in question.  For the Yang-Mills theory the concept of
``phase'' is generalized to a rotation in the internal symmetry group.

A gauge theory is also a theory with a local symmetry.  Gauge
transformations involve arbitrary functions of space time.  Indeed,
with QCD we have an independent $SU(3)$ symmetry at each point of
space time.  With the Wilson action formulated in terms of products of
group elements around closed loops, this symmetry remains exact even
with the cutoff in place.

In perturbative discussions, the local symmetry forces a gauge fixing
to remove a formal infinity coming from integrating over all possible
gauges.  For the lattice formulation, however, the use of a compact
representation for the group elements means that the integration over
all gauges becomes finite.  To study gauge invariant observables, no
gauge fixing is required to define the theory.  Of course gauge fixing
can still be done, and must be introduced to study more conventional
gauge variant quantities such as gluon or quark propagators.  But
physical quantities should be gauge invariant; so, whether gauge
fixing is done or not is irrelevant for their calculation.

One aspect of a continuum gauge theory that the lattice does not
respect is how a gauge field transforms under Lorentz transformations.
In a continuum theory the basic vector potential can change under a
gauge transformation when transforming between frames.  For example,
the Coulomb gauge treats time in a special way, and a Lorentz
transformation can change which direction represents time.  The
lattice, of course, breaks Lorentz invariance, and thus this concept
loses meaning.

Here we provide only a brief introduction to the lattice approach to a
gauge theory.  For more details one should turn to one of the several
excellent books on the subject
\cite{Creutz:1984mg,Gattringer:2010zz,Montvay:1994cy,DeGrand:2006zz,
Rothe:2005nw}.  We postpone until later a discussion of issues related
to lattice fermions.  These are more naturally understood after
exploring some of the peculiarities that must be manifest in any
non-perturbative formulation.

\subsection{Link variables}

Lattice gauge theory is closely tied to two of the above concepts; it
is a theory of phases and it exhibits an exact local symmetry.  Indeed
it is directly defined in terms of group elements representing the
phases acquired by quarks as they hop around the lattice.  The basic
variables are phases associated with each link of a four dimensional
space time lattice.  For non-Abelian case, these variables become an
elements of the gauge group, i.e. $U_{ij}\in SU(3)$ for the strong
interactions.  Here $i$ and $j$ denote the sites being conneted by the
link in question.  We suppress the group indices to keep the notation
under control.  These are three by three unitary matrices satisfying
\begin{equation}
U_{ij}=U_{ji}^{-1}= (U_{ji})^\dagger.
\end{equation}
The analogy with continuum vector fields $A_\mu$ is
\begin{equation}
U_{i,i+e_\mu} = e^{i a g_0 A_\mu}.
\end{equation}
Here $a$ represents the lattice spacing and $g_0$ is the bare
coupling considered at the scale of the cutoff.

In the continuum, a non-trivial gauge field arises when the curl (in a
four dimensional sense) of the potential is non zero.  This in turn
means the phase factor around a small closed loop is not unity.  The
smallest closed path in the lattice is a ``plaquette,'' or elementary
square.  Consider the phase corresponding to one such
\begin{equation}
U_P=U_{12}U_{23}U_{34}U_{41}
\end{equation}
where sites 1 through 4 run around the square in question.  In an
intuitive sense this measures the flux through this plaquette
$U_P\sim \exp(i a^2 g_0 F_{\mu,\nu})$.  This motivates using this
quantity to define an action.  For this,
look at the real part of the trace of $U_P$
\begin{equation}
{\rm Re Tr} U_P =
N - a^4 g_0^2\  {\rm Tr}\ F_{\mu\nu}F_{\mu\nu} + O(a^6).
\end{equation}
The overall added constant $N$ is physically irrelevant.  This leads
directly to the Wilson gauge action
\begin{equation}
S(U)=-\sum_P {\rm Re Tr} U_P.
\end{equation}

Now we have our gauge variables and an action.  To proceed we turn to
a path integral as an integral over all fields of the exponentiated
action.  For a Lie group, there is a natural measure that we will
discuss shortly.  Using this measure, the path integral is
\begin{equation}
Z=\int (dU) e^{-\beta S}
\end{equation}
where $(dU)$ denotes integration over all link variables.  This leads
to the conventional continuum expression ${1\over 2}\int d^4x\
{\rm Tr}\ F_{\mu\nu}F_{\mu\nu}$ if we choose $\beta=2N/g_0^2$ for group $SU(N)$
and use the conventionally normalized bare coupling $g_0$.

Physical correlation functions are obtained from the path integral as
expectation values.  Given an operator $B(U)$ which depends on the
link variables, we have
\begin{equation}
\langle B \rangle = {1\over Z}\int (dU) B(U) e^{-\beta S(U)}.
\end{equation} 
Because of the gauge symmetry, discussed further later, this only
makes physical sense if $B$ is invariant under gauge transformations.

\subsection {Group Integration}

The above path integral involves integration over variables which are
elements of the gauge group.  For this we use a natural measure with a
variety of nice properties.  Given any function $f(g)$ of the group
elements $g\in G$, the Haar measure is constructed so as to be
invariant under ``translation'' by an arbitrary fixed element $g_1$ of
the group
\begin{equation}
\label{groupinvariance}
\int dg\ f(g)=\int dg\ f(g_1g).
\end{equation}
For a compact group, as for the $SU(N)$ relevant to QCD, this is
conventionally normalized so that $\int dg\ 1=1$.  These simple
properties are enough for the measure to be uniquely determined.

An explicit representation for this integration measure is almost
never needed, but fairly straightforward to write down formally.
Suppose a general group element is parameterized by some variables
$\alpha_1, ... \alpha_n$.  Considering here the case $SU(N)$, there
are $n=N^2-1$ such parameters.  Then assume we know some region $R$ in
this parameter space that covers the group exactly once.  Define the
$n$ dimensional fully antisymmetric tensor $\epsilon_{i_1,
\ldots i_n}$ such that, say, $\epsilon_{1,2,...n}=1$.
Now look at the integral
\begin{equation}
\label{measure}
I=A\int_R \{d\alpha\}\ f(g(\vec\alpha))\
\epsilon_{i_1, ...i_n} {\rm Tr}\left( (g^{-1}\partial_{i_1} g)
... (g^{-1}\partial_{i_n} g)\right).
\end{equation}
This has the required invariance properties of
Eq.~(\ref{groupinvariance}).  The properties of a group imply there
should be a set of parameters $\alpha^\prime$ depending on $\alpha$
such that $g_1 g(\vec\alpha) = g(\vec\alpha^\prime)$.  If we change
the integration variables from $\alpha$ to $\alpha^\prime$, then the
epsilon factor generates exactly the Jacobian needed for this variable
change.  The normalization factor $A$ is fixed by the above condition
$\int dg\ 1 = 1.$ Once this is done, we have the invariant measure.
The above form for the measure will appear again when we discuss
topological issues for gauge fields in Section \ref{classical}.

Several interesting properties of the Haar measure are easily found.
If the group is compact, the left and right measures are equal
\begin{equation}
\int d_R g\ f(g)=\int d_R g\ f(gg_1)
=\int d_L g_1\ \int d_Rg_2\ f(g_2g_1)
=\int d_L g\ f(g).
\end{equation}
This also shows the measure is unique since any left invariant measure
could be used.  (For a non-compact group the normalization can
differ.)  A similar argument shows
\begin{equation}
\int dg\ f(g)=\int dg\ f(g^{-1}).
\end{equation}

For a discrete group, $\int dg$ is simply a sum over the elements.
For $U(1)=\{e^{i\theta}|0\le\theta<2\pi\}$ the measure is simply an
integral over the circle
\begin{equation}
\int dg\ f(g)=\int_0^{2\pi} {d\theta\over 2\pi} f(e^{i\theta}).
\end{equation}
For $SU(2)$, group elements take the form
\begin{equation}
g=\{a_0+i\vec a\cdot\vec\sigma | a_0^2+\vec a^2=1\}
\end{equation} 
and the measure is 
\begin{equation}
\int dg\ f(g)={1\over \pi^2}\int d^4a\ f(g)\delta(a^2-1).
\end{equation}
In particular, $SU(2)$ is a 3-sphere.

Some integrals are easily evaluated if we realize that group
integration picks out the ``singlet'' part of a function.  Thus
\begin{equation}
\int dg R_{ab}(g) = 0
\end{equation}
where $R(g)$ is any irreducible matrix representation other than the
trivial one, $R=1$.  For the group $SU(3)$ one can write
\begin{eqnarray}
&&\int dg\ {\rm Tr}g\ {\rm Tr}g^\dagger = 1\\
&&\int dg\  ({\rm Tr}g)^3 = 1
\end{eqnarray}
from the well known formulae $3\otimes\overline 3= 1\oplus 8$ and
$3\otimes 3\otimes 3= 1\oplus 8\oplus 8\oplus 10$.

A simple integral useful for the strong coupling expansion is
\begin{equation}
\int dg\ g_{ij}\ (g^\dagger)_{kl}=I_{ijkl}.
\end{equation}
The group invariance says we can multiply the indices arbitrarily by a
group element on the left or right. There is only one combination of
the indices that can survive for $SU(N)$
\begin{equation}
I_{ijkl}= \delta_{il} \delta_{jk}/N.
\end{equation}
The normalization here is fixed since tracing over $jk$
should give the identity matrix.
Another integral that has a fairly simple form is
\begin{equation}
\int dg\ g_{i_1j_1}\ g_{i_2j_2} \ldots g_{i_Nj_N}
={1\over N!}\epsilon_{i_1\ldots i_N}\epsilon_{j_1\ldots j_N}.
\end{equation}
This is useful for studying baryons in the strong coupling regime.

\subsection {Gauge invariance}

The action of lattice gauge theory has an exact local symmetry.  If we
associate an arbitrary group element $g_i$ with each site $i$ of the
lattice, the action is unchanged if we replace
\begin{equation}
U_{ij}\rightarrow g_i^{-1} U_{ij} g_j.
\end{equation}
One consequence is that no link can have a vacuum expectation
value \cite{Elitzur:1975im}.
\begin{equation}
\langle U_{ij} \rangle= g_i^{-1} \langle U_{ij}\rangle  g_j=0.
\end{equation}
Generalizing this, unless one does some sort of gauge fixing, the
correlation between any two separated $U$ matrices is zero.  Indeed
many things familiar from perturbation theory often vanish without
gauge fixing, including such fundamental objects as quark and gluon
propagators!

An interesting consequence of gauge invariance is that we can forget
to integrate over a tree of links in calculating any gauge invariant
observable \cite{Creutz:1999zy}.  An axial gauge represents fixing all
links pointing in a given direction.\footnote{Using a tree with small
highly-serrated leaves might be called a ``light comb gauge.''}  Note
that this sort of gauge fixing allows the reduction of two dimensional
gauge theories to one dimensional spin models.  To see this, pick the
tree to be a non-intersecting spiral of links starting at the origin
and extending out to the boundary.  Links transverse to this spiral
interact exactly as a one dimensional system.  This also shows that
two dimensional gauge theories are exactly solvable.  Construct the
transfer matrix along this one dimensional system.  The partition
function is the sum of the eigenvalues of this matrix each raised to
the power of the volume of the system.

The trace of any product of link variables around a closed loop is the
famous Wilson loop.  These quantities are by construction gauge
invariant and are the natural observables in the lattice theory.  The well
known criterion for confinement is whether the expectation of the
Wilson loop decreases exponentially in the loop area.  

More general gauges can be introduced using an analogue of the
Fadeev-Popov factor.  If $B(U)$ is gauge invariant, then
\begin{equation}
\langle B \rangle={1\over Z} \int d(U) e^{-S} B(U)
= {1\over Z} \int d(U) e^{-S} B(U) f(U)/\phi(U)
\end{equation}
where $f(U)$ is an arbitrary gauge fixing function and 
\begin{equation}
\phi(U) = \int (dg) f(g_i^{-1} U_{ij} g_j)
\end{equation}
is the integral of the gauge fixing function $f$ over all gauges.  A
possible gauge fixing scheme might be to ask that some function $h$ of
the links vanishes.  In this case we could take $f=\delta(h)$ and then
$\phi=\int (dg) \delta(h)$.  The integral of a delta function of
another function is generically a determinant $\phi=\det (\partial
g/\partial h)$.  A determinant can generally be written as an integral
over a set of auxiliary ``ghost'' fields.  Pursuing this yields the
usual Fadeev-Popov picture \cite{Faddeev:1967fc}.

Gauge fixing in the continuum raises several subtle issues if one
wishes to go beyond perturbation theory.  Given some gauge fixing
condition $h=0$ and the corresponding $f=\delta(h)$, it is desirable
that this function vanish only once on any gauge orbit.  Otherwise one
should correct for the over counting due to what are known of as
``Gribov copies'' \cite{Singer:1978dk}.  This turns out to be
non-trivial with most perturbative gauges in practice, such as the
Coulomb or Landau gauge.  One of the great virtues of the lattice
approach is that by not fixing the gauge, these issues are
sidestepped.

On the lattice gauge fixing is unnecessary and usually not done if one
only cares about measuring gauge invariant quantities such as Wilson
loops.  But this does have the consequence that the basic lattice
fields are far from continuous.  The correlation between link
variables at different locations vanishes.  The locality of the gauge
symmetry literally means that there is an independent symmetry at each
space time point.  If we consider a quark-antiquark pair located at
different positions, they transform under unrelated symmetries.  Thus
concepts such as separating the potential between quarks into singlet and
octet parts are meaningless unless some gauge fixing is imposed.

\subsection {Numerical simulation}

Monte Carlo simulations of lattice gauge theory have come to dominate
the subject.  We will introduce some of the basic algorithms in
Section \ref{montecarlo}.  The idea is to use the analogy to
statistical mechanics to generate in a computer memory sets of gauge
configurations weighted by the exponentiated action of the path
integral.  This is accomplished via a Markov chain of small weighted
changes to a stored system.  Various extrapolations are required to
obtain continuum results; the lattice spacing needs to be taken to
zero and the lattice size to infinity.  Also, such simulations become
increasingly difficult as the quark masses become small; thus,
extrapolations in the quark mass are generally necessary.  It is not
the purpose of this review to cover these techniques; indeed, the
several books mentioned at the beginning of this section are readily
available.  In addition, the proceedings of the annual Symposium on
Lattice Field Theory are available on-line for the latest results.

While confinement is natural in the strong coupling limit of the
lattice theory, we will shortly see that this is not the region of
direct physical interest.  For this a continuum limit is necessary.
The coupling constant on the lattice represents a bare coupling
defined at a length scale given by the lattice spacing.  Non-Abelian
gauge theories possess the property of asymptotic freedom, which means
that in the short distance limit the effective coupling goes to zero.
This remarkable phenomenon allows predictions for the observed scaling
behavior in deeply inelastic processes.  The way quarks expose
themselves in high energy collisions was one of the original
motivations for a non-Abelian gauge theory of the strong interactions.

In addition to enabling perturbative calculations at high energies,
the consequences of asymptotic freedom are crucial for numerical
studies via the lattice approach.  As the lattice spacing goes to
zero, the bare coupling must be taken to zero in a well determined
way.  Because of asymptotic freedom, we know precisely how to adjust
our simulation parameters to take take the continuum limit!

In terms of the statistical analogy, the decreasing coupling takes us
away from high temperature and towards the low temperature regime.
Along the way a general statistical system might undergo dramatic
changes in structure if phase transitions are present.  Such
qualitative shifts in the physical characteristics of a system can
only hamper the task of demonstrating confinement in the non-Abelian
theory.  Early Monte Carlo studies of lattice gauge theory have
provided strong evidence that such troublesome transitions are avoided
in the standard four dimensional $SU(3)$ gauge theory of the nuclear
force \cite{Creutz:1980zw}.

Although the ultimate goal of lattice simulations is to provide a
quantitative understanding of continuum hadronic physics, along the
way many interesting phenomena arise which are peculiar to the
lattice.  Non-trivial phase structure does occur in a variety of
models, some of which do not correspond to any continuum field theory.
We should remember that when the cutoff is still in place, the lattice
formulation is highly non-unique.  One can always add additional terms
that vanish in the continuum limit.  In this way spurious transitions
might be alternatively introduced or removed.  Physical results
require going to the continuum limit.

\subsection {Order parameters}

Formally lattice gauge theory is like a classical statistical
mechanical spin system.  The spins $U_{ij}$ are elements of a gauge
group $G$.  They are located on the bonds of our lattice.  Can this
system become ``ferromagnetic''?  Indeed, as mentioned above, this is
impossible since $\langle U\rangle=0$ follows from the links
themselves not being gauge invariant \cite{Elitzur:1975im}.

But we do expect some sort of ordering to occur in the $U(1)$ theory.
If this is to describe physical photons, there should be a phase with
massless particles.  Strong coupling expansions show that for large
coupling this theory has a mass gap \cite{Wilson:1974sk}. Thus a phase
transition is expected, and has been observed in numerical simulations
\cite{Creutz:1979zg}.  Exactly how this ordering occurs remains
somewhat mysterious; indeed, although people often look for a
``mechanism for confinement,'' it might be interesting to rephrase
this question to ``how does a theory such as electromagnetism avoid
confinement.''

The standard order parameter for gauge theories and confinement
involves the Wilson loop mentioned above.  This is the trace of the
product of link variables multiplied around a closed loop in
space-time.  If the expectation of such a loop decreases exponentially
with the area of the loop, we say the theory obeys an area law and is
confining.  On the other hand, a decrease only as the perimeter
indicates an unconfined theory.  This order parameter by nature is
non-local; it cannot be measured without involving arbitrarily long
distance correlations.  The lattice approach is well known to give the
area law in the strong coupling limit of the pure gauge theory.
Unfortunately, with dynamical quarks this ceases to be a useful
measure of confinement.  As a loop becomes large, it will be screened
dynamically by quarks ``popping'' out of the vacuum.  Thus we always
will have a perimeter law.

Another approach to understanding the confinement phase is to use the
mass gap.  As long as the quarks themselves are massive, a confining
theory should contain no physical massless particles.  All mesons,
glueballs, and nucleons are expected to gain masses through the
dimensional transmutation phenomenon discussed later.  As with the
area law, the presence of a mass gap is easily demonstrated for the
strong coupling limit of the pure glue theory.

If the quarks are massless, this definition also becomes a bit tricky.
In this case we expect spontaneous breaking of chiral symmetry, also
discussed extensively later.  This gives rise to pions as massless
Goldstone bosons.  To distinguish this situation from the unconfined
theory, one could consider the number of massless particles in the
spectrum by looking at how the ``vacuum'' energy depends on
temperature using the Stefan-Boltzmann law.  With $N_f$ flavors we
have $N_f^2-1$ massless scalar Goldstone bosons.  On the other hand,
were the gauge group $SU(N)$ not to confine, we would expect $N^2-1$
massless vector gauge bosons plus $N_f$ massless quarks, all of which
have two degrees of freedom.

\newpage
\Section{Monte Carlo simulation}
\label{montecarlo}

As mentioned earlier, Monte Carlo methods have come to dominate work
in lattice gauge theory.  These are based on the idea that we need not
integrate over all fields, but much information is available already
in a few ``typical configurations.''  For bosonic fields these
techniques work extremely well, while for fermions the methods remain
rather tedious.  Over the years advances in computing power have
brought some such calculations for QCD into the realm of possibility.
Nevertheless in some situations where the path integral involves
complex weightings, the algorithmic issues remain unsolved.  In this
section we review the basics of the method; this is not meant to be an
extensive review, but only a brief introduction.

\subsection{Bosonic fields}

A generic path integral
\begin{equation}
Z=\int (dU) e^{-S}
\end{equation}
on a finite lattice is a finite dimensional integral.  One might try
to evaluate it numerically.  But it is a many dimensional integral.
With $SU(3)$ on $10^4$ lattice we have $4*10^4$ links, each parametrized by 
8 numbers.  Thus it is a $320,000$ dimensional integral.  Taking two
sample points for each direction, this already gives 
\begin{equation}
2^{320,000}=3.8\times 10^{96,329}\qquad {\rm terms.}
\end{equation}
The age of the universe is only $\sim 10^{27}$ nanoseconds, so adding
one term at a time will take a while.

Such big numbers suggest a statistical approach.  The goal of a Monte
Carlo simulation is to 
find a few ``typical'' equilibrium configurations with probability distribution 
\begin{equation}
p(C)\sim e^{-\beta S(C)}.
\end{equation}
On these one can measure observables of choice along with their
statistical fluctuations.

The basic procedure is a Markov process
\begin{equation}
C\rightarrow C^\prime \rightarrow \ldots
\end{equation}
generating a chain of configurations that eventually should approach
the above distribution.  In general we take a configuration $C$ to a
new one with some given probability $P(C\rightarrow C^\prime)$.  As a
probability, this satisfies $0\le P\le 1$ and
$\sum_{C^\prime}P(C\rightarrow C^\prime)=1$.\footnote{For continuous
  groups the sum really means integrals.}  For a Markov process, $P$
should depend only on the current configuration and have no dependence
on the history.

The process should bring us closer to ``equilibrium'' in a sense
shortly to be defined.  This requires at least two things.  First,
equilibrium should be stable; {\it i.e.}  equilibrium is an
``eigen-distribution'' of the Markov chain
\begin{equation}
\sum_{C^\prime}P(C^\prime\rightarrow C) e^{-S(C^\prime)}= e^{-S(C)}.
\end{equation}
Second, we should have ergodicity; {\rm i.e.} all possible states must
in principle be reachable.  

A remarkable result is that these conditions are sufficient for an
algorithm to approach equilibrium, although without any guarantee of
efficiency.  Suppose we start with an ensemble of states, E,
characterized by the probability distribution $p(C)$.  A distance
between ensembles is easily defined
\begin{equation}
D(E,E^\prime)\equiv \sum_C |p(C)-p^\prime(C)|.
\end{equation}
This is positive and vanishes only if the ensembles are equivalent.
A step of our Markov process takes ensemble $E$ into another
$E^\prime$ with 
\begin{equation}
p^\prime(C)=\sum_{C^\prime}P(C^\prime\rightarrow C) p(C^\prime).
\end{equation} 
Now assume that $P$ is chosen so that the equilibrium distribution
$p_{eq}(C)=e^{-S(C)}/Z$ is an eigenvector of eigenvalue 1.  Compare
the new distance from equilibrium with the old
\begin{equation}
D(E^\prime,E_{eq})=\sum_C |p^\prime(C)-p_{eq}(C)|
=\sum_C\left |\sum_{C^\prime} P(C\rightarrow C^\prime)(p(C)-p_{eq}(C))\right|.
\end{equation}
Now the absolute value of a sum is always less than the sum of the
absolute values, so we have
\begin{equation}
D(E^\prime,E_{eq})\le
\sum_C \sum_{C^\prime} P(C\rightarrow C^\prime)|(p(C)-p_{eq}(C))|.
\end{equation}
Since each $C$ must go somewhere, the sum over $C^\prime$ gives
unity and we have
\begin{equation}
D(E^\prime,E_{eq})\le
\sum_C |(p(C)-p_{eq}(C))|=D(E,E_{eq}).
\end{equation}
Thus the algorithm automatically brings one closer to equilibrium.

How can one be sure that equilibrium is an eigen-ensemble?  The usual
way in practice invokes a principle of detailed balance, a sufficient
but not necessary condition.  This states that the forward and
backward rates between two states are equal when one is in equilibrium
\begin{equation}
p_{eq}(C)P(C\rightarrow C^\prime) = p_{eq}(C^\prime)P(C^\prime\rightarrow C).
\end{equation}
Summing this over $C^\prime$ immediately gives the fact that the
equilibrium distribution is an eigen-ensemble.

The famous Metropolis et al.~approach \cite{Metropolis:1953am} is an
elegant and simple way to construct an algorithm satisfying detailed
balance.  This begins with a trial change on the configuration,
specified by a trial probability $P_T(C\rightarrow C^\prime)$.  This
is required to be constructed in a symmetric way, so that
\begin{equation}
P_T(C\rightarrow C^\prime)=P_T(C^\prime\rightarrow C).
\end{equation}
This by itself would just tend to randomize the system.  To restore
the detailed balance, the trial change is conditionally accepted with
probability 
\begin{equation}
A(C,C^\prime)={\rm min}(1, p_{eq}(C^\prime)/p_{eq}(C)).
\end{equation}
In other words, if the Boltzmann weight gets larger, make the change;
otherwise, accept it with probability proportional to the ratio of the
Boltzmann weights.  An explicit expression for the final transition
probability is
\begin{equation}
P(C\rightarrow C^\prime)=P_T(C\rightarrow C^\prime)A(C,C^\prime)
+\delta(C,C^\prime)\left(1-\sum_{C^{\prime\prime}}
P_T(C\rightarrow C^{\prime\prime})A(C,C^{\prime\prime})\right).
\end{equation}
The delta function accounts for the possibility that the change is
rejected.

For lattice gauge theory with its $U$ variables in a group, the trial
change can be most easily set up via a table of group elements $T=\{
g_1, ... g_n\}$.  The trial change consists of picking an element
randomly from this table and using $U_T=gU$.  These can be chosen
arbitrarily with two conditions: (1) multiplying them together in
various combinations should generate the whole group and (2) for each
element in the table, its inverse must also be present, i.e. $g\in T
\Rightarrow g^{-1} \in T$.  The second condition is essential for
having the forward and reverse trial probabilities equal.  An
interesting feature of this approach is that the measure of the group
is not used in any explicit way; indeed, it is generated
automatically.

Generally the group table should be weighted towards the identity.
Otherwise the acceptance gets small and you never go anywhere.  But
this weighting should not be too extreme, because then the motion
through configuraton space becomes slow.  Usually the width of the
table is adjusted to give an acceptance of order 50\%.  For free field
theory the optimum can be worked out, it is a bit less.  In general a
big change with a small acceptance can sometimes be better than small
changes; this appears to be the case with simulating self avoiding
random walks\cite{Madras:1988ei}.

The acceptance criterion involves the ratio
${p_{eq}(C^\prime)\over p_{eq}(C)}$.
An interesting quantity is
the expectation of this ratio in equilibrium.  This is
\begin{equation}
\left\langle {p_{eq}(C^\prime)\over p_{eq}(C)} \right\rangle=
\sum_C p_{eq}(C) \sum_{C^\prime} P_T(C\rightarrow
C^\prime)p_{eq}(C^\prime)/p_{eq}(C)= 1
\end{equation}
since 
\begin{equation}
\sum_C P_T(C\rightarrow C^\prime)=
\sum_C P_T(C^\prime \rightarrow C)=1
\end{equation}
and $\sum_{C^\prime}p_{eq}(C^\prime)=1.$ Of course the average
acceptance is not unity since it is expectation of the minimum of this
ratio and 1.  However monitoring this expectation provides a simple
way to follow the approach to equilibrium.

A full Monte Carlo program consists of looping over all the lattice
links while considering such tentative changes.  To improve
performance there are many tricks that have been developed over the
years.  For example, in a lattice gauge calculation the calculation of
the ``staples'' interacting with a given link takes a fair amount of
time.  This makes it advantageous to apply several Monte Carlo
``hits'' to the given link before moving on.

\subsection{Fermions}

The numerical difficulties with fermionic fields stem from their being
anti-commuting quantities.  Thus it is not immediately straightforward
to place them on a computer, which is designed to manipulate numbers.
Indeed, the Boltzmann factor with fermions is formally an operator in
Grassmann space, and cannot be directly interpreted as a probability.
All algorithms in current use eliminate the fermions at the outset by
a formal analytic integration.  This is possible because most actions
in practice are, or can easily be made, quadratic in the fermionic
fields.  The fermion integrals are then over generalized gaussians.
Unfortunately, the resulting expressions involve the determinant of a
large, albeit sparse, matrix.  This determinant introduces non-local
couplings between the bosonic degrees of freedom, making the path
integrals over the remaining fields rather time consuming.

For this brief overview we will be quite generic and assume we are
interested in a path integral of form
\begin{equation} 
  Z=\int (dA)(d\psi)(d\overline\psi)\ \exp(-S_G(A)-\overline\psi D(A) \psi).
\end{equation} 
Here the gauge fields are formally denoted $A$ and fermionic fields
$\psi$ and $\overline\psi$.  Concentrating on fermionic details, in
this section we ignore the technicality that the gauge fields are
actually group elements. All details of the fermionic formulation are
hidden in the matrix $D(A)$.  While we call $A$ a gauge field, the
algorithms are general, and have potential applications in other field
theories and condensed matter physics.

In the section on Grassmann integration we found the basic formula
for a fermionic Gaussian integral
\begin{equation} 
    \int (d\psi d\overline\psi) \ e^{-\overline\psi D\psi} = \vert D
    \vert
\end{equation} 
where $(d\psi d\overline\psi)=d\psi_1\ d\overline\psi_1 \ldots d\psi_n
\ d\overline\psi_n$.  Using this, we can explicitly integrate out the
fermions to convert the path integral to
\begin{equation} 
    Z=\int (dA)\ \vert D \vert\ e^{-S_G}
=\int (dA) \exp(-S_g+{\rm Tr}\ \log(D)). 
\end{equation} 
This is now an integral over ordinary numbers and therefore in
principle amenable to Monte Carlo attack.

For now we assume that the fermions have been formulated such that
$\vert D\vert$ is positive and thus the integrand can be regarded as
proportional to a probability measure.  This is true for several of
the fermion actions discussed later.  However, if $\vert D\vert$ is
not positive, one can always double the number of fermionic species,
replacing $D$ by $D^\dagger D$.  We will see in later sections that
the case where $D$ is not positive can be rather interesting, but how
to include such situations in numerical simulations is not yet well
understood.

Direct Monte Carlo study of the partition function in this form is
still not practical because of the large size of the matrix $D$.  In
our compact notation, this is a square matrix of dimension equal to
the number of lattice sites times the number of Dirac components times
the number of internal symmetry degrees of freedom.  Thus, it is
typically a hundreds of thousands by hundreds of thousands matrix,
precluding any direct attempt to calculate its determinant.  It is,
however, generally an extremely sparse matrix because most popular
actions do not directly couple distant sites.  All the Monte Carlo
algorithms used in practice for fermions make essential use of this
fact.

Some time ago Weingarten and Petcher \cite{Weingarten:1980hx}
presented a simple ``exact'' algorithm.  By introducing
``pseudofermions'' \cite{Fucito:1980fh,Scalapino:1981qs}, an auxiliary
set of complex scalar fields $\phi$, one can rewrite the path integral
in the form
\begin{equation}
    Z= \int (dA) (d\phi^*\ d\phi) \exp(-S_G-\phi^* D^{-1}\phi). 
\end{equation}
Thus a successful fermionic simulation would be possible if one could
obtain configurations of fields $\phi$ and $A$ with probability
distribution
\begin{equation}
     P(A,\phi) \propto  \exp(-S_G-\phi^* D^{-1}\phi). 
\end{equation} 
To proceed we again assume that $D$ is a positive matrix so this
distribution is well defined.

For an even number of species, generating an independent set of $\phi$
fields is actually quite easy.  If we consider a field $\chi$ that is
gaussianly randomly selected, i.e.  $P(\chi)\sim e^{-\chi^2}$, then
the field $\phi=D\chi$ is distributed as desired for two flavors
$P(\phi)\sim e^{-(D^{-1}\phi)^2}$.  The hard part of the algorithm is
the updating of the $A$ fields, which requires knowledge of how
$\phi^* D^{-1}\phi$ changes under trial changes in $A$.

\subsection{The conjugate-gradient algorithm}

While $D^{-1}$ is the inverse of an enormous matrix, one really only
needs $\phi^* D^{-1}\phi$, which is just one matrix element of this
inverse.  Furthermore, with a local fermionic action the matrix $D$ is
extremely sparse, the non-vanishing matrix elements only connecting
nearby sites.  In this case there exist quite efficient iterative
schemes for finding the inverse of a large sparse matrix applied to a
single vector.  Here we describe one particularly simple approach.

The conjugate gradient method to find $\xi=D^{-1} \phi$ works by
finding the minimum over $\xi$ of the function $\vert
D\xi-\phi\vert^2$.  The solution is iterative; starting with some
$\xi_0$, a sequence of vectors is obtained by moving to the minimum of
this function along successive directions $d_i$.  The clever trick of
the algorithm is to choose the $d_i$ to be orthogonal in a sense
defined by the matrix $D$ itself; in particular $(Dd_i, D d_j)=0$
whenever $i\ne j$.  This last condition serves to eliminate useless
oscillations in undesirable directions, and guarantees convergence to
the minimum in a number of steps equal to the dimension of the
matrix. There are close connections between the conjugate gradient
inversion procedure and the Lanczos algorithm for tridiagonalizing
sparse matrices.

The procedure is a simple recursion.  Select some arbitrary
initial pair of non-vanishing vectors $g_0=d_0$.  For the inversion
problem, convergence will be improved if these are a good guess to
$D^{-1}\phi$. Then generate a sequence of further vectors by
iterating
\begin{eqnarray}
&& g_{i+1}=(Dg_i,Dd_i)g_i-(g_i,g_i)D^\dagger Dd_i   \cr
&& d_{i+1}=(Dd_i,Dd_i)g_{i+1}-(Dd_,Dg_{i+1})d_i.    
\end{eqnarray}
This construction assures that $g_i$ is orthogonal to $g_{i+1}$
and $(Dd_i,Dd_{i+1})=0$.  It should also be clear that the three sets
of vectors $\{d_0,...d_k\}$,
$\{g_0,...g_k\}$, and $\{d_0,...(D^\dagger D)^kd_0\}$ all span the
same space.  

The remarkable core of the algorithm, easily proved by induction, is
that the set of $g_i$ are all mutually orthogonal, as are $Dd_i$. 
For an $N$ dimensional matrix, there can be no more than $N$
independent orthogonal vectors.  Thus, ignoring round-off errors, the
recursion in Eq.~(15) must terminate in $N$ or less steps with the
vectors $g$ and $d$ vanishing from then on.  Furthermore, as the
above sets of vectors all span the same space, in a basis defined by
the $g_i$ the matrix $D^\dagger D$ is in fact tri-diagonal, with
$(Dg_i,Dg_j)$ vanishing unless $i=j\pm 1$.  

To solve $\phi=D\xi$ for $\xi$, simply expand in the $d_i$ 
  \begin{equation}
 \xi=\sum_i \alpha_i d_i.
 \end{equation}
The coefficients are immediately found from the orthogonality
conditions 
 \begin{equation}
\alpha_i=(Dd_i,\phi)/(Dd_i,Dd_i).
\end{equation}
Note that if we start with the solution $d_0=D^{-1}\phi$, then we have
$\alpha_i=\delta_{i0}$.

This discussion applies for a general matrix $D$.  If $D$ is
Hermitean, then one can work with better conditioned matrices by
replacing the orthogonality condition for the $d_i$ with $(d_i,Dd_j)$
vanishing for $i\ne j$.  

In practice, at least when the correlation length is not too large,
this procedure adequately converges in a number of iterations which
does not grow severely with the lattice size.  As each step involves
vector sums with length proportional to the lattice volume, each
conjugate gradient step takes a time which grows with the volume of
the system.  Thus the overall algorithm including the sweep over
lattice variables is expected to require computer time which grows as
the square of the volume of the lattice.  Such a severe growth has
precluded use of this algorithm on any but the smallest lattices.
Nevertheless, it does show the existence of an exact algorithm with
considerably less computational complexity than would be required for
a repeated direct evaluation of the determinant of the fermionic
matrix.

Here and below when we discuss volume dependences, we ignore additional
factors from critical slowing down when the correlation length is
also allowed to grow with the lattice size.  The assumption is that
such factors are common for the local algorithms treated here. In
addition, such slowing occurs in bosonic simulations, and we are
primarily concerned here with the extra problems presented by the
fermions.  

\subsection{Hybrid Monte Carlo}

One could imagine making trial changes of all lattice variables
simultaneously, and then accepting or rejecting the entire new
configuration using the exact action.  The problem with this approach
is that a global random change in the gauge fields will generally
increase the action by an amount proportional to the lattice volume,
and thus the final acceptance rate will fall exponentially with the
volume.  The acceptance rate could in principle be increased by
decreasing the step size of the trial changes, but then the step size
would have to decrease with the volume.  Exploration of a reasonable
region of phase space would thus require a number of steps growing as
the lattice volume.  The net result is an exact algorithm which
still requires computer time growing as volume squared.

So far this discussion has assumed that the trial changes are made in
a random manner.  If, however, one can properly bias these variations,
it might be possible to reduce the volume squared behavior.  The
``hybrid Monte Carlo'' scheme \cite{Duane:1987de} does this with a
global accept/reject step on the entire lattice after a microcanonical
trajectory.

The trick here is to add yet further auxiliary variables in the
form of ``momentum variables'' $p$ conjugate to the gauge fields $A$.
Then we look for a coupled distribution
\begin{equation}
P(p,A,\phi)=e^{-H(p,A,\phi)}
\end{equation}
with 
\begin{equation}
H=p^2/2+V(A)
\end{equation}
and 
\begin{equation}
V(A)=-S_g(A)- \phi^* D^{-1}\phi.
\end{equation} 
The basic observation is that this is a simple classical Hamiltonian
for the conjugate variables $A$ and $p$, and evolution using Newton's
laws will conserve energy.  For the gauge fields one sets up a
``trajectory'' in a fictitious ``Monte Carlo'' time variable $\tau$
and consider the classical evolution
\begin{eqnarray}
&&{dA_i\over d\tau}=p_i\cr
&&{dp_i\over d\tau}= F_i(A)=-{\partial V(A)\over \partial A_i}.
\end{eqnarray}
Under such evolution an equilibrium ensemble will remain in equilibrium.

An approximately energy conserving algorithm is given by a
``leapfrog'' discretization of Newton's law.  With a microcanonical
time discretization of size $\delta$, this involves two half steps in
momentum sandwiching a full step in the coordinate $A$
\begin{eqnarray}
&&p_{1\over 2} = p + \delta\ F(A)/2\cr
&&A^\prime = A+\delta\ p_{1\over 2}\cr
&&p^\prime = p_{1\over 2}+\delta\ F(A^\prime)/2
\end{eqnarray}
or combined
\begin{eqnarray}
&&      A^{\prime}=A+\delta\ p +\delta^2\ F(A)/2  \cr
&&      p^{\prime}=p+\delta\ (F(A)+F(A^\prime))/2. 
\end{eqnarray}
Even for finite step size $\delta$, this is an area preserving map of
the $(A,p)$ plane onto itself.  The scheme iterates this mapping
several times before making a final Metropolis accept/reject decision.
This iterated map also remains reversible and area preserving.  The
computationally most demanding part of this process is calculating the
force term.  The conjugate gradient algorithm mentioned above can
accomplish this.
      
The important point is that after each step the momentum remains
exactly the negative of that which would be required to reverse the
entire trajectory and return to the initial variables.  If at some
point on the trajectory we were to reverse all the momenta, the system
would exactly reverse itself and return to the same set of states from
whence it came.  Thus a final acceptance with the appropriate
probability still makes the overall procedure exact.  After each
accept/reject step, the momenta $p$ can be refreshed, their values
being replaced by new Gaussian random numbers.  The pseudofermion
fields $\phi$ could also be refreshed at this time.  The goal of the
procedure is to use the micro-canonical evolution as a way to restrict
changes in the action so that the final acceptance will remain high
for reasonable step sizes.

This procedure contains several parameters which can be
adjusted for optimization.  First is $N_{mic}$, the number of 
micro-canonical iterations taken before the global accept/reject 
step and refreshing of the momenta $p$.  Then there is the step 
size $\delta$, which presumably should be set to give a 
reasonable acceptance.  Finally, one can also vary the frequency 
with which the auxiliary scalar fields $\phi$ are updated.  

The goal of this approach is to speed flow through phase space by
replacing a random walk of the $A$ field with a coherent motion in the
dynamical direction determined by the conjugate momenta.  A simple
estimate \cite{Creutz:1988wv} suggests a net volume dependence
proportional to $V^{5/4}$ rather the naive volume squared without
these improvements.

As mentioned above, using pseudofermions is simplest if the fermion
matrix is a square, requiring an even number of species.  Users of the
hybrid algorithm without the global accept-reject step have argued for
adjusting the number of fermion species by inserting a factor
proportional to the number of flavors in front of the pseudofermionic
term when the gauge fields are updated.  This modification is simple
to make, but raises some theoretical issues that will be discussed
later.  In particular, it is crucial that the underlying fermion
operator break any anomalous symmetries associated with the reduced
theory.  

Despite the successes of these fermion algorithms, the overall
procedure still seems somewhat awkward, particularly when compared
with the ease of a pure bosonic simulation.  This appears to be tied
to the non-local actions resulting from integrating out the fermions.
Indeed, had one integrated out a set of bosons coupled quadratically
to the gauge field, one would again have a non-local effective action,
indicating that this analytic integration was not a good idea.
Perhaps we should step back and explore algorithms before integrating
out the fermions.

An unsolved problem is to find a practical simulation approach to
fermionic systems where the corresponding determinant is not always
positive.  This situation is of considerable interest because it
arises in the study of quark-gluon thermodynamics when a chemical
potential is present.  All known approaches to this problem are
extremely demanding on computer resources.  One can move the phase of
the determinant into the observables, but then one must divide out the
average value of this sign.  This is a number which is expected to go
to zero exponentially with the lattice volume; thus, such an algorithm
will require computer time growing exponentially with the system size.
Another approach is to do an expansion about zero baryon density, but
again to get to large chemical potential will require rapidly growing
resources.  New techniques are badly needed to avoid this growth;
hopefully this will be a particularly fertile area for future
algorithm development.

\newpage
\Section{Renormalization and the continuum limit}
\label{asymptoticfreedom}

Asymptotic freedom is a signature feature of the theory of the strong
interactions.  Interactions between quarks decrease at very short
distances.  From one point of view this allows perturbative
calculations in the high energy limit, and this has become an industry
in itself.  But the concept is also of extreme importance to lattice
gauge theory.  Indeed, asymptotic freedom tells us precisely how to
take the continuum limit.  This chapter reviews the renormalization
group and this crucial connection to the lattice.  When fermions are
present their masses must also be renormalized, but the
renormalization group also tells us exactly how to do this.

\subsection{Coupling constant renormalization}

At the level of tree Feynman diagrams, relativistic quantum field
theory is well defined and requires no renormalization.  However as
soon as loop corrections are encountered, divergences appear and must
be removed by a regularization scheme.  In general the theory then
depends on some cutoff, which is to be removed with a simultaneous
adjustment of the bare parameters while keeping physical quantities
finite.

For example, consider a lattice cutoff with spacing $a$.  The proton
mass $m_p$ is a finite physical quantity, and on the lattice it will
be some, a priori unknown, function of the cutoff $a$, the bare gauge
coupling $g$ and the bare quark masses.  For the quark-less theory we
could use the lightest glueball mass for this purpose.  The basic idea
is to hold enough physical properties constant to determine how the
coupling and quark masses behave as the lattice spacing is reduced.

As the quark masses go to zero the proton mass is expected to remain
finite; thus, to simplify the discussion, temporarily ignore the quark
masses.  Thus consider the proton mass as a function of the gauge
coupling and the cutoff, $m_p(g,a)$.  Holding this constant as the
cutoff varies determines how $g$ depends on $a$.  This is the basic
renormalization group equation
\begin{equation}
a{d\over da} m_p(g(a),a)=0=a{\partial\over \partial a}m_p(g,a)
+ a\left({ dg\over da}\right)\ {\partial\over\partial g} m_p(g,a).
\end{equation}
By dimensional analysis, the proton mass should scale as $a^{-1}$
at fixed bare coupling.  Thus we know that
\begin{equation}
a{\partial\over \partial a}m_p(g,a)=-m_p(g,a).
\end{equation}
The ``renormalization group function'' 
\begin{equation}
\beta(g)=a{dg\over da}={m_p(g,a)
\over {\partial\over\partial g} m_p(g,a)}
\end{equation} 
characterizes how the bare coupling is to be varied for the continuum
limit.  Note that this particular definition is independent of
perturbation theory or any gauge fixing.

As renormalization is not needed until quantum loops are encountered,
$\beta(g)$ vanishes as $g^3$ when the coupling goes to zero.
Define perturbative coefficients from the asymptotic series
\begin{equation}
\beta(g)=\beta_0 g^3+\beta_1 g^5 +\ldots
\label{AF}
\end{equation}
Politzer \cite{Politzer:1973fx} and Gross and Wilczek \cite
{Gross:1973id,Gross:1973ju} first calculated the coefficient $\beta_0$
for non-Abelian gauge theories, with the result
\begin{equation}
\label{betazero}
\beta_0={1\over 16\pi^2}(11N/3-2N_f/3)
\end{equation}
where the gauge group is $SU(N)$ and $N_f$ denotes the number of
fermionic species.  As long as $N_f < 11 N/2$ this coefficient is
positive.  Assuming we can reach a region where this first term
dominates, decreasing the cutoff corresponds to decreasing the
coupling.  This is the heart of asymptotic freedom, which tells us
that the continuum limit of vanishing lattice spacing requires taking
a limit towards vanishing coupling.  The two loop contribution to
Eq.~(\ref{AF}) is also known \cite{Caswell:1974gg,Jones:1974mm}
\begin{equation}
\label{betaone}
\beta_1=\left({1\over 16\pi^2}\right)^2
\left(34N^2/3-10NN_f/3-N_f(N^2-1)/N\right).
\end{equation} 

In general the function $\beta(g)$ depends on the regularization
scheme in use.  For example it might depend on what physical property
is held fixed as well as details of how the cutoff is imposed.
Remarkably, however, these first two coefficients are universal.
Consider two different schemes each defining a bare coupling as a
function of the cutoff, say $g(a)$ and $g^\prime(a)$.  The expansion
for one in terms of the other will involve all odd powers of the
coupling.  In the weak coupling limit each formulation should reduce
to the classical Yang-Mills theory, and thus to lowest order they
should agree
\begin{equation}
g^\prime=g+cg^3+O(g^5).
\label{gprime}
\end{equation}
We can now calculate the new renormalization group function
\begin{eqnarray}
\beta^\prime(g^\prime)&=&a{dg^\prime\over da}={\partial
  g^\prime\over\partial g} \beta(g)\cr
&=&(1+3cg^3)(\beta_0 g^2+\beta_1 g^3)+O(g^5)\cr
&=&\beta_0 {g^\prime}^3+\beta_1 {g^\prime}^3+O({g^\prime}^5).
\label{universal}
\end{eqnarray}
Through order ${g^\prime}^3$ the dependence on the parameter $c$
cancels.  This, however, does not continue to higher orders, where
alternate definitions of the beta function generally differ.  We will
later comment further on this non-uniqueness.

The renormalization group function determines how rapidly the
coupling decreases with cutoff.  Separating variables
\begin{equation}
d(\log(a))=
{dg\over \beta_0 g^3+\beta_1 g^5+O(g^7)}
\end{equation}
allows integration to obtain
\begin{equation}
\log(a\Lambda)=-{1\over 2 \beta_0 g^2}+
{\beta_1\over \beta_0^2}\log(g)+O(g^2)
\end{equation}
where $\Lambda$ is an integration constant.
This immediately shows that the lattice spacing decreases
exponentially in the inverse coupling
\begin{equation}
\label{rgfinal}
a={1\over \Lambda} e^{-1/2\beta_0 g^2} g^{-\beta_1/\beta_0^2}
(1+O(g^2)).
\end{equation}
Remarkably, although the discussion began with the beta function
obtained in perturbation theory, the right hand side of
Eq.~(\ref{rgfinal}) has an essential singularity at vanishing
coupling.  The renormalization group provides non-perturbative
information from a perturbative result.

Dropping the logarithmic corrections, the coupling as a function of
the cutoff reduces to
\begin{equation}
g^2\sim {1\over 2\beta_0 \log(1/\Lambda a)}
\end{equation}
showing the asymptotic freedom result that the bare coupling goes to
zero logarithmically with the lattice spacing in the continuum limit.  

The integration constant $\Lambda$ is defined from the bare charge and
in a particular cutoff scheme.  Its precise numerical value will
depend on details, but once the scheme is chosen, it is fixed relative
to the scale of the quantity used define the physical scale.  In the
above discussion this was the proton mass.  The existence of a scheme
dependence can be seen by considering two different bare couplings as
related in Eq.~(\ref{gprime}).  The relation between the integration
constants is
\begin{equation}
\log(\Lambda^\prime/\Lambda)={c\over 2\beta_0}.
\end{equation}

The mass $m$ of a physical particle, perhaps the proton used above, is
connected to an inverse correlation length in the statistical analogue
of the theory.  Measuring this correlation length in lattice units, we
can consider the dimensionless combination $\xi=1/am$.  For the
continuum limit, we want this correlation length to diverge.
Multiplying Eq.~(\ref{rgfinal}) by the mass tells us how this
divergence depends on the lattice coupling
\begin{equation}
ma=\xi^{-1}
={m\over \Lambda} e^{-1/2\beta_0 g^2} g^{-\beta_1/\beta_0^2}
(1+O(g^2)).
\label{correlation}
\end{equation}
Conversely, if we know how a correlation length $\xi$ of the
statistical system diverges as the coupling goes to zero, we can read
off the particle mass in units of $\Lambda$ as the coefficient of the
behavior in this equation.  This exemplifies the close connection
between diverging correlation lengths in a statistical system and the
continuum limit of the corresponding quantum field theory.

We emphasize again the exponential dependence on the inverse coupling
appearing in Eq.~(\ref{correlation}).  This is a function that is
highly non-analytic at the origin.  This demonstrates quite
dramatically that QCD cannot be fully described by perturbation theory.

\subsection{A parameter free theory}

This discussion brings us to the remarkable conclusion that, ignoring
the quark masses, the strong interactions have no free parameters.
The cutoff is absorbed into $g(a)$, which in turn is absorbed into
the renormalization group dependence.  The only remaining dimensional
parameter $\Lambda$ serves to set the scale for all other masses.  In
the theory considered in isolation, one may select units such that
$\Lambda$ is unity.  After such a choice, all physical mass ratios are
determined.  Coleman and Weinberg \cite{Coleman:1973jx} have given
this process, wherein a dimensionless parameter $g$ and a
dimensionful one $a$ manage to ``eat'' each other, the marvelous name
``dimensional transmutation.''

In the theory including quarks, their masses represent further
parameters.  Indeed, these are the only parameters in the theory of
the strong interactions.  In the limit where the quark masses vanish,
referred to as the chiral limit, we return to a zero parameter theory.
In this approximation to the physical world, the pion mass is expected
to vanish and all dimensionless observables should be uniquely
determined.  This applies not only to mass ratios, such as of the rho
mass to the proton, but as well to quantities such as the pion-nucleon
coupling constant, once regarded as a parameter for a perturbative
expansion.  As the chiral approximation has been rather successful in
the predictions of current algebra, we expect an expansion in the
small quark masses to be a fairly accurate description of hadronic
physics.  Given a qualitative agreement, a fine tuning of the small
quark masses should give the pion its mass and complete the theory.

The exciting idea of a parameter-free theory is sadly lacking from
most treatments of the other interactions such as electromagnetism or
the weak force.  There the coupling $\alpha\sim 1/137$ is treated as a
parameter.  One might optimistically hope that the inclusion of the
appropriate non-perturbative ideas into a unified scheme would
ultimately render $\alpha$ and the quark and lepton masses calculable.

\subsection{Including quark masses}

Above we concentrated on the flow of the bare coupling as one takes
the continuum limit.  Of course with massive quarks in the theory, the
bare quark mass is also renormalized.  Here we extend the above
discussion to see how the two bare parameters flow together to zero in
a well defined way.

Including the mass flow, the renormalization group equations become
\begin{eqnarray}
&&a{dg\over da}=\beta(g)=\beta_0 g^3+\beta_1 g^5 +\ldots
+{\rm non{\hbox{-}}perturbative}\cr
&&a{dm\over da}=m\gamma(g)=m(\gamma_0 g^2+\gamma_1 g^4 +\ldots)
+{\rm non{\hbox{-}}perturbative.}
\end{eqnarray}
Now we have three perturbative coefficients
$\beta_0,\ \beta_1,\ \gamma_0$ which are scheme independent and known
\cite{Politzer:1973fx,Gross:1973id,Gross:1973ju,
  Caswell:1974gg,Jones:1974mm, Vermaseren:1997fq,Chetyrkin:1997fm }.
For $SU(3)$ we have
\begin{equation}
\matrix{
&\beta_0=&{11-2N_f/3 \over (4\pi)^2}
&\qquad=.0654365977\qquad (N_f=1)\hfill\cr
&\beta_1=&{102-12N_f\over (4 \pi)^4}
&\qquad = .0036091343\qquad (N_f=1)\hfill\cr
&\gamma_0=&{8\over (4 \pi)^2}&\qquad=.0506605918\hfill\cr
}
\end{equation}
For simplicity we work with $N_f$ degenerate quarks, although this is
easily generalized to the non-degenerate case.  It is important to
recognize that the ``non-perturbative'' parts fall faster than any
power of $g$ as $g\rightarrow 0$.  As we will discuss later, unlike
the perturbative pieces, the non-perturbative contributions to
$\gamma$ in general need not be proportional to the quark mass.

\begin{figure*}
\centering
\includegraphics[width=2.5in]{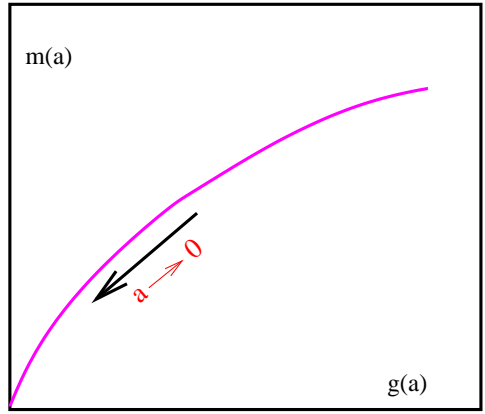}
\caption{ In the continuum limit both the bare coupling and bare mass
  for QCD flow to zero.}
\label{rgflow} 
\end{figure*}

As with the pure gauge theory discussed earlier, these equations are
easily solved to show
\begin{eqnarray}
&&a={1\over \Lambda} e^{-1/2\beta_0 g^2} g^{-\beta_1/\beta_0^2}
(1+O(g^2))\cr
&&m=Mg^{\gamma_0/\beta_0}
(1+O(g^2)).
\end{eqnarray}
The quantities $\Lambda$ and $M$ are ``integration constants'' for the
renormalization group equations.  Rewriting these relations gives the
coupling and mass flow in the continuum limit $a\rightarrow 0$
\begin{eqnarray}
&&g^2\sim {1\over \log(1/\Lambda a)}\rightarrow 0\qquad\qquad
\hbox{``asymptotic freedom''}\cr
&&m\sim M\ \left({1\over \log(1/\Lambda
a)}\right)^{\gamma_0/\beta_0}\rightarrow 0.
\end{eqnarray}
Here $\Lambda$ is usually regarded as the ``QCD scale'' and $M$ as the
``renormalized quark mass.''  The resulting flow is sketched in
Fig.~\ref{rgflow}.

The rate of this flow to the origin is tied to the renormalization
group constants, which can be obtained from the inverted equations
\begin{eqnarray}
&&\Lambda=\lim_{a\rightarrow 0}
\ { e^{-1/2\beta_0 g^2} g^{-\beta_1/\beta_0^2}\over a}\\
&&M=\lim_{a\rightarrow 0}\ m g^{-\gamma_0/\beta_0}.
\end{eqnarray}
Of course, as discussed for $\Lambda$ above, the specific numerical
values of these parameters depend on the detailed renormalization
scheme.

Defining $\beta(g)$ and $\gamma(g)$ is most naturally done by fixing some
physical quantities and adjusting the bare parameters as the cutoff is
removed.  Because of confinement we can't use the quark mass itself, but we
can select several physical particle masses $m_i(g,m,a)$ to hold
fixed.  This leads to the constraint
\begin{equation}
a{dm_i(g,m,a)\over da}=0={\partial m_i \over \partial g}\beta(g)
+{\partial m_i \over \partial m} m\gamma(g)+a{\partial m_i\over
\partial a}.
\end{equation}
For simplicity, continue to work with degenerate quarks of mass $m$.
Then we have two bare parameters $(g,m)$, and we need to fix two
quantities.
\footnote{Actually there is a third parameter related to CP
  conservation.  Here we assume CP is a good symmetry and ignore this
  complication.  This issue will be further discussed in later
  sections.}  Natural candidates are the lightest baryon mass, denoted
here as $m_p$, and the lightest boson mass, $m_\pi$.  Then we can
explicitly rearrange these relations to obtain a somewhat formal but
explicit expression for the renormalization group functions
\begin{eqnarray}
&\beta(g)=
\left(
a{\partial m_\pi \over \partial a}
{\partial m_p \over \partial m}
-a {\partial m_p \over \partial a}
{\partial m_\pi \over \partial m}
\right)
\bigg /
\left(
{\partial m_p \over \partial g} 
{\partial m_\pi \over \partial m}
-{\partial m_\pi \over \partial g} 
{\partial m_p \over \partial m}
\right)
\cr
&\cr
&\gamma(g)=
\left(
a{\partial m_\pi \over \partial a}
{\partial m_p \over \partial g}
-a {\partial m_p \over \partial a}
{\partial m_\pi \over \partial g}
\right)
\bigg /
\left(
{\partial m_p \over \partial m} 
{\partial m_\pi \over \partial g}
-{\partial m_\pi \over \partial m} 
{\partial m_p \over \partial g}
\right).
\end{eqnarray}
Note that this particular definition includes all perturbative and
non-perturbative effects.  In addition, this approach avoids any need
for gauge fixing.

Once given $m_p$, $m_\pi$, and a renormalization scheme, then the
dependence of the bare parameters on the cutoff is completely fixed.
The physical masses are mapped onto the integration constants
\begin{eqnarray}
&&\Lambda=\Lambda(m_p,m_\pi)\\
&&M=M(m_p,m_\pi).
\end{eqnarray}
Formally these relations can be inverted to express the masses as
functions of the integration constants,
$m_i=m_i(\Lambda,M)$.  Straightforward 
dimensional analysis tells us that the masses must take the form
\begin{equation} 
m_i=\Lambda f_i(M/\Lambda).
\end{equation}
As we will discuss in more detail in later sections, for the
multi-flavor theory we expect the pions to be Goldstone bosons with
$m_\pi^2 \sim m_q$.  This tells us that the above function for the
pion should exhibit a square root singularity {$f_\pi(x)\sim
  x^{1/2}$}.  This relation removes any additive ambiguity in defining
the renormalized quark mass $M$.  As will be discussed in more detail
later, this conclusion does not persist if the lightest quark becomes
non-degenerate.

\subsection{Which beta function?}

Thus far our discussion of the renormalization group has been in terms
of the bare charge with a cutoff in place.  This is the natural
procedure in lattice gauge theory; however, there are alternative
approaches to the renormalization group that are frequently used in
the continuum theory.  We now make some comments on connection between
the lattice and the continuum approaches.

An important issue is that there are many different ways to define a
renormalized coupling; it should first of all be an observable that
remains finite in the continuum limit
\begin{equation}
\lim_{a\rightarrow 0} g_r(\mu,a,g(a)) = g_r(\mu).
\end{equation}
Here $\mu$ is a dimensionful energy scale introduced to define the
renormalized coupling.  The subscript $r$ is added to distinguish this
coupling from the bare one.  For perturbative purposes one might use a
renormalized three-gluon vertex in a particular gauge and with all
legs at a given scale of momentum proportional to $\mu$.  But many
alternatives are possible; for example, one might use as an observable
the force between two quarks at separation $1/\mu$.

Secondly, to be properly called a renormalization of the classical
coupling, $g_r$ should be normalized such that it reduces to the bare
coupling in lowest order perturbation theory for the cutoff theory
\begin{equation}
\label{renormalizedg}
g_r(\mu,a,g) = g+O(g^3). 
\end{equation}
Beyond this, the definition of $g_r$ is totally arbitrary.  In
particular, given any physical observable $H$ defined at scale $\mu$ and
satisfying a perturbative expansion
\begin{equation}
H(\mu,a,g)=h_0+h_1 g^2+O(g^4)
\label{obsa}
\end{equation}
we can define a corresponding renormalized coupling
\begin{equation}
g_H^2(\mu)=(H(\mu)-h_0)/h_1.
\end{equation}
As the energy scale goes to infinity, this renormalized charge should
go to zero.  But with a different observable, we will generally obtain
a different functional behavior for this flow.  From this flow of the
renormalized charge we can define a renormalized beta function
\begin{equation}
\beta_r=-\mu{\partial g_r(\mu)\over \partial \mu}.
\end{equation}

We now draw a remarkable connection between the renormalized
renormalization group function $\beta_r(g_r)$ and the function
$\beta(g)$ defined earlier for the bare coupling.  When the cutoff
is still in place, the renormalized coupling is a function of the
scale $\mu$ of the observable, the cutoff $a$, and the bare coupling
$g$.
Since we are working with dimensionless couplings, $g_r$ can depend
directly on $\mu$ and $a$ only through their product.  This simple
application of dimensional analysis implies
\begin{equation}
a\ \left.{\partial g_r\over \partial a}\right |_{g}
=\mu\ \left.{\partial g_r\over \partial \mu}\right |_{g}
=-\beta_r.
\end{equation}
Now, in the continuum limit as we take $a$ to zero and adjust $g$
appropriately, $g_r$ should become a function of the physical scale
$\mu$ alone.  Indeed, we could use $g_r(\mu)$ itself as the physical
quantity to hold fixed for the continuum limit.  Then we obtain
 \begin{equation}
a{\partial g_r\over \partial a}+{\partial g_r\over \partial g}
a{\partial g\over \partial a}=0.
\end{equation}
Using this in an analysis similar to that in Eq.~(\ref{universal}), we
find
\begin{equation}
\beta_r(g_r)=\beta_0 g_r^3+\beta_1 g_r^5 +O(g_r^7).
\label{renormalizedgamma}
\end{equation}
Where $\beta_0$ and $\beta_1$ are the same coefficients that appear in
Eq.~(\ref{AF}).  Both the renormalized and the bare $\beta$ functions
have the same first two coefficients in their perturbative expansions.
Indeed, it was through consideration of the renormalized coupling that
$\beta_0$ and $\beta_1$ were first calculated.

It is important to reiterate the considerable arbitrariness in
defining both the bare and the renormalized couplings.  Far from the
continuum there need be no simple relationship between different
formulations.  Once one leaves the perturbative region, even such
things as zeros in the $\beta$ functions are not universal.  For one
extreme example, it is allowed to force the beta function to consist
of only the first two terms.  In this case, as long as $N_f$ is small
enough that $\beta_1>0$, there is explicitly no other zero of the beta
function except at $g=0$.  On the other hand, one might think it
natural to define the coupling from the force between two quarks.
When dynamical quarks are present, at large distances this falls
exponentially with the pion mass at large distances.  In this case the
beta function must have another zero in the vicinity of where the
screening sets in.  Thus, even the existence of zeros in the beta
function is scheme dependent.  The only exception to this is if a zero
occurs in a region of small enough coupling that perturbation theory
can be trusted.  This has been conjectured to happen for a sufficient
number of flavors \cite{Banks:1981nn}.

The perturbative expansion of $\beta_r$ has important experimental
consequences.  If, as expected, the continuum limit is taken at
vanishing bare coupling and the renormalized coupling is small enough
that the first terms in Eq.~(\ref{renormalizedgamma}) dominate, then
the renormalized coupling will be driven to zero logarithmically as
its defining scale $\mu$ goes to infinity.  Not only does the bare
coupling vanish, but the effective renormalized coupling becomes
arbitrarily weak at short distances.  This is the physical implication
of asymptotic freedom; phenomena involving only short-distance effects
may be accurately described with a perturbative expansion.  Indeed,
asymptotically free gauge theories were first invoked for the strong
interactions as an explanation of the apparently free parton behavior
manifested in the structure functions associated with deeply inelastic
scattering of leptons from hadrons.

The dependence of the integration constant $\Lambda$ on the details of
the renormalization scheme carries over to the continuum
renormalization group as well.  Given a particular definition of the
renormalized coupling $g_r(\mu)$, its behavior as $r$ goes to zero
will involve a scale $\Lambda_r$ in analogy to the scale in the bare
coupling.  Hasenfratz and Hasenfratz
\cite{Hasenfratz:1980kn,Hasenfratz:1981tw} were the first to perform
the necessary one loop calculations to relate $\Lambda$ from the
Wilson lattice gauge theory with $\Lambda_r$ defined from the
three-gluon vertex in the Feynman gauge and with all legs carrying
momentum $\mu^2$.  They found
\begin{equation}
{\Lambda_r\over \Lambda}=\pmatrix{
57.5 & SU(2)\cr
83.5 & SU(3)\cr
}
\end{equation}
for the pure gauge theory.  Note that not only is $\Lambda$ scheme
dependent, but that different definitions can vary by rather large
factors.  The original calculation of these numbers was rather
tedious.  They have been verified with calculationally more efficient
techniques based on quantum fluctuations around a slowly varying
classical background field \cite{Dashen:1980vm}.

\subsection{Flows and irrelevant operators}

\begin{figure*}
\centering
\includegraphics[width=2.5in]{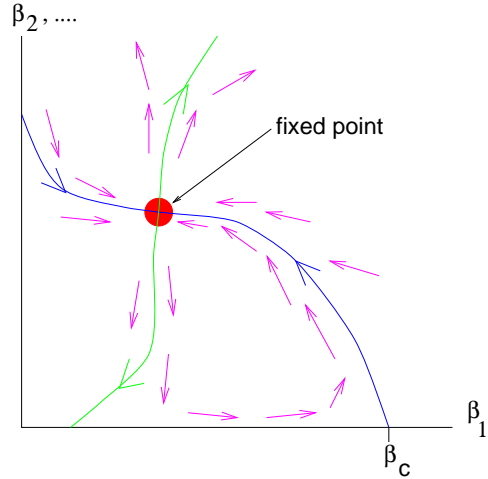}
\caption{\label{flow}
A generic renormalization group flow.  In general this occurs in an
infinite dimensional space.  
}
\end{figure*}

We now briefly discuss another way of looking at the renormalization
group as relating theories with different lattice spacings.  Given one
lattice theory, one could imagine generating another with a larger
lattice spacing by integrating over all links except those on some
subset of the original lattice, thus generating an equivalent theory
with, say, a larger lattice spacing.  While this is conceptually
possible, to do it exactly in more than one dimension will generate an
infinite number of couplings.  If we could keep track of such, the
procedure would be ``exact,'' but in reality we usually need some
truncation.  Continuing to integrate out degrees of freedom, the
couplings flow and might reach some ``fixed point'' in this infinite
space.  With multiple couplings, there can be an attractive ``sheet''
towards which couplings flow, and then they might continue to flow
towards a fixed point, as sketched in Fig.~\ref{flow}.  If the fixed
point has only one attractive direction, then two different models
that flow towards that same fixed point will have the same physics in
the large distance limit.  This is the concept of universality; {\it
  i.e.}  exponents are the same for all models with the same
attractor.

Some hints on this process come from dimensional analysis, although,
in ignoring non-perturbative effects that might occur at strong
coupling, the following arguments are not rigorous.  In $d$ dimensions
a conventional scalar field has dimensions of $M^{d-2\over 2}$.  Thus
the coupling constant $\lambda$ in an interaction of form $\int d^dx\
\lambda \phi^n$ has dimensions of $M^{d-n{d-2\over2}}$.  On a lattice
of spacing $a$, the natural unit of dimension is the inverse lattice
spacing.  Thus without any special tuning, the renormalized coupling
at some fixed physical scale would naturally run as $\lambda\sim
a^{n{d-2\over2}-d}$.  As long as the exponent in this expression is
positive, i.e.
$$
n\ge {2d\over d-2}
$$ 
we expect the coupling to become ``irrelevant'' in the continuum
limit.  The fixed point is driven towards zero in the corresponding
direction.  If $d$ exceeds four, this is the case for all
interactions.  (We ignore $\phi^3$ in 6 dimensions because of
stability problems.)  This suggests that four dimensions is a critical
case, with mean field theory giving the right qualitative critical
behavior for all larger dimensions.  In four dimensions we have
several possible ``renormalizable'' couplings which are dimensionless,
suggesting logarithmic corrections to the simple dimensional
arguments.  Indeed, four-dimensional non-Abelian gauge theories
display exactly such a logarithmic flow; this is asymptotic freedom.

This simple dimensional argument applied to the mass term suggests it
would flow towards infinity in all dimensions.  For a conventional
phase transition, something must be tuned to a critical point.  In
statistical mechanics this is the temperature.  In field theory
language we usually remap this onto a tuning of the bare mass term,
saying that the transition occurs as bare masses go through zero.  For
a scalar theory this tuning for a continuum limit seems unnatural and
is one of the unsatisfying features of the standard model, driving
particle physicists to try to unravel how the Higg's mechanism really
works.

In non-Abelian gauge theories with multiple massless fermions, chiral
symmetry protects the mass from renormalization, avoiding any special
tuning.  Indeed, as we have discussed, because of dimensional
transmutation, all dimensionless parameters in the continuum limit are
completely determined by the basic structure of the initial
Lagrangean, without any continuous parameters to tune.  In the limit
of vanishing pion mass, the rho to nucleon mass ratio should be
determined from first principles; it is the goal of lattice gauge
theory to calculate just such numbers.

As we go below four dimensions, this dimensional argument suggests
that several couplings can become ``relevant,'' requiring the
renormalization group picture of flow towards a non-trivial fixed
point.  Above two dimensions the finite number of renormalizable
couplings corresponds to the renormalization group argument for a
finite number of ``universality classes,'' corresponding to different
basic symmetries.

One might imagine dimensionality as being a continuously variable
parameter.  Then just below four dimensions a renormalizable coupling
becomes ``super-renormalizable'' and a new non-trivial fixed point
breaks away from vanishing coupling.  Near four dimensions this point
is at small coupling, forming the basis for an expansion in $4-d$.
This has become a major industry, making remarkably accurate
predictions for critical exponents in three dimensional
systems \cite{Wilson:1973jj}.

An important consequence of this discussion is that a lattice action
is in general highly non-unique.  One can always add irrelevant
operators and expect to obtain the same continuum limit.
Alternatively, one might hope to improve the approach the continuum
limit by a judicious choice of the lattice action.

The renormalization group is indeed a rich subject.  We have only
touched on a few issues that are particularly valuable for the lattice
theory.  Perhaps the most remarkable result of this section is how a
perturbative analysis of the renormalization-group equation gives rise
to information on the non-perturbative behavior in the particle
masses, as exhibited in Eq.~(\ref{correlation}).

\newpage
\Section{Classical gauge fields and topology}
\label{classical}
The above renormalization group analysis demonstrates that
non-perturbative effects are crucial to understanding the continuum
limit of QCD on the lattice.  However, the importance of going beyond
the perturbation expansion for non-Abelian gauge theories was
dramatically exposed from a completely different direction with the
discovery of non-trivial classical solutions characterized by an
essential singularity at vanishing coupling.  Here we review these
solutions and some of the interesting consequences for the Dirac
operator.

We start with some basic definitions to establish notation in
continuum language.  Being ultimately interested in QCD, we
concentrate on the gauge group $SU(N)$.  This group has $N^2-1$
generators denoted $\lambda^\alpha$.  They are traceless $N$ by $N$
matrices and satisfy the commutation relations
\begin{equation}
[\lambda^\alpha,\lambda^\beta]=if^{\alpha\beta\gamma}\lambda^\gamma,
\end{equation}
involving the group structure constants $f^{\alpha\beta\gamma}$.  By
convention, these generators are orthogonalized and normalized
\begin{equation}
{\rm Tr}\lambda^\alpha\lambda^\beta={1\over 2}\delta^{\alpha\beta}.
\end{equation}
For $SU(2)$ the generators would be the spin matrices
$\lambda^\alpha=\sigma^\alpha/2,$ and the structure constants the
three indexed antisymmetric tensor
$f^{\alpha\beta\gamma}=\epsilon^{\alpha\beta\gamma}.$

Associated with the each of the generators $\lambda^\alpha$ is a gauge
potential $A^\alpha_\mu(x)$.  For the classical theory, assume these
are differentiable functions of space time and vanish rapidly at
infinity.  For the quantum theory this assumption of differentiability
is a subtle issue to which we will later return.  The notation
simplifies a bit by defining a matrix valued field $A$
\begin{equation}
A_\mu=A_\mu^\alpha\lambda^\alpha.
\end{equation}

The covariant derivative is a matrix valued differential operator
defined as
\begin{equation}
D_\mu=
\partial_\mu+igA_\mu.
\end{equation}
Given the gauge potential, the corresponding matrix valued field
strength is
\begin{equation}
F_{\mu\nu}={-i\over g}[D_\mu,D_\nu]=
\partial_\mu A_\nu-\partial_\nu A_\mu
+{ig}[A_\mu,A_\nu]=D_\mu A_\nu-D_\nu A_\mu.
\end{equation}
We define the dual field strength as
\begin{equation}
\tilde F_{\mu\nu}={1\over 2}\epsilon_{\mu\nu\rho\sigma}F_{\rho\sigma}
\end{equation}
with $\epsilon_{\mu\nu\rho\sigma}$ being the antisymmetric tensor with
$\epsilon_{1234}=1.$ 

In terms of the field strength, the classical Yang-Mills action is
\begin{equation}
S={1\over 2}\int d^4x\ {\rm Tr}\ F_{\mu\nu}F_{\mu\nu},
\end{equation}
and the classical equations of motion are
\begin{equation}
D_\mu F_{\mu\nu}=0.
\end{equation}
This defines the classical Yang-Mills theory.

The Jacobi identity
\begin{equation}
[A,[B,C]]+[B,[C,A]]+[C,[A,B]]=0
\end{equation}
applied to the covariant derivative implies that 
\begin{equation}
\epsilon_{\mu\nu\rho\sigma}D_\nu F_{\rho\sigma}=0
\end{equation}
or $D_\mu \tilde F_{\mu\nu}=0$.  This immediately implies that any
self-dual or anti-self-dual field with $F=\pm \tilde F$ automatically
satisfies the classical equations of motion.  This is an interesting
relation since $F=\tilde F$ is linear in derivatives of the gauge
potential.  This leads to a multitude of known
solutions \cite{Rajaraman:1982is}, but here we concentrate on just the
simplest non-trivial one.

This theory is, after all, a gauge theory and therefore has a local
symmetry.  We previously discussed this in the lattice context.  There
it was originally motivated by the continuum gauge transformations of
the classical theory, which we now review.  Let $h(x)$ be a space
dependent element of $SU(N)$ in the fundamental representation.
Assume that $h$ is differentiable.  Now define the gauge transformed
field
\begin{equation}
A_\mu^{(h)}\rightarrow h^\dagger A_\mu h-{i\over g} h^\dagger(\partial_\mu h).
\end{equation}
This transformation takes a simple form for the covariant derivative
\begin{equation}
h^\dagger D_\mu h=D_\mu^{(h)}=\partial_\mu+igA_\mu^{(h)}.
\end{equation}
Similarly for the field strength we have
\begin{equation}
F_{\mu\nu}^{(h)}=h^\dagger F_{\mu\nu} h.
\end{equation}
Thus the action is invariant under this transformation,
$S(A)=S(A^{(h)}).$

\subsection{Surface terms}

A remarkable feature of this formalism is that the combination ${\rm
Tr}\ F\tilde F$ is a total derivative.  To see this first construct
\begin{eqnarray}
&&F\tilde F={1\over 2}\epsilon_{\mu\nu\rho\sigma}
(2\partial_\mu A_\nu+igA_\mu A_\nu)
(2\partial_\rho A_\sigma+igA_\rho A_\sigma)\cr
&&={1\over 2}\epsilon_{\mu\nu\rho\sigma}\left(
4\partial_\mu A_\nu \partial_\rho A_\sigma
+4 ig(\partial_\mu A_\nu) A_\rho A_\sigma
-g^2 A_\mu A_\nu A_\rho A_\sigma
\right).
\end{eqnarray}
If we take a trace of this quantity, the last term will drop out due
to cyclicity.  Thus
\begin{equation}
{\rm Tr}F\tilde F=2\partial_\mu \epsilon_{\mu\nu\rho\sigma} 
{\rm Tr}\left(A_\nu \partial_\rho A_\sigma
+ig A_\nu A_\rho A_\sigma\right)=2\partial_\mu K_\mu
\end{equation}
where we define
\begin{equation}
K_\mu\equiv \epsilon_{\mu\nu\rho\sigma}
{\rm Tr}\left(A_\nu \partial_\rho A_\sigma
+2ig A_\nu A_\rho A_\sigma\right).
\end{equation}
Note that although ${\rm Tr}\ F\tilde F$ is gauge invariant, this is
not true for $K_\mu$.

Being a total derivative, the integral of this quantity
\begin{equation}
\int d^4x\ {1\over 2}{\rm Tr}F\tilde F
\end{equation}
would vanish if we ignore surface terms.  What is remarkable is that
there exist finite action gauge configurations for which this does not
vanish even though the field strengths all go to zero rapidly at
infinity.  This is because the gauge fields $A_\mu$ that appear
explicitly in the current $K_\mu$ need not necessarily vanish as
rapidly as $F_{\mu\nu}$.

These surface terms are closely tied to the topology of the gauge
potential at large distances.  As we want the field strengths to go to
zero at infinity, the potential should approach a pure gauge form
$A_\mu\rightarrow {-i\over g} h^\dagger \partial_\mu h$.  In this case
\begin{equation}
K_\mu \rightarrow -{1\over g^2}\epsilon_{\mu \nu \rho \sigma}
{\rm Tr}
(h^\dagger \partial_\nu h)
(h^\dagger \partial_\rho h)
(h^\dagger \partial_\sigma h).
\end{equation}
Note the similarity of this form to that for the group measure in
Eq.~(\ref{measure}).  Indeed, it is invariant if we take $h\rightarrow
h^\prime h$ with $h^\prime$ being a constant group element.  The
surface at infinity is topologically a three dimensional sphere $S_3$.
If we concentrate on $SU(2)$, this is the same as the topology of the
group space.  For larger groups we can restrict $h$ to an $SU(2)$
subgroup and proceed similarly.  Thus the integral of $K_\mu$ over the
surface reduces to the integral of $h$ over a sphere with the
invariant group measure.  This can give a non-vanishing contribution 
if the mapping of $h$ onto the sphere at infinity covers the entire
group in a non-trivial manner.  Mathematically, one can map the $S_3$
of infinite space onto the $S_3$ of group space an integer number of
times, i.e. $\Pi_3(SU(2))=Z$.  Thus we have
\begin{equation}
\int d^4x\ {1\over 2} {\rm Tr}F\tilde F \propto \nu
\end{equation}
where $\nu$ is an integer describing the number of times $h(x)$ wraps
around the group as $x$ covers the sphere at infinity.  The
normalization involves the surface area of a three dimensional sphere
and can be worked out with the result
\begin{equation}
\label{windingone}
\int d^4x\ {1\over 2}{\rm Tr}F\tilde F ={8\pi^2 \nu\over g^2}.
\end{equation}
For groups larger than $SU(2)$ one can deform $h$ to lie in an $SU(2)$
subgroup, and thus this quantization of the surface term applies to
any $SU(N)$.

If we were to place such a configuration into the path integral for
the quantum theory, we might expect a suppression of these effects by
a factor of $\exp(-8\pi^2/g^2)$.  This is clearly non-perturbative,
however this factor strongly underestimates the importance of
topological effects.  The problem is that we only need to excite
non-trivial fields over the quantum mechanical vacuum, not the
classical one.  The correct suppression is indeed exponential in the
inverse coupling squared, but the coefficient in the exponent can be
determined from asymptotic freedom and dimensional transmutation.  We
will return to this point in Subsection \ref{mixing}.

The combination ${\rm Tr}F\tilde F$ is formally a dimension four
operator, the same as the basic gauge theory action density ${\rm Tr}F
F$.  This naturally leads to the question of what would happen if we
consider a new action which also includes this parity odd term.
Classically it does nothing since it reduces to the surface term
described above.  However quantum mechanically this is no longer the
case.  As we will discuss extensively later, the physics of QCD
depends quite non-trivially on such a term.  An interesting feature of
this new term follows from the quantization of the resulting surface
term.  Because of the above quantization and an imaginary factor in
the path integral, physics is periodic in the coefficient of ${1\over
2}{\rm Tr}F\tilde F$.  Although discussing the consequences directly
with such a term in the action is traditional, we will follow a
somewhat different path in later sections and introduce this physics
through its effects on fermions.

\subsection{An explicit solution}

To demonstrate that non-trivial winding solutions indeed exist,
specialize to $SU(2)$ and find an explicit example.  To start,
consider the positivity of the norm of $F\pm\tilde F$
\begin{equation}
0\le \int d^4x\ (F\pm \tilde F)^2= 2\int d^4x\ F^2 \pm 2\int d^4x\
F\tilde F.
\end{equation}
This means that the action is bounded below by $
\int d^4x\ {1\over 2}{\rm Tr}F\tilde F
$ and this bound is reached only if $F=\pm\tilde F$.  As mentioned
earlier, reaching this is sufficient to guarantee a solution to the
equations of motion.  We will now explicitly construct such a self
dual configuration.

Start with a gauge transformation function which is singular at the
origin but maps around the group at a constant radius
\begin{equation}
h(x_\mu)=
{t+i\vec\tau\cdot \vec x \over \sqrt{x^2}}={T_\mu x_\mu \over |x|}.
\end{equation}
Here we define the four component object $T_\mu=\{1,i\vec\tau\}$.
Considering space with the origin removed, construct the
pure gauge field
\begin{equation}
B_\mu={-i\over g}h^\dagger \partial_\mu h
={-i\over g}h^\dagger (T_\mu x^2 - x_\mu T\cdot x)/|x|^3.
\end{equation}
Because this is nothing but a gauge transformation of a vanishing
gauge field, the corresponding field strength
automatically vanishes
\begin{equation}
\partial_\mu B_\nu -\partial _\nu B_\mu +ig[B_\mu,B_\nu]=0.
\end{equation}

This construction gives a unit winding at infinity.  However this
gauge field is singular at the origin where the winding unwraps.
If we smooth this singularity at $x=0$ with a field of form
\begin{equation}
A_\mu=f(x^2)B_\mu.  
\end{equation}
where $f(0)=0$ and $f(\infty)=1$, this will remove the unwrapping at
the origin and automatically leave a field configuration with
non-trivial winding.  The idea is to find a particular $f(x^2)$
such that $A$ also gives a self dual field strength and thereby is a
solution to the equations of motion.

We have set things up symmetrically under space-time rotations about
the origin.  This connection with $O(4)$ is convenient in that we only
need to verify the self duality along a single direction.  Consider
this to be the time axis, along which self duality requires
\begin{equation}
F_{01}(\vec x=0, t)=\pm F_{23}(\vec x=0, t).
\end{equation}
A little algebra gives
\begin{equation}
F_{\mu\nu}=
(f-f^{2})(\partial_\mu B_\nu -\partial _\nu B_\mu)
+2 f^\prime (x_\mu B_\nu-x_\nu B_\mu).
\end{equation}
So along the time axis we have
\begin{equation}
F_{01}\rightarrow 
2f^\prime t {\tau_1\over gt}
\qquad \qquad
F_{23}\rightarrow (f-f^2) {2\tau_1\over g t^2}.
\end{equation}
Thus the self duality condition reduces to a simple first order
differential equation
\begin{equation}
z f^\prime(z)=\pm (f-f^2).
\end{equation}
This is easily solved to give
\begin{equation}
 f(z)={1\over 1+\rho^2 z^{\pm 1}}
\end{equation}
where $\rho$ is an arbitrary constant of integration.  To have the
function vanish at the origin we take the minus solution.  The
resulting form for the gauge field
\begin{equation}
\label{instantonfield}
A_\mu={-i x^2\over g (x^2+\rho^2)} h^\dagger \partial_\mu h
\end{equation}
is the self dual instanton.  The parameter $\rho$ controls the size
of the configuration.  Its arbitrary value is a consequence of the
conformal invariance of the classical theory.  Switching $h$ and
$h^\dagger$ gives a solution with the opposite winding.

\subsection {Zero modes and the Dirac operator}

A particularly important and intriguing aspect of the above field
configuration is that it supports an exact zero mode for the classical
Dirac operator.  We will later discuss the rigorous connection between
the gauge field winding and the zero modes of the Dirac operator.
Here, however, we will verify this connection explicitly for the above
solution.  Thus we look for a spinor field $\psi(x)$ satisfying
\begin{equation}
\gamma_\mu D_\mu \psi(x)=
\gamma_\mu\left(\partial_\mu+
i g A_\mu\right)
\psi(x)=0
\end{equation}
where we insert the gauge field from Eq.~(\ref{instantonfield}).  The
wave function $\psi$ is a spinor in Dirac space and a doublet in
$SU(2)$ space; i.e. it has 8 components.  Similarly, $\gamma_\mu
A_\mu$ is an 8 by 8 matrix, with a factor of four from spinor space
and a factor of two from the internal gauge symmetry.  The solution
entangles all of these indices in a non-trival manner.

Since we don't want a singularity in $\psi$ at the origin, it is
natural to look for a solution of form
\begin{equation}
\psi(x)= p(|x|) V
\end{equation}
where $p$ is a scalar function of the four dimensional radius and $V$
is a constant vector in spinor and color space.  As before, it is
convenient to look for the solution along the time axis.  There $A_0$
vanishes and we have
\begin{equation}
\vec A=
{1\over g}
\ { t\over t^2+\rho^2}
 \vec \tau. 
\end{equation}
Then the equation of interest reduces to
\begin{equation}
\gamma_0 {d\over dt} \psi(t)=-{ t\over t^2+\rho^2}
\ \vec \tau \cdot \vec \gamma\ \psi(t).
\end{equation} 
The 8 by 8 matrix $\vec \tau \cdot \vec \gamma$ is readily
diagonalized giving the eigenvalues $\{-3,-1,-1,-1,1,1,1,3\}$.  Only the
$+3$ eigenvalue gives a normalizable solution
\begin{equation}
\psi(t)=\psi(0) \exp\left(-3\int_0^t {t\ dt\over t^2+\rho^2}\right).
\end{equation}
For general $x_\mu$ this becomes 
\begin{equation}
\psi(x)
=\psi(0)\left(
{\rho^2 \over x^2+\rho^2}
\right)^{3/2}.
\end{equation}
At large $x$ this goes at $x^{-3}$ so its square is normalizable.
None of the other eigenvalues of $\vec\tau\cdot\vec\gamma$ go to zero
fast enough for normalization; thus, the solution is unique.

We see the appearance of a direct product of two $SU(2)$'s, one from
spin and one from isospin.  As we rotate around the origin, for the
large eigenvalue these rotate together as an overall singlet.  The
other positive eigenvalues of $\vec\tau\cdot\vec\gamma$ represent the
triplet combination while the negative eigenvalues come from
antiparticle states.

This zero eigenvalue of $D$ is robust under smooth deformations of the
gauge field.  This is because the anti-commutation of $D$ with
$\gamma_5$ says that all non-zero eigenvalues of $D$ occur in
conjugate pairs.  Without bringing in another eigenvalue, the isolated
one at zero cannot move.  In the next subsection we will demonstrate
the general result that for arbitrary smooth gauge fields the number
of zero modes of the Dirac operator is directly given by the
topological winding number.

\subsection{The index theorem}

We have seen that associated with a particular topologically
non-trivial gauge configuration is a zero action solution to the Dirac
equation.  Here we will give a simple derivation of the general index
theorem relating zero modes of the Dirac operator to the overall
topology of the gauge field.  We continue to work directly with the
naive continuum Dirac operator.  We assume for this section that the
gauge fields are smooth and differentiable.  While this is unlikely to
be true for typical fields in the path integral, the main purpose here
is to show that robust zero modes must already exist in the classical
theory.  We will later see that the generalization of these zero modes
to the quantum theory is intimately tied to certain quantum mechanical
anomalies crucial to non-perturbative physics.

The combination of the anti-Hermitean character of the classical Dirac
operator $D=\gamma_\mu D_\mu$ along with its anti-commutation with
$\gamma_5$ shows that the non-zero eigenvalues of $D$ all occur in
complex conjugate pairs. In particular, if we have
\begin{equation}
D|\psi\rangle=\lambda |\psi\rangle
\end{equation}
then we immediately obtain the conjugate eigenvector from
\begin{equation}
D \gamma_5\ |\psi\rangle=-\lambda\  \gamma_5|\psi\rangle.
\end{equation}
Since $|\psi\rangle$ and $\gamma_5|\psi\rangle$ have different
eigenvalues under the anti-Hermitean operator $D$, they must be
orthogonal
\begin{equation}
\langle \psi| \gamma_5 |\psi\rangle =0.
\end{equation}
On the other hand, any exact zero eigenmodes need not be paired.
Furthermore, restricted to the space of zero eigenmodes, $\gamma_5$
and $D$ commute and can be simultaneously diagonalized.  The
eigenvalues of $\gamma_5$ are all either plus or minus unity.
Combining all these ideas together gives a simple method to count the
number of zero modes of the Dirac operator weighted by their
chirality.  In particular we have the relation
\begin{equation}
\label{winding}
\nu=n_+-n_-={\rm Tr}\gamma_5 e^{{D}^2/\Lambda^2}
\end{equation}
where $n_\pm$ denotes the number of zero modes with eigenvalue $\pm 1$
under $\gamma_5$.  Here the parameter $\Lambda$ is introduced to
control the behavior of the trace as the eigenvalues go to infinity.
It can be thought of as a regulator, although the above equation is
independent of its value.

To proceed, we first write the square of the Dirac operator appearing
in the above exponential
\begin{equation}
\label{dsquare}
{D}^2=\partial^2-g^2A^2+2igA_\mu\partial_\mu+ig(\partial_\mu A_\mu)
-{g\over 2}\sigma_{\mu\nu}F_{\mu\nu}
\end{equation}
where $[\gamma_\mu,\gamma_\nu]=2i\sigma_{\mu\nu}$.  Expanding
Eq.~(\ref{winding}) for the winding number in powers of the gauge
field, the first non-vanishing term appears in the fourth power of the
Dirac operator.  This involves two powers of the sigma matrices
through the relation
\begin{equation}
{\rm Tr}\ \gamma_5\sigma_{\mu\nu}\sigma_{\rho\sigma}
=4\epsilon_{\mu\nu\rho\sigma}.
\end{equation}
Thus our expression for the winding number becomes
\begin{equation}
\nu={\rm Tr}\gamma_5 e^{{ D}^2/\Lambda^2}
={g^2\over 2\Lambda^4}{\rm Tr}_{x,c}
e^{\partial^2/\Lambda^2}\epsilon_{\mu\nu\rho\sigma}F_{\mu\nu}F_{\rho\sigma}
+O(\Lambda^{-6})
\end{equation}
where ${\rm Tr}_{x,c}$ refers to the trace over space and color, the
trace over the spinor index having been done to give the factor of the
antisymmetric tensor.  It is the trace over the space index that will
give a divergent factor removing the $\Lambda^{-4}$ prefactor.  Higher
order terms go to zero rapidly enough with $\Lambda$ to be ignored.

The factor $e^{\partial^2/\Lambda^2}$ serves to mollify traces over
position space.  Consider some function $f(x)$ as representing a a
diagonal matrix in position space
$M(x,x^\prime)=f(x)\delta(x-x^\prime)$.  The formal trace would be
${\rm Tr}M=\int dx M(x,x)$, but this diverges since it involves a
delta function of zero.  Writing the delta function in terms of its
Fourier transform
\begin{equation}
e^{\partial^2/\Lambda^2} \delta(x-x^\prime)
=\int {d^4p\over (2\pi)^4} e^{ip\cdot (x-x^\prime)}
e^{-p^2/\Lambda^2}
={\Lambda^4\over 16\pi^2}e^{-(x-x^\prime)^2\Lambda^2/4}
\end{equation}
shows how this ``heat kernel'' spreads the delta function.  This
regulates the desired trace
\begin{equation}
{\rm Tr}_x f(x)\equiv
{\Lambda^4\over 16\pi^2}\int d^4x f(x).
\end{equation}
Using this to remove the spatial trace in the above gives the well
known relation
\begin{equation}
\label{windingtwo}
\nu={g^2\over 32\pi^2}{\rm Tr}_c \int d^4x
\epsilon_{\mu\nu\rho\sigma}F_{\mu\nu}F_{\rho\sigma}
={g^2\over 16\pi^2}{\rm Tr}_c \int d^4x F_{\mu\nu}\tilde F_{\mu\nu}.
\end{equation}
As discussed earlier, this integral involves a total derivative that
can be partially integrated into an integral over spatial infinity
that counts the topological winding of the gauge field.  Thus counting
the zero modes of the Dirac operator in a given configuration is an
equivalent way to determine this topology.  The index theorem
represents the fact that Eq.~(\ref{windingone}) and
Eq.~(\ref{windingtwo}) have identical content despite rather different
derivations.

\subsection{Topology and eigenvalue flow}

There is a close connection between the zero modes of the Dirac
operator in the Euclidean path integral and a flow of eigenvalues of
the fermion Hamiltonian in Minkowski space.  To see how this works it
is convenient to work in the temporal gauge with $A_0=0$ and separate
out the space-like part of the Dirac operator
\begin{equation}
D=\gamma_0 \partial_0 + \gamma_0 H(\vec A(t)).
\end{equation}
Consider $\vec A$ as some time dependent gauge field through which the
fermions propagate.  Assume that at large positive or negative times
this background field reduces to a constant.  Without a mass term, the
continuum theory Hamiltonian $H$ commutes with $\gamma_5$ and
anti-commutes with $\gamma_0$.  Therefore its eigenvalues appear in
pairs of opposite energy and opposite chirality; {\it i.e.} if we have
\begin{eqnarray}
H\phi=E\phi\cr
\gamma_5\phi=\pm\phi
\end{eqnarray}
then
\begin{eqnarray}
H\gamma_0\phi=-E\gamma_0\phi\cr
\gamma_5\gamma_0\phi=\mp\gamma_0\phi.
\end{eqnarray}

\begin{figure}
\centering
\includegraphics[width=2.5in]{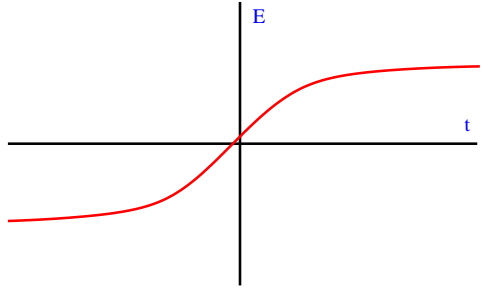}
\caption{An energy eigenvalue that changes in sign between the distant
  past and future.}
\label{eflow} 
\end{figure}

\begin{figure}
\centering
\includegraphics[width=3in]{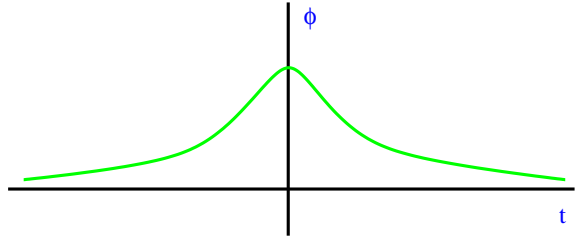}
\caption{The adiabatic evolution gives rise to a normalizable zero
  mode of the four dimensional Dirac operator.}
\label{zeromode} 
\end{figure}

Now suppose we diagonalize $H$
at some given time 
\begin{equation}
H(\vec A(t))\phi_i(t) = E_i(t) \phi_i(t)
\end{equation}
where the wave function $\phi(t)$ implicitly depends on space, spinor,
and color indices.  Suppose further that we can find some eigenvalue
that changes adiabatically from negative to positive in going from
large negative to large positive time, as sketched in
Fig.~\ref{eflow}.  From this particular eigenstate construct the four
dimensional field
\begin{equation}
\psi(t)=e^{-\int_0^t E(t^\prime)\ dt^\prime} \phi(t).
\label{eigenflow}
\end{equation}
Because of the change in sign of the energy, the exponential factor
function goes to zero at both positive and negative large times, as
sketched in Fig.~\ref{zeromode}.

If we now consider the four dimensional Dirac operator applied to this
function we obtain
\begin{equation}
(\gamma_0 \partial_0 + \gamma_0 H(\vec A(t))) \psi(t)
=O(\partial_0 \psi(t)).
\end{equation}
If the evolution is adiabatic, the last term is small and we have an
approximate zero mode.  The assumption of adiabaticity is unnecessary
in the chiral limit of zero mass.  Then the eigenvalues of $D$ are
either real or occur in complex conjugate pairs.  Any unaccompanied
eigenvalue of $D$ occurs robustly at zero.  This is another
manifestation of the index theorem; we can count Euclidean-space zero
modes by studying the zero crossings appearing in the eigenvalues of
the Minkowski-space Hamiltonian.

In the above construction, the evolving eigenmode of $H$ is
accompanied by another of opposite energy and chirality.  Inserted
into Eq.~(\ref{eigenflow}), this will give a non-normalizable form for
the four dimensional field.  Thus we obtain only a single normalizable
zero mode for the Euclidean Dirac operator.  Note that if a small mass
term is included, the up going and down going Hamiltonian eigenstates
will mix and the crossing is forbidden.

This eigenvalue flow provides an intuitive picture of the anomaly
\cite{Ambjorn:1983hp}.  Start at early times with a filled Dirac sea
and all negative-energy eigenstates filled and then slowly evolve
through one of the above crossings.  In the process one of the filled
states moves to positive energy, leaving a non-empty positive energy
state.  At the same time the opposite chirality state moves from
positive to negative energy.  As long as the process is adiabatic, we
wind up at large time with one filled positive-energy state and one
empty negative-energy state.  As these are of opposite chirality,
effectively chirality is not conserved.  The result is particularly
dramatic in the weak interactions, where anomalies are canceled
between quarks and leptons.  This flow from negative to positive
energy states results in baryon non-conservation, although at an
unobservably small rate \cite{'tHooft:1976fv}.

\newpage
\Section{Chiral symmetry}
\label{chiral}
Much older a tool than the lattice, ideas based on chiral symmetry
have historically provided considerable insight into how the strong
interactions work.  In particular, this concept is crucial to our
understanding of why the pion is so much lighter than the rho, despite
them both being made of the same quarks.  Combining these ideas with
the lattice has provided considerable insight into many
non-perturbative issues in QCD.  Here we review the basic ideas of
chiral symmetry for the strong interactions.  A crucial aspect of this
discussion is the famous anomaly and its consequences for the
$\eta^\prime$ meson.

The classical Lagrangean for QCD couples left and right handed quark
fields only through mass terms.  Thus naively the massless theory has
independent conserved currents associated with each handedness.  For
$N_f$ massless flavors, this would be an independent $U(N_f)$ symmetry
associated with each chirality, giving what is often written in terms
of axial and vector fields as an $U(N_f)_V\otimes U(N_f)_A$ symmetry.
As is well known, this full symmetry does not survive quantization,
being broken to a $SU(N_f)_V\otimes SU(N_f)_A\otimes U(1)_B$, where
the $U(1)_B$ represents the symmetry of baryon number conservation.
The only surviving axial symmetries of the massless quantum theory are
non-singlet under flavor symmetry.

This breaking of the classical $U(1)$ axial symmetry is closely tied
to the possibility of introducing into massive QCD a CP violating
parameter, usually called $\Theta$.  For an extensive review, see
Ref.~\cite{Vicari:2008jw}.  While such a term is allowed from
fundamental principles, experimentally it appears to be extremely
small.  This raises an unresolved puzzle for attempts to unify the
strong interactions with the weak.  Since the weak interactions do
violate CP, why is there no residue of this remaining in the strong
sector below the unification scale?

One goal of this section is to provide a qualitative understanding of
the role of the $\Theta$ parameter in meson physics.  We concentrate
on symmetry alone and do not attempt to rely on any specific form for
an effective Lagrangean.  We build on the connection between $\Theta$
and a flavor-singlet $Z_{N_f}$ symmetry that survives the anomaly.  We
will see that, when the lightest quarks are made massive and
degenerate, a first order transition must occur when $\Theta$ passes
through $\pi$.  This transition is quite generic, but can be avoided
under limited conditions with one quark considerably lighter than the
others.  This discussion should also make it clear that the sign of
the quark mass is physically relevant for an odd number of flavors.
This is a non-perturbative effect that is invisible to naive
diagrammatic treatments.

Throughout this section we use the language of continuum field theory.
Of course underlying this we must assume some non-perturbative
regulator has been imposed so that we can make sense of various
products of fields, such as the condensing combination
$\sigma=\overline\psi\psi$.  For a momentum space cutoff, assume that
it is much larger than $\Lambda_{QCD}$.  Correspondingly, for a
lattice cutoff imagine that the lattice spacing is much smaller than
$1/\Lambda_{QCD}$.  In this section we ignore any lattice artifacts
that should vanish in the continuum limit.  We will return to such
issues later when we discuss lattice fermions.

\subsection{Effective potentials}

We begin with an elementary review of the concept of effective
potentials in quantum field theory.  In generic continuum field theory
language, consider the path integral for a scalar field
\begin{equation}
Z=\int d\phi e^{-S(\phi)}.
\end{equation}
After adding in some external sources
\begin{equation}
Z(J)=\int d\phi e^{-S(\phi)+J\phi},
\end{equation}
general correlation functions can be found by differentiating with
respect to $J$.  Here we use a shorthand notation that suppresses the
space dependence; {\it i.e.} $J\phi=\int dx J(x)\phi(x)$ in the
continuum, or $J\phi=\sum_i J_i\phi_i$ on the lattice.  

One can think of $J$ as an external force pulling on the field.  Such
a force will tend to drive the field to have an expectation value
\begin{equation}
\label{fieldforce}
\langle \phi \rangle_J=-{\partial F\over \partial J}
\end{equation}
where the free energy in the presence of the source is defined as
$F(J)=-\log(Z(J))$.

Now imagine inverting Eq.~(\ref{fieldforce}) to determine what value
of the force $J$ would be needed to give some desired expectation
value $\Phi$; {\it i.e.} we want to solve
\begin{equation}
\Phi(J)=\langle \phi \rangle_{J(\Phi)}=-{\partial F\over \partial J}
\end{equation}
for $J(\Phi)$.  In terms of this formal solution, construct the
``Legendre transform''
\begin{equation}
V(\Phi)=F(J(\Phi))+\Phi J(\Phi)
\end{equation}
and look at
\begin{equation}
{\partial V \over \partial \Phi}
=-\Phi {\partial J \over \partial \Phi}+J+\Phi{\partial J \over \partial \Phi}
=J.
\end{equation}
If we now turn off the sources, this derivative vanishes.  Thus the
expectation value of the field in the absence of sources occurs at an
extremum of $V(\Phi)$.  This quantity $V$ is referred to as the
``effective potential.''

An interesting formal property of this construction follows from
looking at the second derivative of $V$
\begin{equation}
{\partial^2 V \over \partial \Phi^2}={\partial J\over \partial \Phi}.
\end{equation}
Actually, it is easier to look at the inverse
\begin{equation}
{\partial \Phi \over \partial J}=
-{\partial^2 F \over \partial J^2}
=\langle \phi^2 \rangle - \langle \phi \rangle ^2
=\langle (\phi-\langle\phi\rangle)^2 \rangle \ge 0. 
\end{equation}
Thus this second derivative is never negative!  This first of all
shows we are actually looking for a minimum and not a maximum of $V$,
but it also implies that $V(\Phi)$ can only have ONE minimum!

This convexity property is usually ignored in conventional
discussions, where phase transitions are signaled by jumps between
distinct minima of the potential..  So what is going on?  Are phase
transitions impossible?  Physically, the more you pull on the field,
the larger the expectation of $\Phi$ will become.  It won't go back.
The proper interpretation is that we must do Maxwell's construction.
If we force the expectation of $\phi$ to lie between two distinct
stable phases, the system will phase separate into a heterogeneous
mixture.  In this region the effective potential is flat.  Note that
there is no large volume limit required in the above discussion.
However other definitions of $V$ can allow a small barrier at finite
volume due to surface tension effects.  A mixed phase must contain
interfaces, and their energy represents a small barrier.

\subsection{Goldstone Bosons}

Now we turn to a brief discussion on some formal aspects of Goldstone
Bosons.  Suppose we have a field theory containing a conserved current
\begin{equation} \partial_\mu j_\mu=0 
\end{equation} 
so the corresponding charge $Q=\int d^3x j_0(x)$ is a constant
\begin{equation} {dQ\over
  dt}=-i[H,Q]=0. 
\end{equation} 
Here $H$ is the Hamiltonian for the system under consideration.
Suppose, however, that for some reason the vacuum is not a singlet
under this charge
\begin{equation} Q|0\rangle \ne 0. 
\end{equation} 
Then there must exist a massless particle in the theory.  Consider the
state
\begin{equation}
\exp(i\theta\int d^3x j_0(x) e^{-\epsilon x^2}) |0\rangle
\end{equation} 
where $\epsilon$ is a convenient cutoff and $\theta$ some parameter.
As epsilon goes to zero this state by assumption is not the vacuum,
but since the Hamiltonian commutes with $Q$, the expectation value of
the Hamiltonian goes to zero (normalize so the ground state energy is
zero).  We can thus find a state that is not the vacuum but with
arbitrarily small energy.  The theory has no mass gap.  This situation
of having a symmetry under which the vacuum is not invariant is
referred to as ``spontaneous symmetry breaking.''  The low energy
states represent massless particles called Goldstone
bosons \cite{Goldstone:1962es}.

Free massless field theory is a marvelous example where everything can
be worked out.  The massless equation of motion
\begin{equation} \partial_\mu\partial_\mu \phi=0 
\end{equation} 
can be written in the form 
\begin{equation} 
\partial_\mu j_\mu=0 
\end{equation}
where 
\begin{equation} j_\mu=\partial_\mu \phi.  
\end{equation} 
The broken symmetry is the invariance of the Lagrangean $L=\int
d^4x(\partial_\mu\phi)^2/2$ under constant shifts of the field
\begin{equation} 
\phi\rightarrow\phi+c. 
\end{equation} 
Note that $j_0=\partial_0\phi=\pi$ is the conjugate variable to
$\phi$.  Since it is a free theory, one could work out explicitly
\begin{equation} \langle 0 | \exp(i\theta\int d^3x j_0(x)
e^{-\epsilon x^2/2}) |0\rangle.  
\end{equation} 
We can, however, save ourselves the work using dimensional analysis.
The field $\phi$ has dimensions of inverse length, while $j_0$ goes as
inverse length squared.  Thus $\theta$ above has units of inverse
length.  These are the same dimensions as $\epsilon^2$.  Now for a
free theory, by Wick's theorem, the answer must be Gaussian in
$\theta$.  We conclude that the above overlap must go as
\begin{equation} 
\exp(-C\theta^2/\epsilon^4) 
\end{equation} 
where $C$ is some non-vanishing dimensionless number.  This expression
rapidly goes to zero as epsilon becomes small, showing that the vacuum
is indeed not invariant under the symmetry.  In the limit of
$\epsilon$ going to zero, we obtain a new vacuum that is not even in
the same Hilbert space.  The overlap of this new state with any local
polynomial of fields on the original vacuum vanishes.

It is perhaps interesting to note that the canonical commutation
relations $[\pi(x),\phi(y)]=i\delta(x-y)$ imply for the currents
\begin{equation}
[j_0(x),j_i(y)]=-i{d\over dx}\delta(x-y). 
\end{equation}
In a Hamiltonian formulation, equal time commutators of different
current components must involve derivatives of delta functions.  This
is a generic property and does not depend on the symmetry being
spontaneously broken \cite{Schwinger:1959xd}.  

\subsection {Pions and spontaneous symmetry breaking}

We now extend the effective potential to a function of several
relevant meson fields in QCD.  Intuitively, $V$ represents the energy
of the lowest state for a given field expectation, as discussed more
formally earlier via a Legendre transformation.  Here we will ignore
the result that effective potentials must be convex functions of their
arguments.  As discussed, this issue is easily understood in terms of
a Maxwell construction involving the phase separation that will occur
if one asks for a field expectation in what would otherwise be a
concave region.  Thus we will use the traditional language of
spontaneous symmetry breaking corresponding to having an effective
potential with more than one minimum.  When the underlying theory
possesses some symmetry but the individual minima do not, spontaneous
breaking comes about when the vacuum selects one of the minima
arbitrarily.  The discussion here closely follows that in
Ref.~\cite{Creutz:2009kx}.

\begin{figure}
\centering
\includegraphics[width=2.5in]{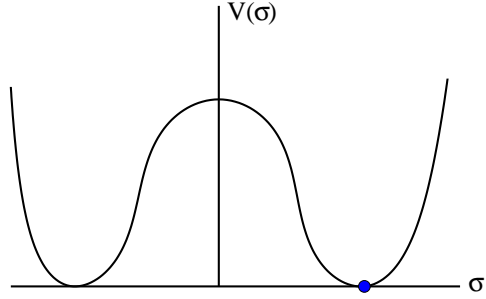}
\caption{\label{v1} 
Spontaneous chiral symmetry breaking is represented by a double well
effective potential with the vacuum settling into one of two possible
minima.  In this minimum chiral symmetry is broken by the selection of
a specific value for the quark condensate. 
}
\end{figure}

We work here with the composite scalar and pseudoscalar fields
\begin{equation}
\matrix{
\sigma&\sim&\overline\psi \psi\cr
\pi_\alpha&\sim& i\overline\psi \lambda_\alpha \gamma_5 \psi \cr
\eta^\prime &\sim& i\overline\psi\gamma_5\psi. \cr
}
\label{mesonfields}
\end{equation}
Here the $\lambda_\alpha$ are the generators for the flavor group
$SU(N_f)$.  They are generalizations of the usual Gell-Mann matrices
from $SU(3)$; however, now we are concerned with the flavor group, not
the internal symmetry group related to confinement.  As mentioned
earlier, we must assume that some sort of regulator, perhaps a
lattice, is in place to define these products of fields at the same
point.  Indeed, most of the quantities mentioned in this section are
formally divergent, although we will concentrate on those aspects that
survive the continuum limit.

To simplify the discussion, consider degenerate quarks with a small
common mass $m$.  Later we will work out in some detail the two flavor
case for non-degenerate quarks.  It is also convenient to initially
restrict $N_f$ to be even, saving for later some interesting
subtleties arising with an odd number of flavors.  And we assume $N_f$
is small enough to maintain asymptotic freedom as well as avoiding any
possible conformal phases.

The conventional picture of spontaneous chiral symmetry breaking in
the limit of massless quarks assumes that the vacuum acquires a quark
condensate with
\begin{equation}
\langle\overline\psi\psi\rangle
=\langle\sigma\rangle =v \ne 0.
\end{equation}
In terms of the effective potential,
$V(\sigma)$ should acquire a double well structure, as sketched in
Fig.~\ref{v1}.  The symmetry under $\sigma\leftrightarrow -\sigma$ is
associated with the invariance of the action under a flavored chiral
rotation.  For example, with two flavors the change of variables
\begin{equation}
\matrix{
\psi\rightarrow e^{i\pi\tau_3\gamma_5/2}\psi=i\tau_3\gamma_5\psi\cr
\overline \psi\rightarrow \overline\psi e^{i\pi\tau_3\gamma_5/2}
=\overline \psi\ i\tau_3\gamma_5\cr
}
\end{equation}
leaves the massless action invariant but takes $\sigma$ to its
negative.  Here $\tau_3$ is the conventional Pauli matrix
corresponding to the third component of isospin.

\begin{figure}
\centering
\includegraphics[width=2.5in]{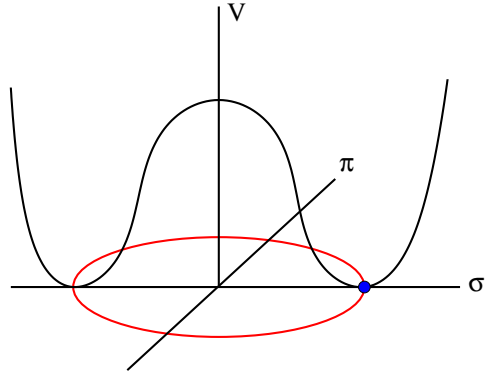}
\caption{\label{v0} 
The flavor non-singlet pseudoscalar mesons are Goldstone bosons
corresponding to flat directions  in the effective potential.
}
\end{figure}

Extending the effective potential to a function of the non-singlet
pseudoscalar fields gives the standard picture of Goldstone bosons.
These are massless when the quark mass vanishes, corresponding to
$N_f^2-1$ ``flat'' directions for the potential.  One such direction
is sketched schematically in Fig.~\ref{v0}.  For the two flavor case,
these rotations represent a symmetry mixing the sigma field with the
pions
\begin{equation}
\matrix{
\sigma\rightarrow\phantom{+}  \sigma \cos(\phi)+\pi^\alpha \sin(\phi)\cr
\pi^\alpha\rightarrow -\sigma \sin(\phi)+\pi^\alpha \cos(\phi).\cr
}
\end{equation}
In some sense the pions are waves propagating through the
non-vanishing sigma condensate.  The oscillations of these waves occur
in a direction ``transverse'' to the sigma expectation.  They are
massless because there is no restoring force in that direction.

\begin{figure}
\centering
\includegraphics[width=2.5in]{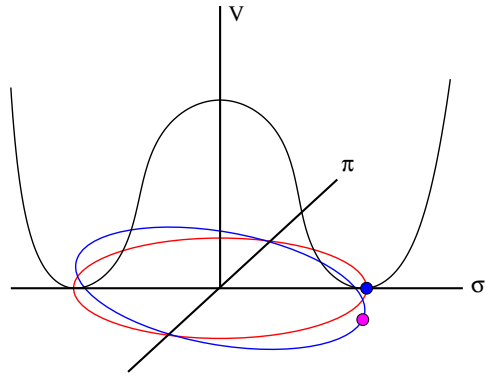}
\caption{\label{v3}
A small quark mass term tilts the effective potential, selecting one
direction for the true vacuum and giving the Goldstone bosons a mass
proportional to the square root of the quark mass.
}
\end{figure}

If we now introduce a small mass for the quarks, this will effectively
tilt the potential $V(\sigma)\rightarrow V(\sigma)-m\sigma$.  This
selects one minimum as the true vacuum.  The tilting of the potential
breaks the global symmetry and gives the Goldstone bosons a small mass
proportional to the square root of the quark mass, as sketched in
Fig.~\ref{v3}.  The standard chiral Lagrangean approach is a
simultaneous expansion in the masses and momenta of these light
particles.

As discussed earlier, in a Hamiltonian approach Goldstone bosons are
associated with conserved currents with charges that do not leave the
vacuum invariant.  In the present case these are the axial currents
formally given by the quark bilinears
\begin{equation}
A_\mu^\alpha=\overline\psi \lambda^\alpha \gamma_\mu\gamma_5 \psi. 
\end{equation} 
Combined with the vector fields
\begin{equation}
V_\mu^\alpha=\overline\psi \lambda^\alpha \gamma_\mu \psi,
\end{equation} 
these give rise to the famous algebra of currents
\begin{eqnarray}
&&[V_0^\alpha(x),V_0^\beta(y)]=if^{\alpha\beta\gamma}V_0^\gamma(x)\delta(x-y)\cr
&&[V_0^\alpha(x),A_0^\beta(y)]=if^{\alpha\beta\gamma}A_0^\gamma(x)\delta(x-y)\cr
&&[A_0^\alpha(x),A_0^\beta(y)]=if^{\alpha\beta\gamma}V_0^\gamma(x)\delta(x-y).
\end{eqnarray}
with the $f^{\alpha\beta\gamma}$ being the structure constants for the
internal symmetry group.  Indeed, it was this algebra that motivated
Bjorken to propose the idea of scaling in deep inelastic lepton
scattering \cite{Bjorken:1967px,Bjorken:1966jh}.

\subsection{The Sigma model}

Much of the structure of low energy QCD is nicely summarized in terms
of an effective chiral Lagrangean formulated in terms of a field which
is an element of the underlying flavor group.  In this section we
review this model for the strong interactions with three quarks,
namely up, down, and strange.  The theory has an approximate SU(3)
symmetry, broken by unequal masses for the quarks.  We work with the
familiar octet of light pseudoscalar mesons $\pi_\alpha$ with
$\alpha=1\ldots 8$ and consider an SU(3) valued field
\begin{equation}
\label{sigma}
\Sigma=\exp(i\pi_\alpha \lambda_\alpha/f_\pi)\in SU(3).
\end{equation}
Here the $\lambda_\alpha$ are the usual Gell-Mann matrices which
generate the flavor group and $f_\pi$ is a dimensional constant with a
phenomenological value of about 93 MeV.  We follow the normalization
convention that ${\rm Tr} \lambda_\alpha \lambda_\beta =
2\delta_{\alpha\beta}$.  The neutral pion and the eta meson will play
a special role later in this review; they are the coefficients of
the commuting generators
\begin{equation}
\lambda_3={1\over \sqrt 3}\pmatrix{1&0&0\cr
0&-1&0\cr 0&0&0\cr}
\end{equation}
and 
\begin{equation}
\lambda_8={1\over \sqrt 3}\pmatrix{1&0&0\cr
0&1&0\cr 0&0&-2\cr},
\end{equation}
respectively.  In the chiral limit of vanishing quark masses, the
interactions of the eight massless Goldstone bosons are modeled with
the effective Lagrangean density
\begin{equation}
\label{kinetic}
L_0={f_\pi^2\over 4}{\rm Tr}(\partial_\mu \Sigma^\dagger \partial_\mu \Sigma).
\end{equation}
The non-linear constraint of $\Sigma$ onto the group SU(3) makes this
theory non-renormalizable.  It is to be understood only as the
starting point for an expansion of particle interactions in powers of
their momenta.  Expanding Eq.~(\ref{kinetic}) to second order in the
meson fields gives the conventional kinetic terms for our eight
mesons.

This theory is invariant under parity and charge conjugation.  These
operators are represented by simple transformations
\begin{equation}
\matrix{
&P:\ \Sigma\ \rightarrow\ \Sigma^{-1}\cr
&CP:\ \Sigma\ \rightarrow\ \Sigma^{*}\cr
}
\end{equation}
where the operation $*$ refers to complex conjugation.  The eight
meson fields are pseudoscalars.  The neutral pion and the eta meson
are both even under charge conjugation.

With massless quarks, the underlying quark-gluon theory has a chiral
symmetry under
\begin{equation}
\matrix{
&\psi_L\rightarrow g_L\psi_L \cr
&\psi_R\rightarrow g_R\psi_R. \cr
}
\end{equation}   
Here $(g_L,g_L)$ is in $SU(3)\otimes SU(3)$ and $\psi_{L,R}$
represent the chiral components of the quark fields, with flavor
indices understood.  This symmetry is expected to be broken
spontaneously to a vector SU(3) via a vacuum expectation value for
$\overline \psi_L \psi_R$.  This motivates the sigma model through the
identification
\begin{equation}
\langle 0 | \overline \psi_L \psi_R | 0 \rangle \leftrightarrow v \Sigma.
\label{vec}
\end{equation}
The quantity $v$, of dimension mass cubed, characterizes the strength
of the spontaneous breaking of this symmetry.  Thus the effective
field transforms under a chiral symmetry of form
\begin{equation}
\Sigma\rightarrow g_L \Sigma g_R^\dagger.
\end{equation} 
The Lagrangean density in Eq.~(\ref{kinetic}) is the simplest
non-trivial expression invariant under this symmetry.

The quark masses break the chiral symmetry explicitly.  From the
analogy in Eq.~(\ref{vec}), these are introduced through a 3 by 3 mass
matrix $M$ appearing in a potential term added to the Lagrangean
density
\begin{equation}
L= L_0 - v {\rm Re\ Tr}(\Sigma M).
\end{equation} 
Here $v$ is the same dimensionful factor appearing in Eq.~(\ref{vec}).
The chiral symmetry of our starting theory shows the physical
equivalence of a given mass matrix $M$ with a rotated matrix
$g_R^\dagger M g_L$.  Using this freedom we can put the mass matrix
into a standard form.  We will assume it is diagonal with increasing
eigenvalues
\begin{equation}
M=\pmatrix{ 
m_u & 0 & 0 \cr
0 & m_d & 0 \cr
0 & 0 & m_s \cr
}
\end{equation} 
representing the up, down, and strange quark masses.  Note that this
matrix has both singlet and octet parts under flavor symmetry
\begin{equation}
M={m_u+m_d+m_s\over 3}
+{m_u-m_d\over 2}\ \lambda_3+{m_u+m_d-2 m_s\over 2\sqrt 3}\ \lambda_8.
\end{equation} 

In general the mass matrix can still be complex.  The chiral symmetry
allows us to move phases between the masses, but the determinant of
$M$ is invariant and physically meaningful.  Under charge conjugation
the mass term would only be invariant if $M=M^*$.  If $|M|$ is not
real, then its phase is the famous CP violating parameter that we will
extensively discuss later.  For the moment, however, we take all quark
masses as real.

To lowest order the pseudoscalar meson masses appear on expanding the
mass term quadratically in the meson fields.  This generates an
effective mass matrix for the eight mesons
\begin{equation}
{\cal M}_{\alpha\beta}\ \propto\ {\rm Re\ Tr}\ \lambda_\alpha\lambda_\beta M.
\end{equation}
The isospin breaking up-down mass difference gives this matrix an off
diagonal piece mixing the $\pi_0$ and the $\eta$
\begin{equation}
{\cal M}_{3,8}\ \propto\ m_u-m_d.
\end{equation}
The eigenvalues of this matrix give the standard mass relations
\begin{eqnarray}
\label{mesons}
&& m_{\pi_0}^2 \propto\  {2\over 3} \bigg(m_u+m_d+m_s
-\sqrt{m_u^2+m_d^2+m_s^2-m_um_d-m_um_s-m_dm_s}\bigg)\cr
&&m_{\eta}^2 \propto \ {2\over 3} \bigg(m_u+m_d+m_s
+\sqrt{m_u^2+m_d^2+m_s^2-m_um_d-m_um_s-m_dm_s}\bigg)\cr
&&m_{\pi_+}^2= \  m_{\pi_-}^2\propto m_u+m_d\cr
&&m_{K_+}^2= \ m_{K_-}^2\propto m_u+m_s\cr
&&m_{K_0}^2= \ m_{\overline K_0}^2\propto m_d+m_s.
\end{eqnarray}  
Here we label the mesons with their conventional names.

Redundancies in these relations test the validity of the model.  For
example, comparing two expressions for the sum of the three quark
masses
\begin{equation}
{2(m_{\pi_+}^2+m_{K_+}^2+m_{K_0}^2)
\over 3(m_{\eta}^2+m_{\pi_0}^2)} \sim 1.07
\end{equation}  
suggests the symmetry should be good to a few percent.  Further ratios
of meson masses then give estimates for the ratios of the quark masses
\cite{Kaplan:1986ru,Weinberg:1977hb,Leutwyler:1996qg}.  For one such
combination, look at
\begin{equation}
{m_u\over m_d}=
{
m_{\pi^+}^2+m_{K_+}^2-m_{K_0}^2\over
m_{\pi^+}^2-m_{K_+}^2+m_{K_0}^2}\sim 0.66
\end{equation}
This particular combination is polluted by electromagnetic effects;
another combination that partially cancels such while ignoring small
$m_um_d/m_s$ corrections in expanding the square root in
Eq.~(\ref{mesons}) is
\begin{equation} 
{m_u\over
m_d}= { 2m_{\pi^0}^2-m_{\pi^0}^2+m_{K_+}^2-m_{K_0}^2\over
m_{\pi^+}^2-m_{K_+}^2+m_{K_0}^2}\sim 0.55
\label{updown}
\end{equation} 
In a moment we will comment on a third combination for this ratio.
For the strange quark, one can take
\begin{equation}
{2 m_s\over m_u+m_d}=
{
m_{K_+}^2+m_{K_0}^2-m_{\pi^+}^2\over
m_{\pi^+}^2
}\sim 26.
\end{equation}

Of course as discussed earlier the quark masses are scale dependent.
While their ratios are more stable, we will see later how these ratios
also acquire some scale dependence.  Nevertheless, from mass
differences such as $m_n-m_p\sim 1.3{\rm MeV} $ and
$m_{K_0}-m_{K_+}\sim 4.0{\rm MeV}$ we conclude that the up and down
quark masses in these effective models are typically of order a few
MeV, while the strange quark mass is of order 100 MeV.  These are what
are known as ``current'' quark masses, related to chiral symmetries
and current algebra.  In contrast, since the proton is made of three
quarks, some simple quark models consider ``constituent'' quark masses
of a few hundred Mev; these are substantially larger because they
include the energy contained in the gluon fields.

While phenomenology, i.e. Eq.~(\ref{updown}), seems to suggest that
the up quark is not massless, there remains a lot of freedom in
extracting that ratio from the pseudoscalar meson masses.  From
Eq.~(\ref{mesons}), the sum of the $\eta$ and $\pi_0$ masses squared
should be proportional to the sum of the three quark masses.
Subtracting off the neutral kaon mass should leave just the up quark.
Thus motivated, look at
\begin{equation}
{m_u\over m_d}=
{
 3(m_{\eta}^2+m_{\pi_0}^2)/2-2m_{K_0}^2
\over
 m_{\pi^+}^2-m_{K_+}^2+m_{K_0}^2
}
\sim-0.8
\end{equation}
This strange result is probably a consequence of $SU(3)$ breaking
inducing eta and eta-prime mixing, thus lowering the eta mass.  But
one might worry that depending on what combination of mesons one uses,
even the sign of the up quark mass is ambiguous.  Attempts to extend
the naive quark mass ratio estimates to higher orders in the chiral
expansion have shown that there are fundamental ambiguities in the
definition of the quark masses \cite{Kaplan:1986ru}.  An important
message of later sections is that this ambiguity is an inherent
property of QCD.

Note that in Eq.~(\ref{mesons}) the neutral pion mass squared can
become negative if
\begin{equation}
m_u< {-m_dm_s\over m_d+m_s}.
\end{equation}
This unphysical situation will result in a condensation of the pion
field and a spontaneous breaking of CP symmetry \cite{Creutz:2003xu}.
This is closely tied to the possibility of a CP violating term in QCD
that we will discuss in later sections.

\newpage
\Section{The chiral anomaly}
The picture of pions as approximate Goldstone bosons is, of course,
completely standard.  It is also common lore that the anomaly prevents
the $\eta^\prime$ from being a Goldstone boson and leaves it with a
mass of order $\Lambda_{QCD}$, even in the massless quark limit.  The
issue is that the effective potential $V$ considered as a function of
the fields in Eq.~(\ref{mesonfields}) must not be symmetric under an
anomalous rotation between $\eta^\prime$ and $\sigma$
\begin{equation}
\label{rotate}
\matrix{
\sigma\rightarrow\phantom{+}  \sigma \cos(\phi)+\eta^\prime \sin(\phi)\cr
\eta^\prime\rightarrow -\sigma \sin(\phi)+\eta^\prime \cos(\phi).\cr
}
\end{equation}
In the next subsection we discuss how this symmetry disappears and its
connection to the zero modes of the Dirac operator.

If we consider the effective potential as a function of the fields
$\sigma$ and $\eta^\prime$, it should have a minimum at $\sigma\sim v$
and $\eta^\prime \sim 0$.  Expanding about that point we expect a
qualitative form
\begin{equation}
V(\sigma,\eta^\prime)\sim m_\sigma^2 (\sigma-v)^2 
+ m_{\eta^\prime}^2 {\eta^\prime}^2
+O((\sigma-v)^3, {\eta^\prime}^4).
\end{equation}
We expect both $m_\sigma$ and $m_{\eta^\prime}$ to remain of order
$\Lambda_{QCD}$, even in the chiral limit.  And, at least with an even
number of flavors as we are currently considering, there should be a
second minimum with $\sigma\sim -v$.  

At this point one can ask whether we know anything else about the
effective potential in the $(\sigma,\eta^\prime)$ plane.  We will
shortly see that indeed we do, and the potential has a total of $N_f$
equivalent minima in the chiral limit.  But first we review how the
above minima arise in quark language.

\subsection{What broke the symmetry?}
\label{quarks}

The classical QCD Lagrangean has a symmetry under a rotation of the
underlying quark fields
\begin{equation}
\matrix{
\psi \rightarrow e^{i\phi\gamma_5/2}\psi\cr
\overline\psi \rightarrow \overline\psi e^{i\phi\gamma_5/2}.\cr
}
\end{equation}
This corresponds directly to the transformation of the composite
fields given in Eq.~\ref{rotate}.  This symmetry is ``anomalous'' in
that any regulator must break it with a remnant surviving in the
as the regulator is removed \cite{Adler:1969gk,Adler:1969er,Bell:1969ts}.
With the lattice this concerns the continuum limit.

The specifics of how the anomaly works depend on the details of the
regulator.  Here we will follow Fujikawa \cite{Fujikawa:1979ay} and
consider the fermionic measure in the path integral.  If we make the
above rotation on the field $\psi$, the measure changes by the
determinant of the rotation matrix
\begin{equation}
\label{fujikawa}
d\psi\rightarrow |e^{-i\phi\gamma_5/2}| d\psi
=e^{-i\phi {\rm Tr}\gamma_5/2} d\psi.
\end{equation}
Here is where the subtlety of the regulator comes in.  Naively
$\gamma_5$ is a simple four by four traceless matrix.  If it is indeed
traceless, then the measure would be invariant.  However, in the
regulated theory this is not the case.  This is intimately tied with
the index theorem for the Dirac operator in topologically non-trivial
gauge fields.

A typical Dirac action takes the form $\overline\psi (D+m)\psi$ with
the kinetic term $D$ a function of the gauge fields.  In the naive
continuum theory $D$ is anti-Hermitean, $D^\dagger=-D$, and
anti-commutes with $\gamma_5$, i.e. $[D,\gamma_5]_+=0$.  What
complicates the issue with fermions is the index theorem discussed
earlier and reviewed in Ref.~\cite{Coleman:1978ae}.  If a background
gauge field has winding $\nu$, then there must be at least $\nu$ exact
zero eigenvalues of $D$.  Furthermore, on the space spanned by the
corresponding eigenvectors, $\gamma_5$ can be simultaneously
diagonalized with $D$.  The net winding number equals the number of
positive eigenvalues of $\gamma_5$ minus the number of negative
eigenvalues.  In this subspace the trace of $\gamma_5$ does not
vanish, but equals $\nu$.

What about the higher eigenvalues of $D$?  We discussed these earlier
when we formulated the index theorem.  Because $[D,\gamma_5]_+=0$,
non-vanishing eigenvalues appear in opposite sign pairs; i.e. if
$D|\psi\rangle =\lambda|\psi\rangle$ then $D\gamma_5|\psi\rangle
=-\lambda\gamma_5|\psi\rangle$.  For an anti-Hermitean $D$, these
modes are orthogonal with $\langle\psi|\gamma_5|\psi\rangle=0$.  As a
consequence, $\gamma_5$ is traceless on the subspace spanned by each
pair of eigenvectors.

So what happened to the opposite chirality states to the zero modes?
In a regulated theory they are in some sense ``above the cutoff.''  In
a simple continuum discussion they have been ``lost at infinity.''
With a lattice regulator there is no ``infinity''; so, something more
subtle must happen.  With the
overlap\cite{Ginsparg:1981bj,Neuberger:1997fp} or
Wilson\cite{Wilson:1975id} fermions, discussed in more detail later,
one gives up the anti-Hermitean character of $D$.  Most eigenvalues
still occur in conjugate pairs and do not contribute to the trace of
$\gamma_5$.  However, in addition to the small real eigenvalues
representing the zero modes, there are additional modes where the
eigenvalues are also real but large.  With Wilson fermions these
appear as massive doubler states.  With the overlap, the eigenvalues
are constrained to lie on a circle.  In this case, for every exact
zero mode there is another mode with the opposite chirality lying on
the opposite side of the circle.  These modes are effectively massive
and break chiral symmetry.  The necessary involvement of both small
and large eigenvalues warns of the implicit danger in attempts to
separate infrared from ultraviolet effects.  When the anomaly is
concerned, going to short distances is not sufficient for ignoring
non-perturbative effects related to topology.

So with the regulator in place, the trace of $\gamma_5$ does not
vanish on gauge configurations of non-trivial topology.  The change of
variables indicated in Eq.~\ref{fujikawa} introduces into the path
integral a modification of the weighting by a factor
\begin{equation}
\label{nu}
e^{-i\phi {\rm Tr}\gamma_5}=e^{-i\phi N_f\nu}.
\end{equation}
Here we have applied the rotation to all flavors equally, thus the
factor of $N_f$ in the exponent.  The conclusion is that gauge
configurations that have non-trivial topology receive a complex weight
after the anomalous rotation.
Although not the topic of discussion here, note that this factor
introduces a sign problem if one wishes to study this physics via
Monte Carlo simulations.  Here we treat all $N_f$ flavors
equivalently; this corresponds to dividing the conventionally defined
CP violation angle, to be discussed later, equally among the flavors,
i.e. effectively $\phi=\Theta/N_f$.

\subsection{A discrete chiral symmetry}
\label{znf}

\begin{figure}
\centering
\includegraphics[width=2.5in]{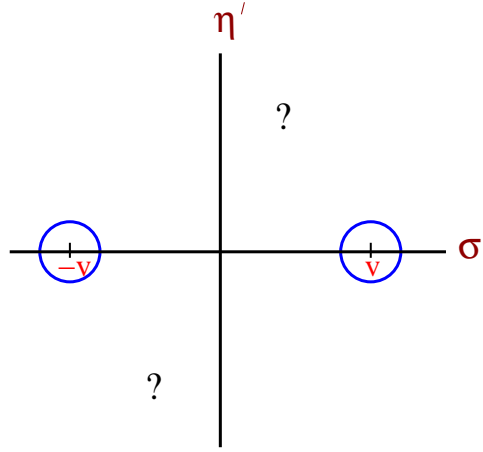}
\caption{\label{potential0} We have two minima in the
  $\sigma,\eta^\prime$ plane located at $\sigma=\pm v$ and
  $\eta^\prime=0$.  The circles represent that the fields will
  fluctuate in a small region about these minima.  Can we find any
  other minima?  }
\end{figure}

We now return to the effective Lagrangean language of before.  For
the massless theory, the symmetry under $\sigma\leftrightarrow
-\sigma$ indicates that we expect at least two minima for the
effective potential considered in the $\sigma,\eta^\prime$ plane.
These are located as sketched in Fig.~\ref{potential0}.  Do we know
anything about the potential elsewhere in this plane?  The answer is
yes, indeed there are actually $N_f$ equivalent minima.

It is convenient to separate the left- and right-hand parts of the
fermion field
\begin{eqnarray}
&&\psi_{L,R}={1\over 2} (1\pm\gamma_5) \psi\cr
&&\overline\psi_{L,R}=\overline\psi {1\over 2} (1\mp\gamma_5). 
\end{eqnarray}
The mass term is thus
\begin{equation}
m\overline\psi \psi=m(\overline\psi_L\psi_R
+\overline\psi_R \psi_L)
\end{equation}
and mixes the left and right components.

\begin{figure}
\centering
\includegraphics[width=2.5in]{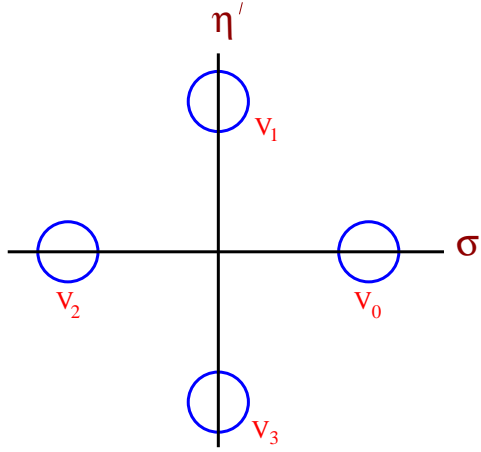}
\caption{\label{potential1}
For four massless flavors we have four equivalent
minima in the  $\sigma,\eta^\prime$ plane.  This generalizes to $N_f$
minima with $N_f$ flavors.
}
\end{figure}

Using this notation, due to the anomaly the singlet rotation
\begin{equation}
\psi_L\rightarrow e^{i\phi}\psi_L
\label{singlet}
\end{equation}
is not a valid symmetry of the theory for generic values of the angle
$\phi$.  On the other hand, flavored chiral symmetries should
survive, and in particular 
\begin{equation}
\psi_L\rightarrow
g_L\psi_L = e^{i\phi_\alpha \lambda_\alpha}\psi_L
\label{flavored}
\end{equation}
is expected to be a valid symmetry for any set of angles
$\phi_\alpha$.  The point of this subsection is that, for special
special discrete values of the angles, the rotations in
Eq.~\ref{singlet} and Eq.~\ref{flavored} can coincide.  At such values
the singlet rotation becomes a valid symmetry.  In particular, note
that
\begin{equation}
g=e^{2\pi i\phi/N_f} \in Z_{N_f} \subset SU(N_f).
\end{equation}
Thus a valid discrete symmetry involving only $\sigma$ and
$\eta^\prime$ is
\begin{equation}
\matrix{
\sigma\rightarrow \phantom{+} \sigma\cos(2\pi/N_f)
+\eta^\prime \sin(2\pi/N_f)\cr
\eta^\prime\rightarrow -\sigma \sin(2\pi/N_f)+\eta^\prime
\cos(2\pi/N_f).\cr
}
\end{equation}
The potential $V(\sigma,\eta^\prime)$ has a $Z_{N_f}$ symmetry
manifested in $N_f$ equivalent minima in the $(\sigma,\eta^\prime)$
plane.  For four flavors this structure is sketched in
Fig.~\ref{potential1}.

This discrete flavor singlet symmetry arises from the trivial fact
that $Z_N$ is a subgroup of both $SU(N)$ and $U(1)$.  At the quark
level the symmetry is easily understood since the quark measure
receives an additional phase proportional to the winding number from
every flavor.  With $N_F$ flavors, these multiply together making
\begin{equation}
\psi_L\rightarrow e^{2\pi i/N_f} \psi_L
\end{equation}
a valid symmetry even though rotations by smaller angles are not.

\begin{figure}
\centering
\includegraphics[width=2.5in]{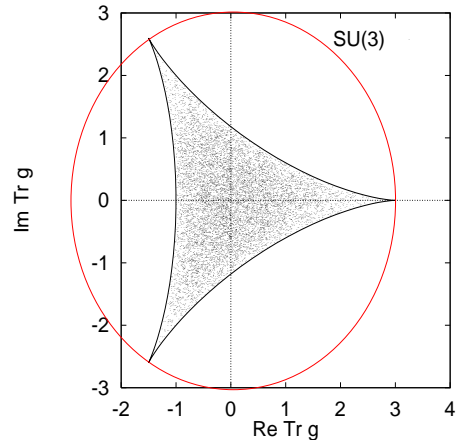}
\caption{\label{su3circle} The real and imaginary parts for the traces
of 10,000 randomly chosen $SU(3)$ matrices.  All points lie within the
boundary representing matrices of the form $\exp(i\phi\lambda_8)$.
The tips of the three points represent the center of the group.  The
outer curve represents the boundary that would be found if the group
was the full $U(1)$.  Taken from Ref.~\cite{Creutz:1995wf}.  }
\end{figure}

\begin{figure}
\centering
\includegraphics[width=2.5in]{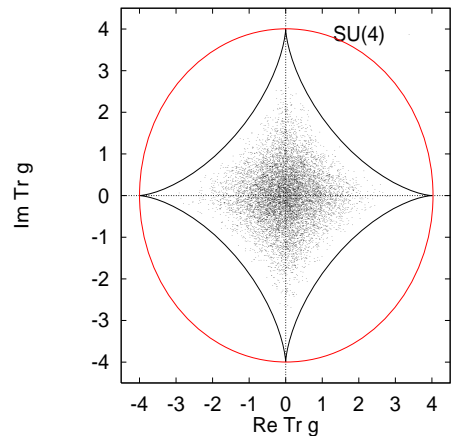}
\caption{\label{su4circle} The generalization of Fig.~\ref{su3circle}
to $SU(4)$.  The real and imaginary parts for the traces of 10,000
randomly chosen $SU(4)$ matrices.  Taken from
Ref.~\cite{Creutz:1995wf}.  }
\end{figure}

The role of the $Z_N$ center of $SU(N)$ is illustrated graphically in
Fig.~\ref{su3circle}, taken from Ref.~\cite{Creutz:1995wf}.  Here we
plot the real and the imaginary parts of the traces of 10,000 $SU(3)$
matrices drawn randomly with the invariant group measure.  The region
of support only touches the $U(1)$ circle at the elements of the
center.  All elements lie on or within the curve mapped out by
elements of form $\exp(i\phi\lambda_8)$.  Figure \ref{su4circle} is a
similar plot for the group $SU(4)$.

\subsection{The 't Hooft vertex}
The consequences of non-trivial gauge topology and the connections to
the anomaly are often described in terms of an effective multi fermion
interaction referred to as the `` `t Hooft vertex.''  To understand
the 't Hooft interaction in path integral language, we begin with a
reminder of the underlying strategy of lattice simulations.  Consider
the generic path integral, or ``partition function,'' for quarks and
gluons
\begin{equation}
Z=\int (dA)(d\psi\ d\overline\psi) 
\exp\left(-S_g(A)-\overline\psi D(A)\psi\right). 
\end{equation}
Here $A$ denotes the gauge fields and $\overline\psi,\psi$ the quark
fields.  The pure gauge part of the action is $S_g(A)$ and the matrix
describing the fermion part of the action is $D(A)$.  Since direct
numerical evaluation of the fermionic integrals appears to be
impractical, the Grassmann integrals are conventionally evaluated
analytically, reducing the partition function to
\begin{equation}
Z=\int (dA)\ e^{-S_g(A)}\ |D(A)|.
\end{equation}
Here $|D(A)|$ denotes the determinant of the Dirac matrix evaluated in
the given gauge field.  Thus motivated, the basic lattice approach is
to generate a set of random gauge configurations weighted by
$\exp(-S_g(A))\ |D(A)|$.  Given an ensemble of such configurations,
one then estimates physical observables by averages over this
ensemble.

This procedure seems innocent enough, but it can run into trouble when
one has massless fermions and corresponding zero modes associated with
topology.  To see the issue, write the determinant as a product of the
eigenvalues $\lambda_i$ of the matrix $D$.  In general $D$ may not be
a normal matrix; so, one should pick either left or right eigenvectors
at one's discretion.  This is a technical detail that will not play
any further role here.  In order to control infrared issues with
massless quarks, introduce a small explicit mass $m$ and reduce the
path integral to
\begin{equation}
Z=\int (dA)\ e^{-S_g(A)}\ \prod (\lambda_i+m). 
\end{equation}

Now suppose we have a configuration where one of the eigenvalues of
$D(A)$ vanishes, {\it i.e.} assume that some $\lambda_i=0$.  This, of
course, is what happens with non-trivial topology present.  As we take
the mass to zero, any configurations involving such an eigenvalue will
drop out of the ensemble.  At first one might suspect this would be a
set of measure zero in the space of all possible gauge fields.
However, as discussed above, the index theorem ties gauge field
topology to such zero modes.  In general these modes are robust under
small deformations of the fields.  Under the traditional lattice
strategy the corresponding configurations would then have zero weight
in the massless limit.  The naive conclusion is that such
configurations are irrelevant to physics in the chiral limit.

It was this reasoning that 't Hooft showed to be incorrect.  Indeed,
he demonstrated that it is natural for some observables to have $1/m$
factors when zero modes are present.  These can cancel the terms
linear in $m$ from the determinant, leaving a finite contribution.

As a simple example, consider the quark condensate in one flavor QCD 
\begin{equation}
\langle \overline\psi \psi\rangle=
{1\over VZ}
\int (dA)\ 
e^{-S_g}\ |D|\ \ {\rm Tr}D^{-1}.
\end{equation}
Here $V$ represents the system volume, inserted to give an intensive
quantity.  Expressing the fermionic factors in terms of the
eigenvalues of $D$ reduces this to
\begin{equation}
\langle \overline\psi \psi\rangle=
{1\over VZ}
\int (dA)\ 
e^{-S_g}\ 
\left( \prod_i (\lambda_i+m)\right)
\ \ \sum_j {1\over \lambda_j+m}.
\end{equation}
Now if there is a mode with $\lambda_i=0$, the factor of $m$ is
canceled by a $1/m$ piece in the trace of $D^{-1}$.  Configurations
containing a zero mode give a constant contribution to the condensate
and this contribution survives in the massless limit.  Note that this
effect is unrelated to spontaneous breaking of chiral symmetry and
appears even with finite volume.

This contribution to the condensate is special to the one-flavor
theory.  Because of the anomaly, this quark condensate is not an order
parameter for any symmetry.  With more fermion species there will be
additional factors of $m$ from the determinant.  Then the effect of
the 't Hooft vertex is of higher order in the fermion fields and does
not appear directly in the condensate.  For two or more flavors the
standard Banks-Casher picture \cite{Banks:1979yr} of an eigenvalue
accumulation leading to the spontaneous breaking of chiral symmetry
should apply.

The conventional discussion of the 't Hooft vertex starts by inserting
fermionic sources into the path integral
\begin{equation}
Z(\eta,\overline\eta)=\int (dA)\ (d\psi)\ (d\overline\psi)\
e^{-S_g-\overline\psi (D+m) \psi +\overline\psi \eta+
\overline\eta\psi}.
\end{equation}
Differentiation, in the Grasmannian sense, with respect to these
sources will generate the expectation for an arbitrary product of
fermionic operators.  Integrating out the fermions reduces this to
\begin{equation}
Z=\int (dA)\ 
e^{-S_g{+\overline\eta(D+m)^{-1}\eta}}\ 
\prod (\lambda_i+m). 
\end{equation}
Consider a zero mode $\psi_0$ satisfying {$D\psi_0=0$}.  In general
there is also a left zero mode satisfying $\overline\psi_0 D=0$.  If
the sources have an overlap with the mode, that is
$(\overline\eta|\psi_0)\ne 0$, then a factor of {$1/m$} in the source
term can cancel the {$m$} from the determinant.  Although non-trivial
topological configurations do not contribute to $Z$, their effects can
survive in correlation functions.  For the one-flavor theory the
effective interaction is bilinear in the fermion sources and is
proportional to
\begin{equation}
(\overline\eta|\psi_0)(\overline\psi_0|\eta).
\label{bilinear}
\end{equation}
As discussed earlier, the index theorem tells us that in general the
zero mode is chiral; it appears in either {$\overline\eta_L\eta_R$} or
{$\overline\eta_R\eta_L$}, depending on the sign of the gauge field
winding.

\begin{figure}
\centering
\includegraphics[width=2.5in]{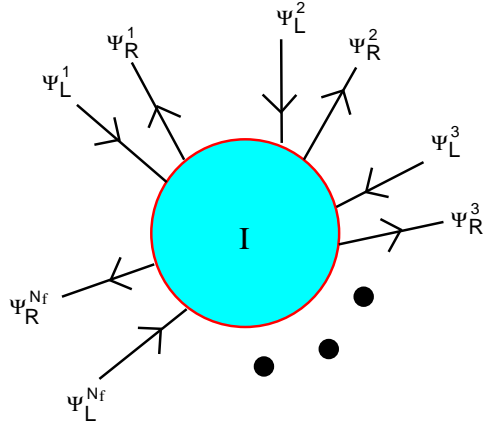}
\caption{ The 't Hooft vertex for $N_f$ flavors is a $2N_f$ effective
fermion operator that flips the spin of every flavor.}
\label{instanton} 
\end{figure}

With $N_f\ge 2$ flavors, the cancellation of the mass factors in the
determinant requires source factors from each flavor.  This
combination is the 't Hooft vertex.  It is an effective $2N_f$ fermion
operator.  In the process, every flavor flips its spin, as sketched in
Fig.~\ref{instanton}.  Indeed, this is the chiral anomaly; left and
right helicities are not separately conserved.

Because of Pauli statistics, the multi-flavor vertex can be written in
the form of a determinant.  This clarifies how the vertex preserves
flavored chiral symmetries.  With two flavors, call their sources $u$
and $d$, Eq.~\ref{bilinear} generalizes to
\begin{equation}
\left\vert
\matrix{
(\overline u |\psi_0)(\overline\psi_0| u) 
& (\overline u |\psi_0)(\overline\psi_0| d)\cr
(\overline d |\psi_0)(\overline\psi_0| u) 
& (\overline d |\psi_0)(\overline\psi_0| d)\cr
}
\right\vert.
\end{equation}

Note that the effect of the vertex is non-local.  In general the zero
mode $\psi_0$ is spread out over the finite region of the
``instanton'', {\it i.e.} the size parameter $\rho$ from the explicit
solution given earlier.  This means there is an inherent position space
uncertainty on where the fermions are interacting.  A particular
consequence is that fermion conservation is only a global symmetry.
In Minkowski space language, this non-locality can be thought of in
terms of states sliding in and out of the Dirac sea at different
locations.

\subsection{Fermions in higher representations}
\label{reps}

When the quarks are massless, the classical field theory corresponding
to the strong interactions has a $U(1)$ axial symmetry under the
transformation
\begin{equation} 
\psi\rightarrow
e^{i\theta\gamma_5}\psi\qquad
\overline\psi\rightarrow \overline\psi e^{i\theta\gamma_5}.
\label{thetarot}
\end{equation}
It is the 't Hooft vertex that explains how this symmetry does not
survive quantization.  In this subsection we discuss how when the
quarks are in non-fundamental representations of the gauge group,
discrete subgroups of this symmetry can remain because of additional
zeros in the Dirac operator.

While these considerations do not apply to the usual theory of the
strong interactions where the quarks are in the fundamental
representation, there are several reasons to study them anyway.  At
higher energies, perhaps as being probed at the Large Hadron Collider,
one might well discover new strong interactions that play a
substantial role in the spontaneous breaking of the electroweak
theory.  Also, many grand unified theories involve fermions in
non-fundamental representations.  As one example, we will see that
massless fermions in the 10 representation of $SU(5)$ possess a $Z_3$
discrete chiral symmetry.  Similarly the left handed 16 covering
representation of $SO(10)$ gives a chiral gauge theory with a
surviving discrete $Z_2$ chiral symmetry.  Understanding these
symmetries may eventually play a role in a discretization of chiral
gauge theories on the lattice.

Here we are generalizing the index theorem relating gauge field
topology to zero modes of the Dirac operator.  In particular, fermions
in higher representations can involve multiple zero modes for a
given winding.  Being generic, consider representation $X$ of a gauge
group $G$.  Denote by $N_X$ the number of zero modes that are required
per unit of winding number in the gauge fields.  That is, suppose the
index theorem generalizes to
\begin{equation}
n_r-n_l=N_X\nu
\end{equation}
where $n_r$ and $n_l$ are the number of right and left handed zero
modes, respectively, and $\nu$ is the winding number of the associated
gauge field.  The basic 't Hooft vertex receives contributions from
each zero mode, resulting in an effective operator which is a product
of $2N_X$ fermion fields.  Schematically, the vertex is modified along
the lines $\overline\psi_L \psi_R \longrightarrow (\overline\psi_L
\psi_R)^{N_X}$.  While this form still breaks the $U(1)$ axial
symmetry, it is invariant under $\psi_R\rightarrow e^{2\pi
i/N_X}\psi_R$.  In other words, there is a $Z_{N_X}$ discrete axial
symmetry.

There are a variety of convenient tools for determining $N_X$.
Consider building up representations from lower ones.  Take two
representations $X_1$ and $X_2$ and form the direct product
representation $X_1\otimes X_2$.  Let the matrix dimensions for $X_1$
and $X_2$ be $D_1$ and $D_2$, respectively.  Then for the product
representation we have
\begin{equation}
N_{X_1\otimes X_2}= N_{X_1} D_{X_2}+N_{X_2} D_{X_1}.
\end{equation}
To see this, start with $X_1$ and $X_2$ representing two independent
groups $G_1$ and $G_2$.  With $G_1$ having winding, there will be a
zero mode for each of the dimensions of the matrix index associated
with $X_2$.  Similarly there will be multiple modes for winding in
$G_2$.  These modes are robust and all should remain if we now
constrain the groups to be the same.

As a first example, denote the fundamental representation of $SU(N)$
as $F$ and the adjoint representation as $A$.  Then using $\overline F
\otimes F = A\oplus 1$ in the above gives $N_A=2N_F$, as noted some
time ago \cite{Witten:1982df}.  With $SU(3)$, fermions in the
adjoint representation will have six-fold degenerate zero modes.

For another example, consider $SU(2)$ and build up towards arbitrary
spin $s\in\{0,{1\over 2}, 1, {3\over 2},\ldots\}$.  Recursing the
above relation gives the result for arbitrary spin
\begin{equation}
N_s=s(2s+1)(2s+2)/3.
\end{equation}

Another technique for finding $N_X$ in more complicated groups begins
by rotating all topological structure into an $SU(2)$ subgroup and
then counting the corresponding $SU(2)$ representations making up the
larger representation of the whole group.  An example to illustrate
this procedure is the antisymmetric two indexed representation of
$SU(N)$.  This representation has been extensively used in
\cite{Corrigan:1979xf,Armoni:2003fb,Sannino:2003xe,Unsal:2006pj} for
an alternative approach to the large gauge group limit.  The basic
$N(N-1)/2$ fermion fields take the form
\begin{equation}
\psi_{ab}=-\psi_{ba}, \qquad a,b\in 1,2,...N.
\end{equation}
Consider rotating all topology into the $SU(2)$ subgroup involving the
first two indices, i.e. 1 and 2.  Because of the anti-symmetrization,
the field $\psi_{12}$ is a singlet in this subgroup.  The field pairs
$(\psi_{1,j},\psi_{2,j})$ form a doublet for each $j\ge 3$.  Finally,
the $(N-2)(N-3)/2$ remaining fields do not transform under this
subgroup and are singlets.  Overall we have $N-2$ doublets under the
$SU(2)$ subgroup, each of which gives one zero mode per winding
number.  We conclude that the 't Hooft vertex leaves behind a
$Z_{N-2}$ discrete chiral symmetry.  Specializing to the 10
representation of $SU(5)$, this is the $Z_3$ mentioned earlier.

Another example is the group $SO(10)$ with fermions in the 16
dimensional covering group.  This forms the basis of a rather
interesting grand unified theory, where one generation of fermions is
placed into a single left handed 16 multiplet \cite{Georgi:1979dq}.
This representation includes two quark species interacting with the
$SU(3)$ subgroup of the strong interactions, Rotating a topological
excitation into this subgroup, we see that the effective vertex will
be a four fermion operator and preserve a $Z_2$ discrete chiral
symmetry.

It is unclear whether these discrete symmetries are expected to be
spontaneously broken.  Since they are discrete, such breaking is not
associated with Goldstone bosons.  But the quark condensate does
provide an order parameter; so, when $N_X>1$, any such breaking would
be conceptually meaningful.  This could be checked in numerical
simulations.

\newpage
\Section{Massive quarks and the Theta parameter}
\label{mass}

As discussed earlier and illustrated in Fig.~\ref{potential1}, a quark
mass term $-m\overline\psi\psi\sim -m\sigma$ is represented by a
``tilting'' of the effective potential.  This selects one of the
multiple minima in the $\sigma,\eta^\prime$ plane as the true vacuum.
For masses small compared to the scale of QCD, the other minima will
persist as extrema, although due to the flat flavor non-singlet
directions, some of them will become unstable under small
fluctuations.  Counting the minima sequentially with the true vacuum
having $n=0$, each is associated with small excitations in the
pseudo-Goldstone directions having an effective mass of $m_\pi^2 \sim
m \cos(2\pi n/N_f)$.  Note that when $N_f$ exceeds four, there will be
more than one meta-stable state.  However, in the usual case of
considering two or three quarks as light, only one minimum remains
locally stable.

\subsection{Twisted tilting}

Conventionally the mass tilts the potential downward in the positive
$\sigma$ direction.  However, it is an interesting exercise to
consider tilts in other directions in the $\sigma,\eta^\prime$ plane.
This is accomplished with an anomalous rotation on the mass term
\begin{eqnarray}
-m\overline\psi\psi&\rightarrow&
-m\cos(\phi)\overline\psi\psi
-im\sin(\phi)\overline\psi\gamma_5\psi\cr
&\sim& -m\cos(\phi)\sigma+m\sin(\phi)\eta^\prime.
\end{eqnarray}
Were it not for the anomaly, this would just be a redefinition of
fields.  However the same effect that gives the $\eta^\prime$ its mass
indicates that this new form for the mass term gives an inequivalent
theory.  As $i\overline\psi\gamma_5\psi$ is odd under CP, this theory
is explicitly CP violating.

\begin{figure*}
\centering
\includegraphics[width=2.5in]{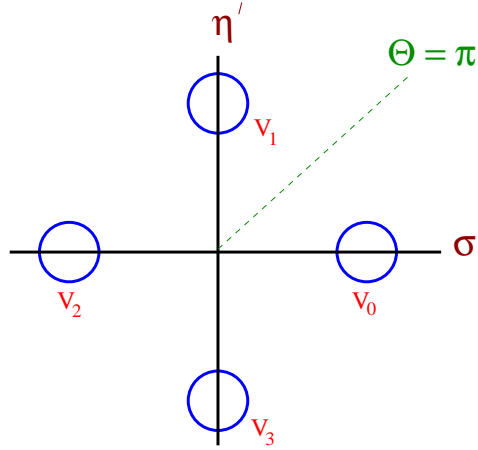}
\caption{\label{potential}
With massive quarks and a twisting angle of $\phi=\pi/N_f$, two of the
minima in the $\sigma,\eta^\prime$ plane become degenerate.  This
corresponds to a first order transition at $\Theta=\pi$.
}
\end{figure*}

The conventional notation for this effect involves the angle
$\Theta=N_f\phi$.  Then the $Z_{N_f}$ symmetry amounts to a $2\pi$
periodicity in $\Theta$.  As Fig.~\ref{potential} indicates, at
special values of the twisting angle $\phi$, there will exist two
degenerate minima.  This occurs, for example, at $\phi=\pi/N_f$ or
$\Theta=\pi$.  As the twisting increases through this point, there
will be a first order transition as the true vacuum jumps from the
vicinity of one minimum to the next.

Because of the $Z_{N_f}$ symmetry of the massless theory, all the
$N_f$ separate minima are physically equivalent.  This means that if
we apply our mass term in the direction of any of them, we obtain the
same theory.  In particular, for four flavors the usual mass term
$m\overline \psi \psi$ is equivalent to using the alternative mass
term $i m\overline\psi\gamma_5\psi$.  This result, however, is true if
and only if $N_f$ is a multiple of four.

\subsection{Odd $N_f$}

\begin{figure*}
\centering
\includegraphics[width=2.5in]{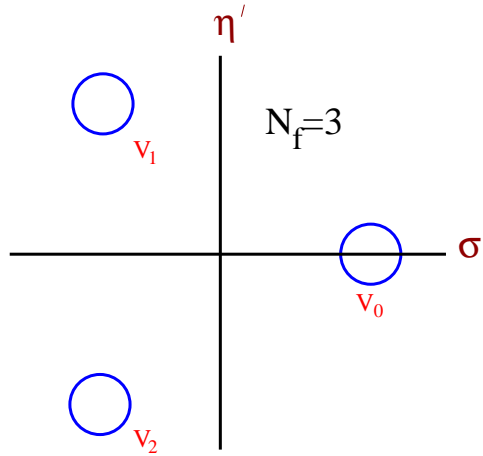}
\caption{\label{potential2} For odd $N_f$, such as the three flavor
  case sketched here, QCD is not symmetric under changing the sign of
  the quark mass.  Negative mass corresponds to taking $\Theta=\pi$.
}
\end{figure*}

One interesting consequence of this picture concerns QCD with an odd
number of flavors.  The group $SU(N_f)$ with odd $N_f$ does not
include the element $-1$.  In particular, the $Z_{N_f}$ structure is
not symmetric under reflections about the $\eta^\prime$ axis.
Figure \ref{potential2} sketches the situation for $SU(3)$.  One
immediate conclusion is that positive and negative mass are not
equivalent.  Indeed, a negative mass with three degenerate flavors
corresponds to the $\Theta=\pi$ case and a spontaneous breaking of CP
is expected.  In this case there is no symmetry under taking
$\sigma\sim \overline\psi \psi$ to its negative.  The simple picture
sketched in Fig.~\ref{v1} no longer applies.

At $\Theta=\pi$ the theory lies on top of a first order phase
transition line.  A simple order parameter for this transition is the
expectation value for the $\eta^\prime$ field.  As this field is odd
under CP symmetry, this shows that negative mass QCD with an odd
number of flavors spontaneously breaks CP.\footnote{Dashen's
  original paper \cite{Dashen:1970et} speculates that this might be
  related to the parity breaking seen in nature.  This presumably
  requires a new ``beyond the standard model'' interaction rather than
  QCD.}  This does not contradict the Vafa-Witten theorem
\cite{Vafa:1984xg} because in this regime the fermion determinant is
not positive definite.

Note that the asymmetry in the sign of the quark mass is not easily
seen in perturbation theory.  Any quark loop in a perturbative diagram
can have the sign of the quark mass flipped by a $\gamma_5$
transformation.  It is only through the subtleties of regulating the
divergent triangle diagram
\cite{Adler:1969gk,Adler:1969er,Bell:1969ts} that the sign of the mass
enters.  

A remarkable conclusion of these observations is that two physically
distinct theories can have identical perturbative expansions.  For
example, with flavor $SU(3)$ the negative mass theory has spontaneous
$CP$ violation, while the positive mass theory does not.  Yet both
cases have exactly the same perturbation theory.  This dramatically
demonstrates what we already knew: non-perturbative effects are
essential to understanding QCD.

\begin{figure*}
\centering
\includegraphics[width=2.5in]{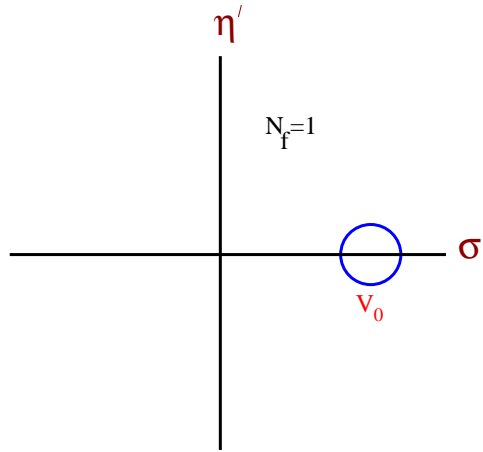}
\caption{ The effective potential for one-flavor QCD with small quark
  mass has a unique minimum in the $\sigma,\eta^\prime$ plane.  The
  minimum is shifted from zero due to the effect of the 't Hooft
  vertex.  }
\label{potential3} 
\end{figure*}

A special case of an odd number of flavors is one-flavor QCD
\cite{Creutz:2006ts}.  In this case the anomaly removes all chiral
symmetry and there is a unique minimum in the $\sigma,\eta^\prime$
plane, as sketched in Fig.~\ref{potential3}.  This minimum does not
occur at the origin, being shifted to $\langle \overline\psi
\psi\rangle > 0$ by the 't Hooft vertex, which for one flavor is just
an additive mass shift \cite{Creutz:2007yr}.  Unlike the case with
more flavors, the resulting expectation value for $\sigma$ is not from
a spontaneous symmetry breaking; indeed, there is no chiral symmetry
to break in one flavor QCD.  Any regulator that preserves a remnant of
chiral symmetry must inevitably fail \cite{Creutz:2008nk}.  Note also
that for one-flavor QCD there is no longer the necessity of a first
order phase transition at $\Theta=\pi$.  It has been argued
\cite{Creutz:2006ts} that for finite quark mass such a transition
should still occur if the mass is sufficiently negative, but the region
around vanishing mass is not expected to show any singularity.

An unusual feature of one-flavor QCD is that the renormalization of
the quark mass is not multiplicative when non-perturbative effects are
taken into account.  The additive mass shift is generally scheme
dependent since the details of the instanton effects depend on scale.
This is the basic reason that a massless up quark is not a possible
solution to the strong CP problem \cite{Creutz:2003xc}.  Later we will
discuss this in more detail in the context of the two flavor theory
with non-degenerate masses.

Because of this shift in the mass, the conventional parameters
$\Theta$ and $m$ are singular coordinates for the one-flavor theory.
A cleaner set of variables would be the coefficients of the two
possible mass terms $\overline\psi \psi$ and $i\overline\psi
\gamma_5\psi$ appearing in the Lagrangean.  The ambiguity in the quark
mass is tied to rough gauge configurations with ambiguous winding
number.  This applies even to the formally elegant overlap operator
that we will discuss later; when rough gauge fields are present, the
existence of a zero mode can depend on the detailed fermion operator
in use.  Smoothness conditions imposed on the gauge fields to remove
this ambiguity appear to conflict with fundamental principles, such as
reflection positivity \cite{Creutz:2004ir}.

The $Z_{N_f}$ symmetry discussed here is a property of the fermion
determinant and is independent of the gauge field dynamics.  In Monte
Carlo simulation language, this symmetry appears configuration by
configuration.  With $N_f$ flavors, we always have $|D|=|e^{2\pi
  i/N_f} D|$ for any gauge field.  This discrete chiral symmetry is
inherently discontinuous in $N_f$.  This non-continuity lies at the
heart of the issues with the rooted staggered quark approximation.  We
will return to this topic in a later section.

\subsection{Quark scattering and mass mixing}
\label{mixing}
So far we have worked with degenerate quarks.  In general each species
introduces another complex mass parameter.  Using flavored chiral
rotations we can move the phases of the masses around arbitrarily,
leaving only one overall phase, the Theta parameter.  Thus once the
overall scale has been set, QCD depends on $N_f+1$ parameters.

Here we explore the rich phase diagram of two-flavor QCD as a function
of the most general quark masses, including the $\Theta$ parameter.
This section closely follows the discussion in
Ref.~\cite{Creutz:2010ts}.  This theory involves three independent
parameters.  One is CP violating; its strong experimental limit is the
strong CP problem.  Here we will characterize the parameters by
distinguishing their transformations under various symmetries.  As we
define them, the resulting variables are each multiplicatively
renormalized.  However non-perturbative effects are not universal,
leaving individual quark mass ratios with a renormalization scheme
dependence.  This exposes ambiguities in matching lattice results with
perturbative schemes and the tautology involved in approaches that
attack the strong CP problem via a vanishing mass for the lightest
quark.

Before turning on the masses, we reemphasize the qualitative
properties expected in massless two-flavor QCD.  Of course, being an
interacting quantum field theory, nothing has been proven rigorously.
While the classical theory is conformally invariant, as discussed
earlier, confinement and dimensional transmutation generate a
non-trivial mass scale.  The theory should, of course, contain massive
stable nucleons.  On the other hand, spontaneous chiral symmetry
breaking should give rise to three massless pions as Goldstone bosons.
Bound states of glue in general will acquire a width due to decays
into pions.  In addition, the two-flavor analog of the eta-prime meson
should acquire its mass from the anomaly.

In this standard picture, the eta-prime and neutral pion involve
distinct combinations of quark-antiquark bound states.  In the simple
quark model the neutral pseudoscalars involve the combinations
\begin{eqnarray}
&&\pi_0 \sim \overline u \gamma_5 u -\overline d\gamma_5 d\cr
&&\eta^\prime \sim \overline u\gamma_5 u
 +\overline d\gamma_5 d +\hbox{glue}.
\end{eqnarray}
Here we include a gluonic contribution from mixing between the
$\eta^\prime$ and glueball states.  When the quarks are degenerate,
isospin forbids such mixing for the pion.

\begin{figure}
\centering
{\includegraphics[width=2.5in]{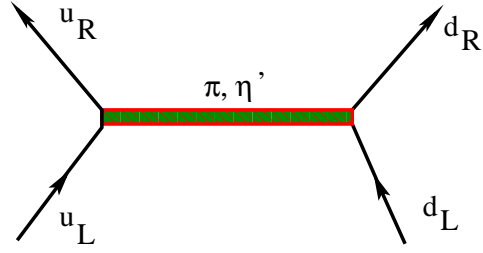}}
\caption{Both pion and eta-prime exchange contribute to spin
  flip scattering between up and down quarks. Figure from
  Ref.~\cite{Creutz:2010ts}.}
\label{scattering}
\end{figure}

Projecting out helicity states for the quarks $q_{L,R} =
(1\pm\gamma_5)q/2$, the pseudoscalars are combinations of left with
right handed fermions, {\it i.e.} $\overline q_L q_R-\overline q_R
q_L$.  Thus, as shown schematically in Fig.~\ref{scattering}, meson
exchange will contribute to a a spin flip process in a hypothetical
quark scattering experiment.  More precisely, the four point function
$\langle \overline u_R u_L \overline d_R d_L\rangle$ is not expected
to vanish.  Scalar meson exchange will also contribute to this
process, but this is not important for the qualitative argument below.
Of course we must assume that some sort of gauge fixing has been done
to eliminate a trivial vanishing of this function from an integral
over gauges.  We also consider this four point function at a scale
before confinement sets in.

It is important that the $\pi_0$ and $\eta^\prime$ are not degenerate.
This is due to the anomaly and the fact that the $\eta^\prime$ is not
a Goldstone boson.  As we discussed earlier, the $\pi_0$-$\eta^\prime$
mass difference can be ascribed to topological structures in the gauge
field.  Because the mesons are not degenerate, their contributions to
the above diagram cannot cancel.  The conclusion of this simple
argument is that helicity-flip quark-quark scattering is not
suppressed in the chiral limit.

\begin{figure}
\centering
\includegraphics[width=2.5in]{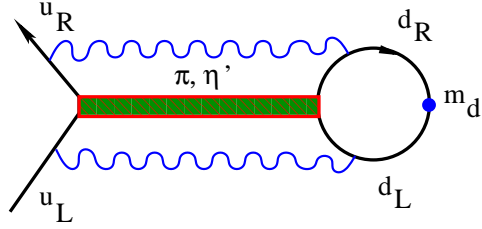}
\caption{Through physical meson exchange, a down quark mass can induce
  an effective mass for the up quark.  The gluon exchanges can
  compensate for the pseudoscalar nature of the meson fields.  Figure
  from Ref.~\cite{Creutz:2010ts}.}
\label{induced}
\end{figure}

Now consider turning on a small down quark mass while leaving the up
quark massless.  Formally such a mass allows one to connect the
ingoing and outgoing down quark lines in Fig.~\ref{scattering} and
thereby induce a mixing between the left and right handed up quark.
Such a process is sketched in Fig.~\ref{induced}.  Here we allow for
additional gluon exchanges to compensate for turning the pseudoscalar
field into a traditional mass term.

So the presence of a non-zero {$d$}-quark mass will induce an
effective mass for the {$u$} quark, even if the latter initially
vanishes. As a consequence, non-perturbative effects renormalize
{$m_u/ m_d$}.  If this ratio is zero at some scale, it cannot remain
so for all scales.  Only in the isospin limit are quark mass ratios
renormalization group invariant.  As lattice simulations include all
perturbative and non-perturbative effects, this phenomenon is
automatically included in such an approach.

This cross talk between the masses of different quark species is a
relatively straightforward consequence of the chiral anomaly and has
been discussed several times in the past, usually in the context of
gauge field topology and the index theorem \cite{
Georgi:1981be,Banks:1994yg, Creutz:2003cj, Creutz:2003xc}.  This
result is, however, frequently met with consternation from the
community well versed in perturbation theory.  Indeed, Feynman
diagrams tend to suppress spin-flip processes as the quark masses go
to zero.  The above argument shows that this lore need not apply when
anomalous processes come into play.  In particular, mass
renormalization is not flavor blind and the concept of mass
independent regularization is problematic.  Since the quark masses
influence each other, there are inherent ambiguities defining $m_u=0$.
This has consequences for the strong CP problem, discussed further
below.  Furthermore, a traditional perturbative regulator such as
$\overline{MS}$ is not complete when $m_u\ne m_d$.  Because of this,
the practice of matching lattice calculations to $\overline{MS}$ is
also problematic.

Given the simplicity of the above argument, it is perhaps somewhat
surprising that it continues to receive criticism.  The first
complaint sometimes made is that one should work directly with bare
quark masses.  This ignores the fact that the bare masses all vanish
under renormalization.  We discussed earlier the renormalization
group equation for a quark mass
\begin{equation}
a {dm_i\over da}=\gamma(g) m_i = \gamma_0 g^2 + O(g^4).
\label{mrg}
\end{equation}
As asymptotic freedom drives the bare coupling to zero, the bare
masses behave as
\begin{equation}
m\sim g^{\gamma_0/\beta_0} (1+O(g^2))\rightarrow 0
\label{mflow}
\end{equation}
where $\beta_0$ is the first term in the beta function controlling the
vanishing of the bare coupling in the continuum limit.  Since all bare
quark masses are formally zero, one must address these questions in
terms of a renormalization scheme at a finite cutoff.

A second frequent objection is that in a mass independent
regularization scheme, mass ratios are automatically constant.  Such
an approach asks that the renormalization group function $\gamma(g)$
in Eq.~(\ref{mrg}) be chosen to be independent of the quark species
and mass.  This immediately implies the constancy of all quark mass
ratios.  As only the first term in the perturbative expansion of
$\gamma(g)$ is universal, a mass independent scheme is indeed an
allowed procedure.  However, such a scheme obscures the off-diagonal
$m_d$ effect on $m_u$ discussed above.  In particular, by forcing
constancy of bare mass ratios, the ratios of physical particle masses
must vary as a function of cutoff.  This will be in a manner that
cancels the flow from the process discussed above.  The fact that
physical particle mass ratios are not just a function of quark mass
ratios is shown explicitly in subsection
\ref{phasesection}, where we observe that in the chiral limit the
combination $1-{m_{\pi_0}^2 / m_{\pi_\pm}^2}$ is proportional to
$(m_d-m_u)^2\over (m_d+m_u)\Lambda_{qcd}$.

From a non-perturbative point of view, having physical mass ratios
vary with the cutoff seems rather peculiar; indeed, the particle
masses are physical quantities that would be natural to hold fixed.
And, even though a mass independent approach is theoretically
possible, there is no guarantee that any given quark mass ratio
will be universal between schemes.  Finally, the lattice approach
itself is usually implemented with physical particle masses as input.
As such it is not a mass independent regulator, making a perturbative
matching to lattice results rather subtle.

A third complaint against the above argument is that one should simply
do the matching at some high energy, say 100 GeV, where ``instanton''
effects are exponentially suppressed and irrelevant.  This point of
view has several problems.  First, current lattice simulations are not
done at miniscule scales and non-perturbative effects are present and
substantial.  Furthermore, the exponential suppression of topological
effects is in the inverse coupling, which runs logarithmically with
the scale.  As such, the non-perturbative suppression is a power law
in the scale and straightforward to estimate.

Since the eta-prime mass is expected to be of order $\Lambda_{qcd}$,
we know from the previous renormalization group discussion how it
depends on the bare coupling in the continuum limit
\begin{equation}
\label{rg}
m_{\eta^\prime} \propto {1\over a}{ e^{-1/(2\beta_0 g_0^2)}
g_0^{-\beta_1/\beta_0^2}}.
\end{equation}
While this formula indeed shows an exponential suppression in
  $1/g_0^2$, this is cancelled by the inverse cutoff factor in just
  such a way that the mass of this physical particle remains finite.
  The ambiguity in the quark mass splitting is controlled by the mass
  splitting $m_{\eta^\prime}-m_{\pi_0}$ as well as being proportional
  to $m_d-m_u$.  Considering $m_d=5$ MeV at a scale of $\mu=2$ GeV, a
  rough estimate of the order of the $u$ quark mass shift is
\begin{equation}
\Delta m_u(\mu) \sim 
\left({m_{\eta^\prime}-m_{\pi_0}\over \Lambda_{qcd}}\right)
\ (m_d-m_u)
= O(1\ \hbox{MeV}),
\end{equation} 
a number comparable to typical phenomenological estimates.  This
result depends on the scale $\mu$, but that dependence is only
logarithmic and given by Eq.~(\ref{mflow}).  Additional flavors will
reduce the size of this effect; with the strange quark present, it
should be proportional to $m_d m_s$.

A particularly important observation is that the exponent controlling
the coupling constant suppression in Eq.~\ref{rg} is substantially
smaller than the classical instanton action
\begin{equation}
{1\over 2\beta_0 g_0^2}={ 8\pi^2\over (11-2n_f/3) g_0^2}<< 
{ 8\pi^2\over g_0^2}.
\end{equation} 
This difference arises because one needs to consider topological
excitations above the quantum, not the classical, vacuum.  Zero modes
of the Dirac operator are still responsible for the bulk of the eta
prime mass, but naive semi-classical arguments strongly underestimate
their effect.

\subsection{General masses in two-flavor QCD}

Given the confusion over the meaning of quark masses, it is
interesting to explore how two-flavor QCD behaves as these quantities
are varied, including the possibility of explicit CP violation through
the Theta parameter.  The full theory has a rather rich phase diagram,
including first and second order phase transitions, some occuring when
none of the quark masses vanish.

We consider the quark fields $\psi$ as carrying implicit isospin,
color, and flavor indices.  Assume as usual that the theory in the
massless limit maintains the $SU(2)$ flavored chiral symmetry under
\begin{eqnarray}
\psi \longrightarrow e^{i\gamma_5 \tau_\alpha\phi_\alpha/2}\psi\cr
\overline \psi \longrightarrow \overline\psi 
e^{i\gamma_5 \tau_\alpha\phi_\alpha/2}.
\end{eqnarray}
Here $\tau_\alpha$ represents the Pauli matrices generating isospin
rotations.  The angles $\phi_\alpha$ are arbitrary rotation
parameters.

We wish to construct the most general two-flavor mass term to add to
the massless Lagrangean.  Such should be a dimension 3 quadratic form
in the fermion fields and should transform as a singlet under Lorentz
transformations.  For simplicity, only consider quantities that are
charge neutral as well.  This leaves four candidate fields, giving the
general form for consideration
\begin{equation}
m_1\overline\psi\psi+
m_2\overline\psi\tau_3\psi+
im_3\overline\psi\gamma_5\psi+
im_4\overline\psi\gamma_5\tau_3\psi.
\label{genmass}
\end{equation}
The first two terms are naturally interpreted as the average quark
mass and the quark mass difference, respectively.  The remaining two
are less conventional.  The $m_3$ term is connected with the CP
violating parameter of the theory.  The final $m_4$ term has been used
in conjunction with the Wilson discretization of lattice fermions,
where it is referred to as a ``twisted mass''
\cite{Frezzotti:2000nk,Boucaud:2007uk}. Its utility in this context is
the ability to reduce lattice discretization errors.  We will return
to this term later when we discuss the effect of lattice artifacts on
chiral symmetry.

These four terms are not independent.  Indeed, consider the above
flavored chiral rotation in the $\tau_3$ direction, $\psi\rightarrow
e^{i\theta\tau_3\gamma_5/2}\psi$.  Under this the composite fields
transform as
\begin{eqnarray}
\overline\psi\psi\ &\longrightarrow\  
\cos(\theta)\overline\psi\psi
+\sin(\theta)i\overline\psi\gamma_5\tau_3\psi\cr
\overline\psi\tau_3\psi\ &\longrightarrow\
\cos(\theta)\overline\psi\tau_3\psi
+\sin(\theta)i\overline\psi\gamma_5\psi\cr
i\overline\psi\tau_3\gamma_5\psi\ &\longrightarrow\  
\cos(\theta)i\overline\psi\tau_3\gamma_5\psi
-\sin(\theta)\overline\psi\psi\cr
i\overline\psi\gamma_5\psi\ &\longrightarrow\
\cos(\theta)i\overline\psi\gamma_5\psi
-\sin(\theta)\overline\psi\tau_3\psi.
\end{eqnarray}
This rotation mixes $m_1$ with $m_4$ and $m_2$ with $m_3$.  Using this
freedom, we can select any one of the $m_i$ to vanish and a second to
be positive.

The most common choice is to set $m_4=0$ and use $m_1$ as controlling
the average quark mass.  Then $m_2$ gives the quark mass difference,
while CP violation appears in $m_3$.  This, however, is only a
convention.  The alternative ``twisted mass'' scheme
\cite{Frezzotti:2000nk,Boucaud:2007uk}, makes the choice $m_1=0$.
This uses {$m_4>0$} for the average quark mass and {$m_3$} becomes the
up-down mass difference.  In this case $m_2$ becomes the CP violating
term.  It is amusing to note that an up down quark mass difference in
such a formulation involves the naively CP odd
$i\overline\psi\gamma_5\psi$.  The strong CP problem has been rotated
into the smallness of the $\overline\psi\tau_3\psi$ term, which with
the usual conventions is the mass difference.  But because of the
flavored chiral symmetry, both sets of conventions are physically
equivalent.

For the following, take the arbitrary choice $m_4=0$, although one
should remember that this is only a convention and we could have
chosen any of the four parameters in Eq.~(\ref{genmass}) to vanish.
With this choice, two-flavor QCD, after scale setting, depends on
three mass parameters
\begin{equation}
m_1\overline\psi\psi+
m_2\overline\psi\tau_3\psi+
im_3\overline\psi\gamma_5\psi.
\end{equation}
It is the possible presence of $m_3$ that represents the strong CP
problem.  As all the parameters are independent and transform
differently under the symmetries of the problem, there is no
connection between the strong CP problem and $m_1$ or $m_2$.

As discussed extensively above, the chiral anomaly is responsible for
the iso-singlet rotation
\begin{eqnarray}
\psi \longrightarrow e^{i\gamma_5 \phi/2}\psi\cr
\overline \psi \longrightarrow \overline\psi 
e^{i\gamma_5 \phi/2}
\end{eqnarray}
not being a valid symmetry, despite the fact that $\gamma_5$ naively
anti-commutes with the massless Dirac operator.
Subection \ref{quarks} showed this anomaly is nicely summarized via
Fujikawa's \cite{Fujikawa:1979ay} approach where the fermion measure
in the path integral picks up a non-trivial factor.  In any given gauge
configuration only the zero eigenmodes of $\slashchar D$ contribute,
and by the index theorem they are connected to the winding number of
the gauge configuration.  The conclusion is that the above rotation
changes the fermion measure by an amount depending non-trivially on
the gauge field configuration.

Note that this anomalous rotation allows one to remove any topological
term from the gauge part of the action.  Naively this would have been
yet another parameter for the theory, but by including all three mass
terms for the fermions, this can be absorbed.  For the following we
consider that any topological term has thus been rotated away. After
this one is left with the three mass parameters above, all of which
are independent and relevant to physics. 

These parameters are a complete set for two-flavor QCD; however, this
choice differs somewhat from what is often discussed.  Formally we can
define the more conventional variables as
\begin{eqnarray}
 &&m_u=m_1+m_2+im_3\cr
 &&m_d=m_1-m_2+im_3\cr
&&e^{i\Theta}={m_1^2-m_2^2-m_3^2+2im_1m_3\over
\sqrt{m_1^4+m_2^4+m_3^4+2m_1^2m_3^2+2m_2^2m_3^2-2m_1^2m_2^2}.
}
\end{eqnarray}
Particularly for $\Theta$, this is a rather complicated change of
variables.  For non-degenerate quarks in the context of the phase
diagram discussed below, the variables $\{m_1,m_2,m_3\}$ are more
natural.

\subsection {The strong CP problem and the up quark mass}

Strong interactions preserve CP to high accuracy \cite{Baluni:1978rf}.
Thus only two of the three possible mass parameters seem to be needed.
With the above conventions, it is natural to ask why is $m_3$ so
small?  It is the concept of unification that brings this question to
the fore.  We know that the weak interactions violate CP.  Thus, if
the electroweak and the strong interactions separate at some high
scale, shouldn't some remnant of this breaking survive?  How is CP
recovered for the strong force?

One possible solution is that there is no unification and one should
just consider the weak interactions as a small perturbation.  Another
approach involves adding a new dynamical ``axion'' field that couples
to the quarks through a coupling to $i\overline\psi\gamma_5\psi$.
Shifts in this field make $m_3$ essentially dynamical, and potentially
the theory could relax to $m_3=0$.

There is a third proposed solution, being criticized here, that the up
quark mass might vanish.  This would naively allow a flavored chiral
rotation to remove any phases from the quark mass matrix.  Why is a
vanishing up quark mass not a sensible approach?  From the above, one
can define the up quark mass as a complex number
\begin{equation}
m_u\equiv m_1+m_2+im_3.
\end{equation}
But the quantities $m_1$, $m_2$, and $m_3$ are independent parameters
with different symmetry properties.  With our conventions, $m_1$
represents an iso-singlet mass contribution, $m_2$ is isovector in
nature, and $m_3$ is CP violating.  And, as discussed earlier, the
combination $m_1+m_2=0$ is scale and scheme dependent.  The strong CP
problem only requires small $m_3$. So while it may be true formally
that
\begin{equation}
 m_1+m_2+im_3=0\ \Rightarrow\ m_3=0,
\end{equation}
this would depend on scale and might well be regarded as ``not even
wrong.''

\subsection {The two-flavor phase diagram}
\label{phasesection}

\begin{figure}
\centering
{\includegraphics[width=2.5in] {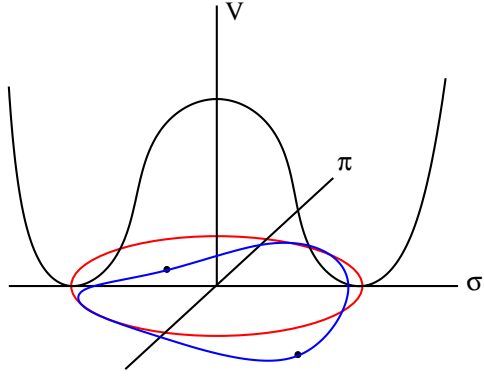}}
\caption{ The $m_2$ and $m_3$ terms warp the Mexican hat potential
  into two separate minima.  The direction of the warping is
  determined by the relative size of these parameters.  Figure taken
  from Ref.~\cite{Creutz:2010ts}.}
\label{warping}
\end{figure}

As a function of the three mass parameters, QCD has a rather intricate
phase diagram that we now discuss. Using simple chiral Lagrangean
arguments, this can be qualitatively mapped out.  To begin we consider
the composite fields similar to those used in the earlier discussion
of pions as Goldstone bosons
\begin{eqnarray}
&\sigma\sim \overline\psi\psi \qquad
&\eta\sim i\overline\psi\gamma_5\psi\cr
&\vec\pi\sim i\overline\psi\gamma_5\vec\tau\psi \qquad
&{\vec a_0}\sim \overline\psi\vec\tau\psi.
\end{eqnarray}
In terms of these, a natural model for a starting effective potential
is
\begin{eqnarray}
V=&\lambda(\sigma^2+\vec\pi^2-v^2)^2-{m_1}\sigma-{m_2}{a_0}_3
-{m_3}\eta\cr
&+\alpha ({\eta}^2+{\vec a_0}^2)
-\beta (\eta\sigma+ {\vec a_0} \cdot \vec \pi)^2.
\end{eqnarray}
Here $\alpha$ and $\beta$ are ``low energy constants'' that bring in a
chirally symmetric coupling of $(\sigma,\vec\pi)$ with $(\eta,\vec
a_0)$.  As discussed in Ref.~\cite{Creutz:1995wf}, the sign of the
$\beta$ term is suggested so that $m_{\eta} < m_{a_0}$.

This potential augments the famous ``Mexican hat'' or ``wine bottle''
potential discussed earlier, in which the Goldstone pions are
associated with the flat directions running around at constant
$\sigma^2+\vec\pi^2=v^2$.  The $m_2$ and $m_3$ terms do not directly
affect the $\sigma$ and $\pi$ fields, but induce an expectation value
for ${a_0}_3$ and $\eta$, respectively.  This in turn results in the
$\alpha$ and $\beta$ terms inducing a warping of the Mexican hat into
two separate minima, as sketched in Fig.~\ref{warping}.  The direction
of this warping is determined by the relative size of $m_2$ and $m_3$;
$m_2$ ($m_3$) warps downward in $\pi_0$ ($\sigma$) direction.  If we
now turn on $m_1$, this will select one of the two minimum as
favored.  This gives rise to a generic first order transition at
$m_1=0$.

\begin{figure}
\centering
\includegraphics[width=3.5in]{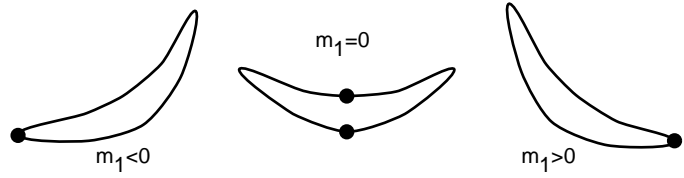} 
\caption{In the $m_1,m_2$ plane, $m_{\pi_0}^2$ can pass through zero,
giving rise to pion condensation at an Ising-like transition.  Figure
taken from \cite{Creutz:1995wf}. }
\label{ising}
\end{figure}

There is additional structure in the $m_1,m_2$ plane when $m_3$
vanishes.  In this situation the quadratic warping is downward in the
$\sigma$ direction.  For large $|m_1|$ only $\sigma$ will have an
expectation, with sign determined by the sign of $m_1$.  The pion will
be massive, but with $m_2$ reducing the neutral pion mass below that
of the charged pions.  If now $m_1$ is decreased in magnitude at fixed
$m_2$, eventually the neutral pion becomes massless and condenses.
How this occurs is sketched in Fig.~\ref{ising}.  An order parameter
for the transition is the expectation value of the $\pi_0$ field, with
the transition being Ising-like.

In this simple model the ratio of the neutral to charged pion masses
can be estimated from a quadratic expansion about the minimum of the
potential.  For $m_3=0$ and $m_1$ above the transition line, this
gives
\begin{equation}
{m_{\pi_0}^2\over m_{\pi_\pm}^2}
=1-{\beta v m_2^2\over 2\alpha^2 m_1}+O(m^2).
\end{equation}
The second order transition is located where this vanishes, and thus
occurs for $m_1$ proportional to $m_2^2$.  Note that this equation
verifies the important result that a constant quark mass ratio does
not correspond to a constant meson mass ratio and vice versa.  This is
the ambiguity discussed at the beginning of this section.

\begin{figure}
\centering
\includegraphics[width=3in]{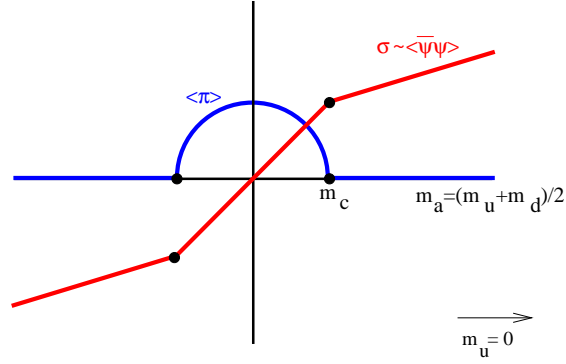}
\caption{ With a constant up-down quark mass difference, the jump in
the chiral condensate splits into two second order transitions.  The
order parameter distinguishing the intermediate phase is the
expectation value of the neutral pion field.  Figure taken from
Ref.~\cite{Creutz:2005gb}. }
\label{condensate1}
\end{figure}

This structure can also be observed in the expectation values for the
pion and sigma fields as functions of the average quark mass while
holding the quark mass difference fixed.  This is sketched in
Fig.~\ref{condensate1}.  The jump in $\sigma$ as we go from large
positive to large negative masses is split into two transitions with
the pion field acquiring an expectation value in the intermediate
region.

This second order transition occurs when both $m_u$ and $m_d$ are
non-vanishing but of opposite sign, i.e. $|m_1|<|m_2|$.  This is
required to avoid the Vafa-Witten theorem \cite{Vafa:1984xg}, which
says that no parity breaking phase transition can occur if the fermion
determinant is positive definite.  At the transition the correlation
length diverges.  This shows that it is possible to have significant
long distance physics without the presence of small Dirac eigenvalues.
In contrast, we see that there is no transition at the point where
only one of the quark masses vanishes.  In this situation there is no
long distance physics despite the possible existence of small Dirac
eigenvalues.

\begin{figure}
\centering
{\includegraphics[width=3.5in] {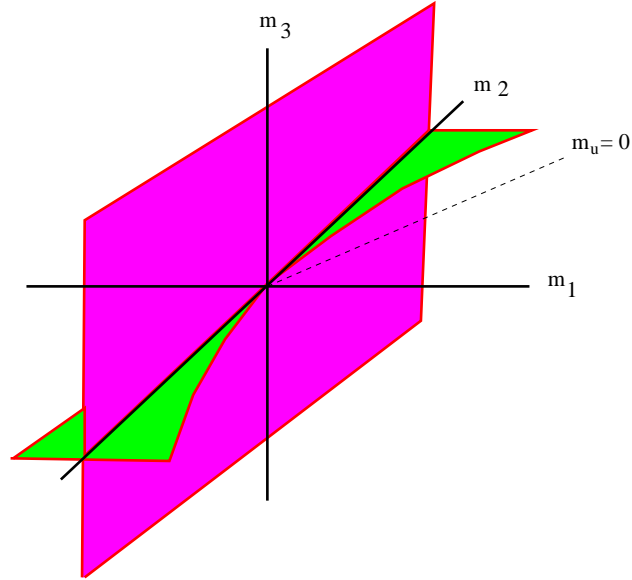}}
\caption{The full phase diagram for two-flavor QCD as a function of
  the three mass parameters.  It consists of two intersecting first
  order surfaces with a second order edge along curves satisfying
  $m_3=0$, $|m_1|<|m_2|$.  There is no structure along the $m_u=0$
  line except when both quark masses vanish.  Figure from
  Ref.~\cite{Creutz:2010ts}}
\label{phasediagram}
\end{figure}

Putting this all together, we obtain the phase diagram sketched in
Fig.~\ref{phasediagram}.  There are two intersecting first order
transition surfaces, one at $(m_1=0$, $m_3\ne 0)$ and the second at
$(m_1<m_2$,\ $m_3=0)$.  These each occur where $\Theta=\pi$. However,
note that with non-degenerate quarks there is also a $\Theta=\pi$
region at $m_2=m_1+\epsilon$ for small but non-vanishing $\epsilon$
where there is no transition.  The absence of a physical singularity
at $m_u=0$ when $m_d\ne 0$ lies at the heart of the problem in
defining a vanishing up quark mass.

In the next section we will see that the structure in the $m_1,m_2$
plane is closely related to an interesting lattice artifact in the
degenerate quark limit.  Aoki \cite{Aoki:1983qi} discussed a possible
phase with spontaneous parity violation with the Wilson fermion
formulation.  Indeed, lattice artifacts can modify the effective
potential in a similar way to the $m_2$ term and allow the CP
violating phase at finite cutoff to include part of the $m_1$ axis as
well.  

\newpage
\Section{Lattice fermions}
We now have a fairly coherent picture of how the spectrum of
pseudoscalar mesons is connected with chiral symmetry in the continuum
theory.  The anomaly plays a crucial role in introducing the Theta
parameter into the theory and contributing to the $\eta^\prime$
mass. Throughout we have assumed that we have in hand a regulator to
define the various composite fields, but we have not been specific in
how that regulator is formulated.  Early sections indicated the
lattice should provide a natural route to a non-perturbative
formulation, but we have postponed the details until some of the
desired continuum features were elucidated.

The lattice can be regarded as a fully non-perturbative definition of
a quantum field theory.  As such, the entire structure explored in
previous sections should follow as we approach the continuum limit.
But there are a variety of interesting and subtle issues concerning
how this comes about.  When the lattice is in place, all infinities in
the theory are automatically removed.  However we have argued that the
anomaly is closely tied to the divergences in the theory.  As such the
physics associated with the anomaly must appear somewhere in any valid
lattice formulation.  If we try to formulate a lattice version of QCD
with all classical symmetries in place, there is no way for this to
happen.  In particular, this imposes subtleties for how the action for
the quarks is formulated.  Here we go into this problem in some detail
and explore some of the methods for dealing with it.

\subsection{Hopping and doublers}

The essence of lattice doubling already appears in the quantum
mechanics of the simplest fermion Hamiltonian in one space dimension
\begin{equation}
H=iK\sum_j a_{j+1}^\dagger a_j - a_j^\dagger a_{j+1}.
\end{equation}
Here $j$ is an integer labeling the sites of an infinite
chain and the $a_j$ are fermion annihilation operators satisfying
standard anti-commutation relations
\begin{equation}
\left[ a_j, a_k^\dagger\right]_+\equiv a_j a_k^\dagger+ a_k^\dagger a_j
=\delta_{j,k}. 
\end{equation}
The fermions hop from site to neighboring site with amplitude $K$;
thus, we refer to $K$ as the ``hopping parameter'' and by convention
take it to be positive.  The bare vacuum $\vert 0 \rangle$ satisfies
$a_j \vert 0 \rangle=0.$ This vacuum is not the physical one, which
requires constructing a filled Dirac sea.  Energy eigenstates in the
single fermion sector
\begin{equation}
\vert \chi\rangle=\sum_j \chi_j a_j^\dagger \vert 0 \rangle 
\end{equation}
can be easily found in momentum space
\begin{equation}
\chi_j(q)= e^{iqj} \chi_0
\end{equation}
where we can restrict $-\pi < q \le \pi$.  The resulting energy is
\begin{equation}
E(q)=2K \sin(q). 
\end{equation}
The physical vacuum fills all the negative energy states, {\it i.e.}
those with $-\pi<q < 0.$ 

On this vacuum, consider constructing a fermionic wave packet by
exciting a few modes of small momentum $q$.  This packet will have a
group velocity $dE/dq\sim 2K$ that is positive.  Thus it moves to the
right and represents a right-moving fermion.  On the other hand, a
wave packet of low energy can also be produced by exciting momenta in
the vicinity of $q\sim \pi$.  This packet will have group velocity
$\left . {dE\over dq}\right|_{q=\pi}\sim -2K$ and therefore be left moving.
The essence of the Nielsen Ninomiya theorem \cite{Nielsen:1980rz} is
that we must have both types of excitation.  We will go into this in
more detail later, but for this one dimensional case the periodicity
in $q$ requires the dispersion relation to have an equal number of
zeros with positive and negative slopes.  If we now consider a two
component spinor to describe the fermion, we will have independent
states corresponding to each component.  This is the so called
``doubling'' issue.

As a preliminary to later discussion, here we concentrate on a
Hamiltonian version of the Wilson approach to remove the doublers.
Continuing to work in one dimension, consider a two component spinor
\begin{equation}
\psi=\pmatrix{a\cr b\cr}.
\end{equation}
where $a$ and $b$ are distinct fermion annihilation operators on the
lattice sites.  The so called ``naive'' lattice Hamiltonian begins
with the simple hopping case of above and adds in the lower components
and a mass term that mixes the upper and lower components
\begin{equation}
H=iK\sum_j 
a_{j+1}^\dagger a_j - a_j^\dagger a_{j+1}
-b_{j+1}^\dagger b_j + b_j^\dagger b_{j+1}
 +M\sum_j a^\dagger_j b_j + b^\dagger_j a_j.
\end{equation} 
Introducing two by two Dirac matrices 
\begin{equation}
\gamma_0=\sigma_1=\pmatrix{0&1\cr
                   1&0\cr},
\ \ \gamma_1=\sigma_2=\pmatrix{0&-i\cr
                             i&0\cr},
\ \ \gamma_5=\sigma_3=\pmatrix{1&0\cr
                             0& -1\cr},
\end{equation}
and defining
\def \psibar{\overline\psi}
\def \chibar{\overline\chi}
$\psibar=\psi^\dagger\gamma_0$,  
we write the Hamiltonian compactly as
\begin{equation}
H=\sum_j K(\psibar_{j+1} \gamma_1 \psi_j-\psibar_j \gamma_1 \psi_{j+1})
+M\sum_j\psibar_j \psi_j.
\end{equation}  
This looks very much like the continuum Dirac Hamiltonian with the
derivative term represented on the lattice by a nearest neighbor
difference.  Chiral symmetry is manifest in the possibility of
independent rotations of the $a$ and $b$ type particles when the mass
term is absent.  The latter mixes these components and opens a gap in
the spectrum.

As before, the single particle states are found by Fourier
transformation and satisfy
\begin{equation}
E^2=4K^2 \sin^2(q)+M^2.
\end{equation}
At each momentum there is one positive and one negative energy state.
Again, we are to fill the negative energy sea to form the physical
vacuum.  The doubling issue is that there are low energy excitations
that satisfy the Dirac equation appearing both at $q\sim 0$ and $q\sim
\pi$.  The physical momenta $k$ of the latter excitations appear in
expanding about pi, i.e $k=q-\pi$.  These states have a smooth spatial
dependence in a redefined field $\chi_j=(-1)^j \psi_j$.  The doublers
at $q\sim\pi$ are still with us.

\subsection{Wilson fermions}

One way to remove the degeneracy of the doublers is to make the mixing
of the upper and lower components momentum dependent.  A simple way of
doing this was proposed by Wilson \cite{Wilson:1975id}.  To our
Hamiltonian model, we add one more term 
\begin{eqnarray}
H&=&iK\sum_j 
a_{j+1}^\dagger a_j - a_j^\dagger a_{j+1}
-b_{j+1}^\dagger b_j + b_j^\dagger b_{j+1}+\cr
&& M\sum_j a^\dagger_j b_j + b^\dagger_j a_j
-rK \sum_j a_j^\dagger b_{j+1}+b_j^\dagger a_{j+1}
 +b_{j+1}^\dagger a_j+a_{j+1}^\dagger b_j \cr 
&=&\sum_j K(\psibar_{j+1} (\gamma_1-r) \psi_j-\psibar_j (\gamma_1+r)
 \psi_{j+1})+\sum_jM\psibar_j \psi_j.
\end{eqnarray}
Now the spectrum satisfies
\begin{equation}
E^2=4K^2 \sin^2(q)+(M-2rK\cos(q))^2.
\end{equation}
The doublers at $q\sim \pi$ are increased in energy relative
to the states at $q\sim 0$.  The physical particle mass is now $
m=M-2rK $ while that of the doubler is at $M+2rK$.

When $r$ becomes large, the dip in the spectrum of near $q=\pi$
actually becomes a maximum.  This is irrelevant for our discussion,
although we note that the case $r=1$ is somewhat special.  For this
value, the matrices $(\gamma_1\pm 1)/2$, which determine how the
fermions hop along the lattice, are projection operators.  In a sense,
the doubler is removed because only one component can hop.  This
choice $r=1$ has been the most popular in practice.

The hopping parameter has a critical value at
\begin{equation}
K_c={M \over 2r}. 
\end{equation}
At this point the gap in the spectrum closes and one species of
fermion becomes massless.  The Wilson term, proportional to $r$, still
mixes the $a$ and $b$ type particles; so, there is no exact chiral
symmetry.  Nevertheless, in the continuum limit this represents a
candidate for a chirally symmetric theory.  Before the limit, chiral
symmetry does not provide a good order parameter.

Now we generalize this approach to the Euclidean path integral
formulation in four space-time dimensions.  In the continuum, one
usually writes for the free fermion action density
\begin{equation}
\psibar D \psi=\psibar (\slashchar\partial +m) \psi
\end{equation}
or in momentum space
\begin{equation}
\psibar (i\slashchar p +m) \psi.
\end{equation}
By convention we use Hermitean gamma matrices.  Note that $D$ is the
sum of Hermitean and anti-Hermitean parts.  In the continuum the
former is just a constant, the mass.  A Hermitean operator appears in
the combination $\gamma_5 D$, but we don't need that just now.

A matrix can be diagonalized when it commutes with its adjoint; then
it is called ``normal.''  For the naive continuum operator this is the
case, and we see that all eigenvalues of $D$ lie along a line parallel
to the imaginary axis intersecting the real axis at $m$.  This simple
structure will be lost on the lattice.

As discussed earlier, a simple transcription of derivatives onto the
lattice replaces factors of $p$ with trigonometric functions.  Thus
the naive lattice action becomes
\begin{equation}
\psibar\left ({i\over a}\sum_\mu \gamma_\mu\sin(p_\mu a) +m\right) \psi
\end{equation} 
where we have explicitly included the lattice spacing $a$.  For small
momentum this reduces to the continuum result $\psibar(i\gamma_\mu
p_\mu+m)\psi$.  Now let one component of $p$ get large and be near
$\pi/a$.  Then we again have small eigenvalues and a nearby pole in
the propagator.  As any of the four components of momentum can be near
$0$ or $\pi$ there are a total of 16 places in momentum space that
give rise to a Dirac like behavior.  The naive fermion action gives
rise to 16 doublers.

\begin{figure}
\centering
\includegraphics[width=3in]{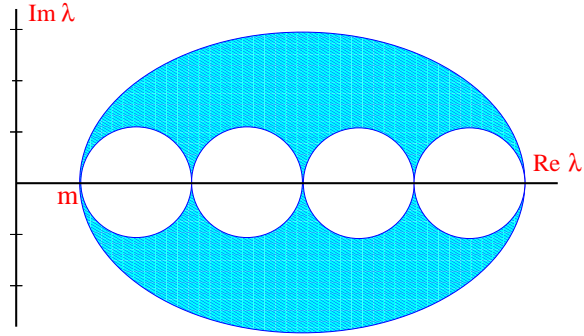}
\caption{ The eigenvalue spectrum of the free Wilson fermion operator
is a set of nested circles.  On turning on the gauge fields, some
eigenvalues drift into the open regions.  Some complex pairs can
collide and become real.  These are connected to gauge field topology.
Figure taken from Ref.~\cite{Creutz:2007fe}.}
\label{eigen1} 
\end{figure}

As in the earlier example, the Wilson solution adds a momentum
dependent mass.  Wishing to maintain only nearest neighbor terms, it
also involves trigonometric functions.  To maintain hyper-cubic
symmetry, we put in the Wilson term symmetrically for all space-time
directions.  For simplicity we set the Wilson parameter $r$ from
before to unity.  Explicitly for free fields we consider the momentum
space form
\begin{equation}
\psibar D_W \psi = \psibar \left(
{1\over a} \sum_\mu(i\gamma_\mu\sin(p_\mu a)+1-\cos(p_\mu a)) 
+m\right) \psi.
\end{equation} 
Now for a momentum component near $\pi$ the eigenvalues are of order
$1/a$ in size.  Note that the lattice artifacts in the propagator
start at order $p^2 a$, rather than $O(a^2)$ as for naive fermions.
The eigenvalue structure of $D_W$ is rather interesting.  The
eigenvalues for the free Wilson theory occur at
\begin{equation}
\label{freewilson}
\lambda=\pm {i\over a} \sqrt{\sum_\mu \sin^2(p_\mu a)} + 
{1\over a}\sum_\mu 1-\cos(p_\mu a) 
+m.
\end{equation}
The eigenvalues of this free operator lie on a set of ``nested
circles,'' as sketched in Fig.~\ref{eigen1}.  Note that
$m\leftrightarrow -m$ is not a symmetry. Naively it would be in the
continuum, but as we discussed earlier, it cannot be so in the quantum
theory when one has an odd number of flavors.

Note that to obtain real eigenvalues in Eq.~(\ref{freewilson}), each
component of the momentum must be an integer multiple of $\pi$.  There
are actually several critical values that can give rise to massless
fermions.  For $m=0,-2,-4,-6,-8$ we have $1,4,6,4,1$ massless species.
When interactions are present these values of the mass will also be
renormalized.\footnote{Actually the 6 flavor case at m=-4 does have a
discrete symmetry that will protect against additive mass
renormalization.}  Whether a continuum limit at these alternative
points is useful has not been investigated.

Rescaling to lattice units and restoring the hopping parameter, the
Wilson fermion action with the site indices explicit becomes
\begin{equation}
{D_W}_{ij}=\delta_{i,j}+K\sum_\mu 
(1-\gamma_\mu )\delta_{i,j+e_\mu}
+(1+\gamma_\mu )\delta_{i,j-e_\mu}.
\end{equation}
By taking the coefficient $r$ of the Wilson term as unity we have
projection operators in the hoppings.  The physical fermion mass
is read off from the small momentum behavior as $m={1\over
  2a}(1/K-8)$.  This vanishes at at $K=K_c=1/8$.

Here we consider that the gauge fields are formulated as usual with
group valued matrices on the lattice links.  These are to be inserted
into the above hopping terms.  One could use the simple Wilson gauge
action as a sum over plaquettes
\begin{equation}
S_g={\beta\over 3}\sum_p {\rm Re\ Tr\ } U_p
\end{equation}
although this specific form is not essential the qualitative nature of
the phase diagram.  When the gauge fields are turned on, the dynamics
will move the fermion eigenvalues around, partially filling the holes
in eigenvalue pattern of Fig.~\ref{eigen1}.  Some eigenvalues can
become real and are related to gauge field topology
\cite{Creutz:2006ts}.

For the free theory the Hermitean and anti-Hermitean parts of the
action commute.  This ceases to be true in the interacting case
since both terms contain gauge matrices that themselves do not
commute.  Thus the left eigenvalues are generally different from right
ones.  Nevertheless, it is still true that the eigenvalues 
either appear in complex conjugate pairs or they are real.  This
follows from $\gamma_5$ Hermiticity, $D^\dagger=\gamma_5 D \gamma_5$.
Since $\gamma_5$ has unit determinant, $|D-\lambda|=0$ implies 
$|D^\dagger-\lambda|=|D-\lambda^*|^*=0$.

A technical difficulty with this approach is that gauge interactions
will renormalize the parameters.  To obtain massless pions one must
finely tune $K$ to $K_{c}$, an {\it a priori} unknown function of
the gauge coupling.  Despite the awkwardness of such tuning, this is
how numerical simulations with Wilson quarks generally proceed.  The
hopping parameter is adjusted to get the pion mass right, and one
assumes that the remaining predictions of current algebra reappear in
the continuum limit.

Note that the basic lattice theory has two parameters, $\beta$ and
$K$.  These are related to bare coupling, $\beta\sim 6/g_0^2$, and
quark mass, $(1/K-1/K_c)\sim m_q$.  We will now turn to a discussion
of this relation in more detail.

\subsection{Lattice versus continuum parameters}
\label{parameters}

As emphasized earlier, QCD is a remarkably economical theory in terms
of the number of adjustable parameters.  First of these is the overall
strong interaction scale, $\Lambda_{qcd}$.  This is scheme dependent,
but once a renormalization procedure has been selected, it is well
defined.  It is not independent of the coupling constant, the
connection being fixed by asymptotic freedom.  In addition, the theory
depends on the renormalized quark masses $m_i$, or more precisely the
dimensionless ratios $m_i/\Lambda_{qcd}$.  As with the overall scale,
the definition of $m_i$ is scheme dependent.  The two flavor theory
with degenerate quarks and $\Theta=0$ has one such mass parameter.  As
we wish to formulate the theory with a lattice cutoff in place, there
is a scale for this cutoff.  As with everything else, it is convenient
to measure this in units of the overall scale; so, a third parameter
for the cutoff theory is $a\Lambda_{qcd}$, where one can regard $a$ as
the lattice spacing.

How the bare parameters behave as the continuum limit is taken was
discussed rather abstractly in Section \ref{asymptoticfreedom}.  The
goal here is to explore some of the the lattice artifacts that arise
with Wilson fermions \cite{Wilson:1975id}.  On the lattice it is
generally easier to work directly with lattice parameters.  One of
these is the plaquette coupling $\beta$, which, with the usual
conventions, is related to the bare coupling $\beta=6/g_0^2$.  For the
quarks, the natural lattice quantity is the ``hopping parameter'' $K$.
And finally, the connection with physical scales appears via the
lattice spacing $a$.

\begin{figure}
\centering
\includegraphics[width=3in]{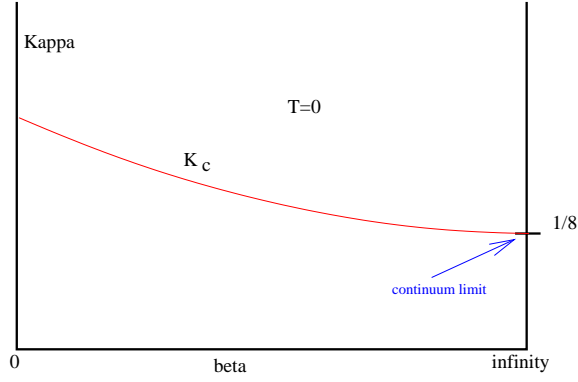}
\caption{ The continuum limit of lattice gauge theory with Wilson
fermions occurs at $\beta\rightarrow\infty$ and $K\rightarrow 1/8$.
Coming in from this point to finite beta is the curve $K_{c}(\beta)$,
representing the lowest phase transition in $K$ for fixed beta.  The
nature of this phase transition is a delicate matter, discussed in the
text.  Figure taken from Ref.~\cite{Creutz:2007fe}.}
\label{kbeta1}
\end{figure}

The set of physical parameters and the set of lattice parameters are,
of course, equivalent, and there is a well understood non-linear
mapping between them
\begin{equation}
\left\{
a\Lambda_{qcd},
{m\over\Lambda_{qcd}}
\right\}
 \longleftrightarrow \{\beta,K\}.
\end{equation}
Of course, to extract physical predictions we are interested in the
continuum limit $a\Lambda_{qcd}\rightarrow 0$.  For this, asymptotic
freedom tells us we must take $\beta\rightarrow \infty$ at a rate tied
to $\Lambda_{qcd}$.  Simultaneously we must take the hopping parameter
to a critical value.  With normal conventions, this takes
$K\rightarrow K_c\rightarrow 1/8$ at a rate tied to desired quark mass
$m$.  Figure \ref{kbeta1} sketches how the continuum limit is taken in
the $\beta,K$ plane.  Here we wish to further explore this phase
diagram with particular attention to hopping parameters larger than
$K_c$.  This discussion is adapted from Ref.~\cite{Creutz:2007fe} and
adds the possible twisted mass term to the exposition from
Ref.~\cite{Creutz:1996bg}.

\subsection{Artifacts and the Aoki phase}

We previously made extensive use of an effective field theory to
describe the interactions of the pseudo-scalar mesons.  Here we will
begin with the simplest form for the two flavor theory and then add
terms to mimic possible lattice artifacts.  The language is framed as
before in terms of the isovector pion field $\vec\pi\sim
i\overline\psi \gamma_5 \vec \tau \psi$ and the scalar sigma
$\sigma\sim \overline\psi \psi$.  The starting point for this
discussion is the canonical ``Mexican hat'' potential
\begin{equation}
V_0=\lambda (\sigma^2+\vec\pi^2-v^2)^2
\end{equation}
schematically sketched earlier in Fig.~\ref{v0}.  The potential has a
symmetry under $O(4)$ rotations amongst the pion and sigma fields
expressed as the four vector $\Sigma=(\sigma,\vec\pi)$.  This
represents the axial symmetry of the underlying quark theory.

As discussed before, the massless theory is expected to spontaneously
break chiral symmetry with the minimum for the potential occurring at
a non-vanishing value for the fields.  As usual, we take the vacuum to
lie in the sigma direction with $\langle\sigma\rangle > 0$.  The pions
are then Goldstone bosons, being massless because the potential
provides no barrier to oscillations of the fields in the pion
directions.  Also as discussed before, we include a quark mass by
adding a constant times the sigma field
\begin{equation}
 V_1=-m\sigma.
\end{equation}
This explicitly breaks the chiral symmetry by ``tilting'' the
potential as sketched in Fig.~\ref{v3}.  That selects a unique
vacuum which, for $m>0$, gives a positive expectation for sigma.  In
the process the pions gain a mass, with $m_\pi^2\sim m$.

Because of the symmetry of $V_0$, it does not matter physically in
which direction we tilt the vacuum.  In particular, a mass term of
form
\begin{equation}
m\sigma\rightarrow m\cos(\theta) \sigma + m\sin(\theta)\pi_3
\label{mrot}
\end{equation}
should give equivalent physics for any $\theta$.  In the earlier
continuum discussion we used this freedom to rotate the second term
away.  However, as we will see, lattice artifacts can break this
symmetry, introducing the possibility of physics at finite lattice
spacing depending on this angle.  As mentioned before, the second term
in this equation is what is usually called a ``twisted mass.''

The Wilson term inherently breaks chiral symmetry.  This will give
rise to various modifications of the effective potential.  The first
correction is expected to be an additive contribution to the quark
mass, i.e. an additional tilt to the potential.  This means that the
critical kappa, defined as the smallest kappa where a singularity is
found in the $\beta,K$ plane, will move away from the limiting value
of $1/8$.  Thus we introduce the function $K_c(\beta)$ and imagine
that the mass term is modeled with the form
\begin{equation}
m\rightarrow c_1(1/K-1/K_c(\beta)).
\label{c1}
\end{equation}

In general the lattice modification of the effective potential will
have further corrections of higher order in the effective fields.  A
natural way to include such is as an expansion in the chiral fields.
With this motivation we include a term in the potential of form $c_2
\sigma^2$.  Including these ideas in the effective model, we are led
to
\begin{equation}
V(\vec\pi,\sigma)=
\lambda(\sigma^2+\vec\pi^2-v^2)^2-c_1(1/K-1/K_c(\beta))\sigma
+c_2\sigma^2.
\end{equation}
Such a term was considered in
Refs.~\cite{Creutz:1996bg,Sharpe:1998xm}.  The predicted phase
structure depends qualitatively on the sign of $c_2$, but a priori we
have no information on this.\footnote{Ref.~\cite{Damgaard:2010cz} has
argued that $c_2$ should be positive.  We will return to this argument
a bit later in this section.} Indeed, as it is a lattice artifact, it
is expected that this sign might depend on the choice of gauge action.
Note that we could have added a term like $\vec\pi^2$, but this is
essentially equivalent since
$\vec\pi^2=(\sigma^2+\vec\pi^2)-\sigma^2$, and the first term here can
be absorbed, up to an irrelevant constant, into the starting Mexican
hat potential.

\begin{figure}
\centering
\includegraphics[width=3in]{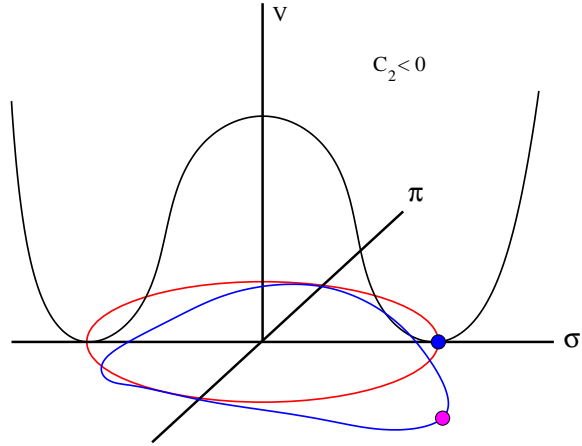}
\caption{
Lattice artifacts could quadratically warp the effective potential.
If this warping is downward in the sigma direction, the chiral
transition becomes first order without the pions becoming massless.
Figure taken from Ref.~\cite{Creutz:2007fe}.}
\label{cltzero}
\end{figure}

First consider the case when $c_2$ is less than zero, thus lowering
the potential energy when the field points in the positive or negative
sigma direction.  This quadratic warping helps to stabilize the sigma
direction, as sketched in Fig.~\ref{cltzero}, and the pions cease to be
true Goldstone bosons when the quark mass vanishes.  Instead, as the
mass passes through zero, we have a first order transition as the
expectation of $\sigma$ jumps from positive to negative.  This jump
occurs without any physical particles becoming massless.

\begin{figure}
\centering
\includegraphics[width=3in]{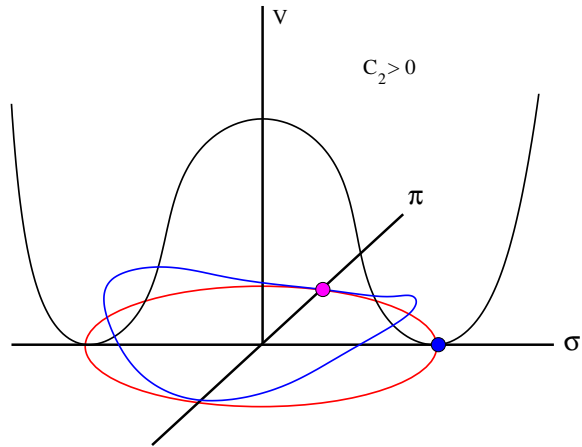}
\caption{
If the lattice artifacts warping the potential upward in the sigma
direction, the chiral transition splits into two second order
transitions separated by a phase where the pion field has an
expectation value.
Figure taken from Ref.~\cite{Creutz:2007fe}.}
\label{cgtzero}
\end{figure}

\begin{figure}
\centering
\includegraphics[width=3in]{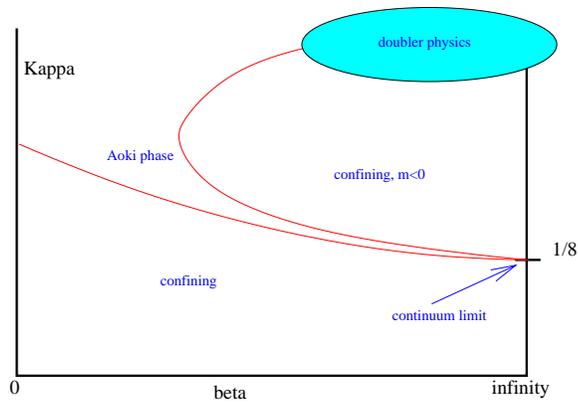}
\caption{The qualitative structure of the $\beta,K$ plane including
the possibility of an Aoki phase.  }
\label{kbeta3}
\end{figure}

Things get a bit more complicated if $c_2>0$, as sketched in
Fig.~\ref{cgtzero}.  In that case the chiral transition splits into
two second order transitions separated by a phase with an expectation
for the pion field, i.e. $\langle \vec\pi \rangle \ne 0$.  The
behavior is directly analogous to that shown in
Fig.~\ref{condensate1}, the main difference being that now the two
quarks are degenerate.  Since the pion field has odd parity and charge
conjugation as well as carries isospin, all of these symmetries are
spontaneously broken in the intermediate phase.  As isospin is a
continuous group, this phase will exhibit Goldstone bosons.  The
number of these is two, representing the two flavor generators
orthogonal to the direction of the expectation value.  If higher order
terms do not change the order of the transitions, there will be a
third massless particle exactly at the transition endpoints.  In this
way the theory reveals three massless pions exactly at the
transitions, as discussed by Aoki
\cite{Aoki:1983qi}.  The intermediate phase is usually referred to as
the ``Aoki phase.''  Assuming this $c_2>0$ case, Fig.~\ref{kbeta3}
shows the qualitative phase diagram expected.

Note the similarity of this discussion to that leading to the phase
diagram in Fig.~\ref{phasediagram}.  Indeed, lattice artifacts can
lead to the spontaneously broken CP region found there for the
$(m_1,m_2)$ plane to open up and remain present for degenerate quarks.
The Aoki phase is closely related to the possibility of CP violation
at $\Theta=\pi$ for unequal mass quarks.  Note also that this
connection with the earlier continuum discussion shows that with an
odd number of flavors, the spontaneous breaking of parity is the
normal expectation whenever the hopping parameter exceeds its critical
value.  Indeed, in this case the Aoki phase is less a lattice artifact
than a direct consequence of the CP violation expected in the
continuum theory at $\theta=\pi$.

\subsection{Twisted mass}
\label{twisted}

The $c_2$ term breaks the equivalence of different chiral directions.
This means that physics will indeed depend on the angle $\theta$ if
one takes a mass term of the form in Eq.~(\ref{mrot}).
Consider complexifying the fermion mass in the usual way
\begin{equation}
m\overline\psi\psi\rightarrow m\overline\psi_L\psi_R
+m^*\overline\psi_R\psi_L
={\rm Re}\ m\ \overline\psi\psi+i{\rm Im}\ m\ \overline\psi\gamma_5\psi.
\end{equation}
The rotation of  Eq.~(\ref{mrot}) is equivalent to giving the up and
down quark masses opposite phases
\begin{eqnarray}
m_u\rightarrow e^{+i\theta}m_u\cr 
m_d\rightarrow e^{-i\theta}m_d.
\end{eqnarray}
Thus motivated, we can consider adding a new mass term to the lattice theory
\begin{equation}
\mu\ i\overline\psi \tau_3\gamma_5\psi \sim \mu\pi_3.
\end{equation}
This extends our effective potential to
\begin{equation}
V(\vec\pi,\sigma)=\lambda(\sigma^2+\vec\pi^2-v^2)^2-c_1(1/K-1/K_c(\beta))\sigma
+c_2\sigma^2-\mu \pi_3.
\end{equation}
The twisted mass represents the addition of a ``magnetic field''
conjugate to the order parameter for the Aoki phase.

\begin{figure}
\centering
\includegraphics[width=3in]{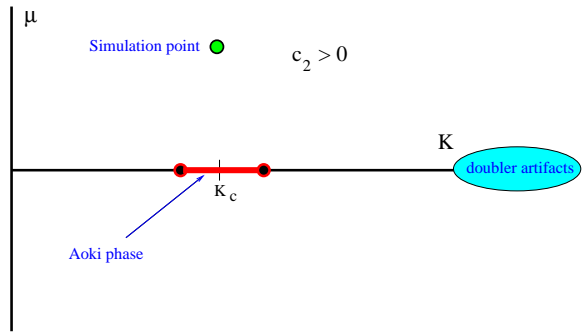}
\caption{
Continuing around the Aoki phase with twisted mass.  This sketch
considers the case $c_2>0$ where the parity broken phase extends over
a region along the kappa axis.
Figure taken from Ref.~\cite{Creutz:2007fe}.}
\label{tm1}
\end{figure}

\begin{figure}
\centering
\includegraphics[width=3in]{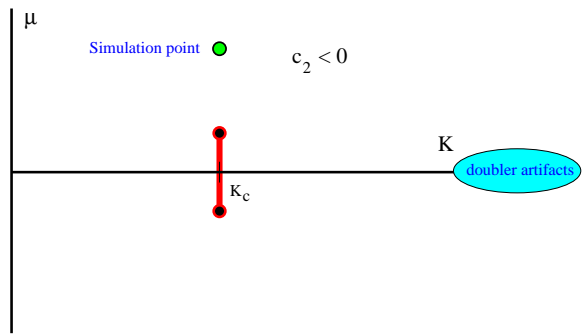}
\caption{
As in Fig.~\ref{tm1}, but for the case $c_2<0$ so the chiral
transition on the kappa axis becomes first order.
Figure taken from Ref.~\cite{Creutz:2007fe}.}
\label{tm2}
\end{figure}

There are a variety of motivations for adding such a term to our
lattice action \cite{Frezzotti:2003ni,Munster:2004wt}.  Primary among
them is that $O(a)$ lattice artifacts can be arranged to cancel.  With
two flavors of conventional Wilson fermions, these effects change sign
on going from positive to negative mass, and if we put all the mass
into the twisted term we are half way between.  It should be noted
that this cancellation only occurs when all the mass comes from the
twisted term; for other combinations with a traditional mass term,
some lattice artifacts of $O(a)$ will survive.  Also, although it
looks like we are putting phases into the quark masses, these cancel
between the two flavors.  The resulting fermion determinant remains
positive and we are working at $\Theta=0$.  Furthermore, the algorithm
is considerably simpler and faster than either overlap
\cite{Narayanan:1994gw,Neuberger:1997fp} or domain wall
\cite{Kaplan:1992bt,Furman:1994ky} fermions while avoiding the
diseases of staggered quarks \cite{Creutz:2007yg}.  Another nice
feature of adding a twisted mass term is that it allows a better
understanding of the Aoki phase and shows how to continue around it.
Figures\ref{tm1} and \ref{tm2} show how this works for the case
$c_2>0$ and $c_2<0$, respectively.

Of course some difficulties come along with these advantages.  First,
one needs to know $K_c$.  Indeed, with the Aoki phase present, the
definition of this quantity is a bit fuzzy.
And second, the mass needs to be larger than the
$c_2$ artifacts.  Indeed, as Figs.~\ref{tm1} and \ref{tm2} suggest, if
it is not, then one is really studying the physics of the Aoki phase,
not the correct continuum limit.  This also has implications for how
close to the continuum one must be to study this structure; in
particular, one must have $\beta$ large enough so the Aoki phase does
not extend into the doubler region.

The question of the sign of $c_2$ remains open.  Simulations suggest
that the usual Aoki phase with $c_2>0$ is the situation with the
Wilson gauge action.  Recently Ref.~\cite{Damgaard:2010cz} has pointed
out that with the twisted mass term present, all eigenvalues of the
product of gamma five with the Dirac operator will have non-zero
imaginary parts.  Thus to have $c_2<0$, the phase transition at
non-vanishing twisted mass must occur where the fermion determinant
does not vanish on any configuration.  This contrasts with the $c_2>0$
case where small eigenvalues of $D$ are expected in the vicinity of
the critical hopping.  This at first sight makes $c_2<0$ seem somewhat
unnatural; however, this is not a proof since we saw in
Subsection \ref{phasesection} that phase transitions without small
eigenvalues of the Dirac operator do occur in the continuum theory for
two flavors with non-degenerate quarks.

This picture of the artifacts associated with Wilson fermions raises
some interesting questions.  One concerns the three flavor theory.  As
discussed previously, in this case a parity broken phase becomes
physical with negative mass.  Indeed, three degenerate quarks of
negative mass represent QCD with a strong CP angle $\theta=\pi$, for
which spontaneous breaking of CP is expected.  In some sense the Aoki
phase becomes physical.  Also, with three flavors the twisting process
is not unique, with possible twists in the $\lambda_3$ or $\lambda_8$
directions.  For example, using only $\lambda_3$ would suggest a
possible twisted mass of form $m_u\sim e^{2\pi i/3}$, $m_d\sim
e^{-2\pi i/3}$, $m_s\sim 1$.  Whether there is an optimum twisting
procedure for three flavors is unclear.

Another special case is one flavor QCD \cite{Creutz:2006ts}.  In this
situation the anomaly removes all chiral symmetry, and the quark
condensate loses meaning as an order parameter.  The critical value of
kappa where the mass gap disappears is decoupled from the point of
zero physical quark mass.  There is a parity broken phase, but it
occurs only at sufficiently negative mass.  And from the point of view
of twisting the mass, without chiral symmetry there is nothing to
twist other than turning on the physical parameter $\Theta$.

\newpage
\Section{Lattice actions preserving chiral symmetry}

\subsection{The Nielsen-Ninomiya theorem}

As discussed some time ago \cite{Nielsen:1980rz}, the doubling issue
is closely tied to topology in momentum space.  To see how this works,
let us first establish a gamma matrix convention
\begin{eqnarray}
&&\vec\gamma=\sigma_1\otimes \vec\sigma=
\pmatrix{0 & \vec\sigma\cr \vec\sigma & 0}\\
&&\gamma_0=\sigma_2\otimes I=
\pmatrix{0 & -i\cr i & 0}\\
&&\gamma_5=\sigma_3\otimes I=
\pmatrix{1 & 0\cr 0 & -1}.
\end{eqnarray} 
Now suppose we have an anti-Hermitean Dirac operator $D$ that anti-commutes
with $\gamma_5$
\begin{equation}
D=-D^\dagger=-\gamma_5 D \gamma_5.
\end{equation}
Considering this quantity in momentum space, its most general form is
\begin{equation}
D(p)=\pmatrix{0&z(p) \cr -z^*(p)&0\cr}
\end{equation}
where $z(p)$ is a quaternion
\begin{equation}
z(p)=z_0(p)+i\vec\sigma\cdot\vec z(p).
\end{equation}
Thus we see that any chirally symmetric Dirac operator maps momentum
space onto the space of quaternions.

The Dirac equation is obtained by expanding the momentum space
operator around a zero, i.e. $D(p)\simeq i\slashchar p =
i\gamma_\mu p_\mu$.  Now consider a three dimensional sphere embedded
in four-dimensional momentum-space and surrounding the zero with a
constant $D^2\sim p^2$.  The above quaternion wraps non-trivially
about the the origin as we cover this sphere.  Here is where the
topology comes in \cite{Nielsen:1980rz,Creutz:2007af}.  Momentum space
is periodic over Brillouin zones.  We must have $z(p)=z(p+2\pi n)$
where $n$ is an arbitrary four vector with integer components.
Because of this, we can restrict the momentum components to lie in the
range $-\pi<p_\mu\le\pi$, and we cannot have any non-trivial topology
on the surface of this zone.  Any mapping associated with a zero in
$z(p)$ must unwrap somewhere else before we get to the surface.
Assuming $D(p)$ remains finite, any zero must be accompanied by
another wrapping in the opposite sense.  Because of doubling, the 16
species with naive fermions split up into 8 zeros of each sense.

The above argument only tells us that a chiral lattice theory must
have an even number of species.  The case of a minimal doubling with
only two species is in fact possible, although all methods presented
so far \cite{Misumi:2010ea} appear to involve a breaking of
hyper-cubic symmetry.  This breaking is associated with the direction
between the zeros; this makes one direction special, although it might
be possible to avoid it by having the zeros form a symmetric lattice
using the periodicity of momentum space.  This has not yet been
demonstrated.

In earlier sections we discussed how an odd number of flavors raised
some interesting issues; in particular the sign of the mass becomes
relevant.  In spite of this, there seems to be no contradiction with
having, say, three light flavors in the continuum with a well defined
chiral limit.  The above lattice argument, however, seems to indicate
troubles with maintaining an exact chiral symmetry with an odd number
of flavors.  Whether this apparent conflict is serious is unclear.
One could always start with a multiple fermion theory and then, with
something like a Wilson term, give a few species masses while leaving
behind an odd number of massless fermions.  This will involve some
parameter tuning, but presumably can give a reasonable chiral limit
for odd $N_f>2$.  This does not obviate the fact emphasized earlier
that with only one flavor there must not be any remaining chiral
symmetry even in the continuum.

\subsection{Minimal doubling}

Several chiral lattice actions satisfying the minimal condition of
$N_f=2$ flavors are known.  Some time ago Karsten
\cite{Karsten:1981gd} presented a simple form by inserting a factor of
$i\gamma_4$ into a Wilson like term for space-like hoppings.  A slight
variation appeared in a discussion by Wilczek \cite{Wilczek:1987kw} a
few years later.  More recently, a new four-dimensional action was
motivated by the analogy with two dimensional graphene
\cite{Creutz:2007af}.  Since then numerous variations have been
presented \cite{Borici:2007kz, Bedaque:2008jm,Kimura:2009di,
  Creutz:2010cz,Misumi:2010ea}.

The main potential advantage with these approaches lies in their
ultra-locality.  They all involve only nearby neighbor hoppings for
the fermions.  Thus they should be extremely fast in simulations while
still protecting masses from additive renormalization and helping
control mixing of operators with different chirality.  The approach
also avoids the uncontrolled errors associated with the rooting
approximation discussed later
\cite{Creutz:2007yg,Creutz:2007rk,Creutz:2009kx,Creutz:2009zq}.  On
the other hand, all minimally-doubled actions presented so far have
the above mentioned disadvantage of breaking lattice hyper-cubic
symmetry.  With interactions, this will lead to the necessity of
renormalization counter-terms that also violate this symmetry
\cite{Capitani:2010nn}.  The extent to which this will complicate
practical simulations remains to be investigated.

Minimally-doubled chiral fermions have the unusual property of a
single local field creating two distinct fermionic species.  Here we
discuss a point-splitting method for separating the effects of the two
flavors which can be created by a single fermion field.  For this we
will work with one specific form for the fermion action, but the method
should be easily extended to other minimally-doubled formulations.

We concentrate on a minimally-doubled fermion action which
is a slight generalization of those presented by Karsten
\cite{Karsten:1981gd} and Wilczek \cite{Wilczek:1987kw}.  The fermion
term in the lattice action takes the form $\overline\psi D\psi$.  For
free fermions we start in momentum space with
\begin{equation}
\label{pspace}
D(p)=
i\sum_{i=1}^3 \gamma_i \sin(p_i)
+{i\gamma_4\over \sin(\alpha)}\left(
\cos(\alpha)+3
-\sum_{\mu=1}^4\cos(p_\mu)
\right).
\end{equation}
This includes a Wilson like term for the space-like hoppings but
containing an extra factor of $i\gamma_4$.  As a function of the
momentum $p_\mu$, the propagator $D^{-1}(p)$ has two poles, located at
$\vec p = 0$, $p_4=\pm \alpha$.  Relative to the naive fermion action,
the other doublers have been given a large ``imaginary chemical
potential'' by the Wilson like term.  The parameter $\alpha$ allows
adjusting the relative positions of the poles.  The original
Karsten/Wilczek actions correspond to $\alpha=\pi/2$.

This action maintains one exact chiral symmetry, manifested in the
anti-commutation relation $[D,\gamma_5]_+=0$.  The two species,
however, are not equivalent, but have opposite chirality.  To see
this, expand the propagator around the two poles and observe that one
species, that corresponding to $p_4=+\alpha$, uses the usual gamma
matrices, but the second pole gives a proper Dirac behavior using
another set of matrices $\gamma_\mu^\prime=\Gamma^{-1}\gamma_\mu
\Gamma$.  The Karsten/Wilczek formulation uses
$\Gamma=i\gamma_4\gamma_5$, although other minimally-doubled actions
may involve a different transformation.  After this transformation
$\gamma_5^\prime=-\gamma_5$, showing that the two species rotate
oppositely under the exact chiral symmetry, and this symmetry should
be regarded as ``flavored.''  One can think of the physical chiral
symmetry as that generated in the continuum theory by
$\tau_3\gamma_5$.

It is straightforward to transform the momentum space action in
Eq.~(\ref{pspace}) to position space and insert gauge fields
$U_{ij}=U_{ji}^\dagger$ on the links connecting lattice sites.
Explicitly indicating the site indices, the Dirac operator becomes
\begin{eqnarray}
 D_{ij}&=&
U_{ij}\sum_{\mu=1}^3 \gamma_i{\delta_{i,j+e_\mu}
-\delta_{i,j-e_\mu}\over
2}\cr
&+&
{i\gamma_4\over \sin(\alpha)}
\bigg(
(\cos(\alpha)+3)\delta_{ij}
-U_{ij} \sum_{\mu=1}^4{\delta_{i,j+e_\mu}
+\delta_{i,j-e_\mu}\over 2}
\bigg).
\end{eqnarray}
Again we see analogy with Wilson fermions \cite{Wilson:1975id} for the
space directions but augmented with an {$i\gamma_4$} inserted in the
Wilson term.

Perturbative calculations \cite{Capitani:2010nn} have shown that
interactions with the gauge fields can shift the relative positions of
the poles along the direction between them.  In other words, the
parameter $\alpha$ receives an additive renormalization.  Furthermore,
the form of the action treats the fourth direction differently than
the spatial coordinates, this is the breaking of hyper-cubic symmetry
mentioned above.  There arise three possible new counter-terms for the
renormalization of the theory.  First there is a possible
renormalization of the on-site contribution to the action proportional
to $i\overline\psi\gamma_4\psi$.  This provides a handle on the shift
of the parameter $\alpha$.  Secondly, the breaking of the hyper-cubic
symmetry indicates one may need to adjust the fermion ``speed of
light.''  This involves a combination of the above on-site term and
the strength of temporal hopping proportional to ${\delta_{i,j+e_4}
+\delta_{i,j-e_4}}$.  Finally, the breaking of hyper-cubic symmetry
can feed back into the gluonic sector, suggesting a possible
counter-term of form $F_{4\mu}F_{4\mu}$ to maintain the gluon ``speed
of light.''  In lattice language, this corresponds to adjusting the
strength of time-like plaquettes relative to space-like ones.

Of these counter-terms, $i\overline\psi\gamma_4\psi$ is of dimension 3
and is probably the most essential.  Quantum corrections induce the
dimension 4 terms, suggesting they may be small and could partially be
absorbed by accepting a lattice asymmetry.  How difficult these
counter-terms are to control awaits simulations.

Note that all other dimension 3 counter-terms are forbidden by basic
symmetries.  For example, chiral symmetry forbids $\overline\psi\psi$
and $i\overline\psi\gamma_5\psi$ terms, and spatial cubic symmetry
removes $\overline\psi\gamma_i\psi$,
$\overline\psi\gamma_i\gamma_5\psi$, and
$\overline\psi\sigma_{ij}\psi$ terms.  Finally, commutation with
$\gamma_4$ plus space inversion eliminates $\overline\psi
\gamma_4\gamma_5\psi$.

The fundamental field $\psi$ can create either of the two species.
For a quantity that creates only one of them, it is natural to combine
fields on nearby sites in such a way as to cancel the other.  In other
words, one can point split the fields to separate the poles which
occur at distinct ``bare momenta.''  For the free theory, one
construction that accomplishes this is to consider
\begin{eqnarray} 
&&u(q)={1\over 2}\left(1+{\sin(q_4+\alpha)\over \sin(\alpha)}
\right)\psi(q+\alpha e_4)\cr
&&d(q)={1\over 2}\ {\Gamma}\left(1-{\sin(q_4-\alpha)\over \sin(\alpha)}
\right)\psi(q-\alpha e_4)
\end{eqnarray}
where $\Gamma=i\gamma_4\gamma_5$ for the Karsten/Wilczek formulation.
Here we have inserted factors containing zeros cancelling the undesired
pole.  This construction is not unique, and specific details will
depend on the particular minimally-doubled action in use.  The factor
of $\Gamma$ inserted in the $d$ quark field accounts for the fact that
the two species use different gamma matrices.  This is required since
the chiral symmetry is flavored, corresponding to an effective minus
sign in $\gamma_5$ for one of the species.

It is now straightforward to proceed to position space and insert
gauge field factors to keep gauge transformation properties simple
\begin{eqnarray}
&&u_x
=
{1\over 2}{e^{i\alpha x_4}}\left(\psi_x+i\ 
{U_{x,x-e_4}\psi_{x-e_4}
-U_{x,x+e_4}\psi_{x+e_4}\over 2 \sin(\alpha)}\right)\cr
&&d_x
=
{1\over 2}{\Gamma e^{-i\alpha x_4}}\left(\psi_x-i\ {
U_{x,x-e_4}\psi_{x-e_4}-U_{x,x+e_4}\psi_{x+e_4}\over 2 \sin(\alpha)}\right).
\end{eqnarray}
The various additional phase factors serve to remove the oscillations
associated with the bare fields having their poles at non-zero
momentum.

Given the basic fields for the individual quarks, one can go on to
construct mesonic fields, which then also involve point splitting.  To
keep the equations simpler, we now consider the case $\alpha=\pi/2$.
For example, the neutral pion field becomes
\begin{eqnarray}
&\pi_0(x)
={i\over 2}(\overline u_x\gamma_5 u_x-\overline d_x \gamma_5 d_x)=\cr
&{i\over 16}\bigg(4\overline\psi_x\gamma_5\psi_x
+\overline\psi_{x-e_4}\gamma_5\psi_{x-e_4}
+\overline\psi_{x+e_4}\gamma_5\psi_{x+e_4}
\cr&
-\overline\psi_{x+e_4}UU\gamma_5\psi_{x-e_4}
-\overline\psi_{x-e_4}UU\gamma_5\psi_{x+e_4}\bigg).
\end{eqnarray}
Note that this involves combinations of fields at sites separated 
by either 0 or 2 lattice spacings.  In contrast, the $\eta^\prime$ takes
the form
\begin{eqnarray}
&\eta^\prime(x)
={i\over 2}(\overline u_x \gamma_5 u_x+\overline d_x \gamma_5 d_x)=
\cr
&{1\over 8}\bigg(
\overline\psi_{x-e_4}U\gamma_5\psi_x
-\overline\psi_xU\gamma_5\psi_{x-e_4}
-\overline\psi_{x+e_4}U\gamma_5\psi_x
+\overline\psi_xU\gamma_5\psi_{x+e_4}\bigg)
\end{eqnarray}
where all terms connect even with odd parity sites.  In a recent
paper, Tiburzi \cite {Tiburzi:2010bm} has discussed how the anomaly,
which gives the $\eta^\prime$ a mass of order $\Lambda_{qcd}$, can be
understood in terms of the necessary point splitting.

\subsection{Domain wall and overlap fermions}

The overlap fermion was originally developed \cite{Narayanan:1993sk}
as a limit of a fermion formulation using four dimensional surface
modes on a five dimensional lattice.  This effectively amounts to
using Shockley surface states as the basis for a theory maintaining
chiral symmetry \cite{Kaplan:1992bt}.  For a review see
Ref.~\cite{Jansen:1994ym}.  The idea is to set up a theory in one
extra dimension so that surface modes exist, and our observed world is
an interface with our quarks and leptons being these surface modes.
Particle hole symmetry naturally gives the basic fermions zero mass.
In the continuum limit the extra dimension becomes unobservable due to
states in the interior requiring a large energy to create.  In this
picture, opposing surfaces carry states of opposite helicity, and the
anomalies are due to a tunnelling through the extra dimension.

Ref. \cite{Creutz:1994ny} discussed the general conditions for surface
modes to exist.  Normalized solutions are bound to any interface
separating a region with supercritical from sub-critical hopping.
Kaplan's original paper \cite{Kaplan:1992bt} considered not a surface,
but an interface with $M=M_{cr}+m \epsilon(x)$, where $M_{cr}$ is the
critical value for the mass parameter where the five dimensional
fermions would be massless.  Shamir \cite{Shamir:1993zy} presented a
somewhat simpler picture where the hopping vanishes on one side, which
then drops out of the problem and we have a surface.

To couple gluon fields to this theory without adding unneeded degrees
of freedom, the gauge fields are taken to lie in the four physical
space-time directions and be independent of the fifth coordinate.  In
this approach, the extra dimension is perhaps best thought of as a
flavor space \cite{Narayanan:1992wx}.  With a finite lattice this
procedure gives equal couplings of the gauge field to the fermion
modes on opposing walls in the extra dimension.  Since the left and
right handed modes are separated by the extra dimension, they only
couple through the gauge field.  The result is an effective light
Dirac fermion.  In the case of the strong interactions, this provides
an elegant scheme for a natural chiral symmetry without the tuning
inherent in the Wilson approach.  The breaking of chiral symmetry
arises only through finiteness of the extra dimension.\footnote{The
  anomaly, however, shows that some communication between the surfaces
  survives even as the extra dimension becomes infinite.  This is
  possible since the same gauge fields are on each surface.}

The name ``overlap operator'' comes from the overlap of eigenstates of
the different five dimensional transfer matrices on each side of the
interface.  Although originally derived from the infinite limit of the
five dimensional formalism, one can formulate the overlap operator
directly in four dimensions.  We begin with the fermionic part of some
generic action as a quadratic form
$
S_f= \overline\psi  D \psi.
$ 
The usual ``continuum'' Dirac operator $D=\sum\gamma_\mu D_\mu$
naively anti-commutes with $\gamma_5$, i.e. $[\gamma_5, D]_+=0$.  Then
the change of variables $\psi \rightarrow e^{i\theta\gamma_5} \psi$
and $\overline\psi \rightarrow \overline\psi e^{i\theta\gamma_5}$
would be a symmetry of the action.  This, however, is inconsistent
with the chiral anomalies.  The conventional continuum discussion
presented earlier maps this phenomenon into the fermionic measure
\cite{Fujikawa:1979ay}.  

On the lattice we work with a finite number of degrees of freedom;
thus, the above variable change is automatically a symmetry of the
measure.  To parallel the continuum discussion, it is necessary to
modify the symmetry transformation on the action so that the measure
is no longer invariant.  Remarkably, it is possible to construct a
modified symmetry under which corresponding actions are exactly
invariant.

To be specific, one particular variation \cite{Neuberger:1997fp,
  Neuberger:1998my,Neuberger:1998wv,Chiu:1998gp,Chandrasekharan:1998wg}
modifies the change of variables to
\begin{eqnarray}
&&\psi \longrightarrow 
e^{i\theta\gamma_5}
\psi\cr
&&\overline\psi \longrightarrow \overline\psi 
e^{i\theta(1-aD)\gamma_5} 
\end{eqnarray}
where $a$ represents the lattice spacing.  Note the asymmetric way in
which the independent Grassmann variables $\psi$ and $\overline\psi$
are treated.  Requiring the action to be unchanged gives the relation
\cite{Ginsparg:1981bj,Hasenfratz:1998ri,Hasenfratz:1998jp}.
\begin{equation}
\label{relation}
D \gamma_5  = -\gamma_5 D+a D\gamma_5 D=
-\hat\gamma_5 D
\end{equation} 
with $\hat\gamma_5=(1-aD)\gamma_5$.  To proceed, we also assume the
Hermeticity condition $\gamma_5 D \gamma_5 = D^\dagger$.  We see that
the naive anticommutation relation receives a correction of order the
lattice spacing.  The above ``Ginsparg-Wilson relation'' along with
the Hermeticity condition is equivalent to the unitarity of the
combination $V=1-aD$.

Neuberger \cite{Neuberger:1998my,Neuberger:1998wv} and Chiu and Zenkin
\cite{Chiu:1998gp} presented an explicit operator with the above
properties.  They first construct $V$ via a unitarization of an
undoubled chiral violating Dirac operator, such as the Wilson operator
$D_w$.  This operator should also satisfy the above Hermeticity
condition $\gamma_5 D_w \gamma_5 = D_w^\dagger$.  Specifically, they
consider
\begin{equation}
\label{projection}
V=-D_w(D_w^\dagger D_w)^{-1/2}.
\end{equation} 
The combination $(D_w^\dagger D_w)^{-1/2}$ is formally defined by
finding a unitary operator to diagonalize the Hermitean combination
$D_w^\dagger D_w$, taking the square root of the eigenvalues, and then
undoing the unitary transformation.

\begin{figure}
\centering
\includegraphics[width=4in]{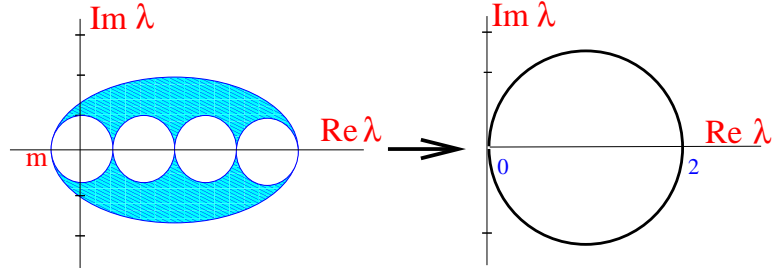}
\caption{
\label{eigen3} 
The overlap operator is constructed by projecting the eigenvalues of
the Wilson operator onto a circle.  Figure taken from
Ref.~\cite{Creutz:2002qa}. }
\end{figure}

Directly from $V$ we construct the overlap operator as
\begin{equation}
D=(1-V)/a.
\end{equation}
The Ginsparg-Wilson relation of Eq.~(\ref{relation}) is most
succinctly written as the unitarity of $V$ coupled with its $\gamma_5$
Hermeticity
\begin{equation}
\gamma_5 V\gamma_5 V=1.
\end{equation}
The basic projection process is illustrated in Fig.~\ref{eigen3}.

The overlap operator has several nice properties.  Being constructed
from a unitary operator, the normality of $D$ is guaranteed.  But,
most important, it exhibits a lattice version of an exact chiral
symmetry.\cite{Luscher:1998pqa} The fermionic action $\overline\psi
D\psi$ is invariant under the transformation
\begin{eqnarray}
&\psi\rightarrow e^{i\theta\gamma_5}\psi\cr
&\overline\psi\rightarrow \overline\psi e^{i\theta\hat\gamma_5}
\label{symmetry}
\end{eqnarray}
where
\begin{equation}
\hat\gamma_5 =V\gamma_5.
\end{equation}
As with $\gamma_5$, this quantity which appeared in
Eq.~(\ref{relation}) is Hermitean and its square is unity.
Thus its eigenvalues are all plus or minus unity.  The trace
defines an index
\begin{equation}
\label{tracehat}
\nu={1\over 2}{\rm Tr}\hat\gamma_5
\end{equation}
which plays exactly the role of the index in the continuum.  If the
gauge fields are smooth, this counts the topology of the gauge
configuration.  The factor of $1/2$ in Eq.~(\ref{tracehat}) appears
because the exact zero modes of the overlap operator have partners on
the opposite side of the unitarity circle that also contribute to the
trace.

At this point the hopping parameter in $D_w$ is a parameter.  To have
the desired single light fermion per flavor of the theory, the hopping
parameter should be appropriately adjusted to lie above the critical
value where $D_w$ describes a massless flavor, but not so large that
additional doublers come into play \cite{Neuberger:1999jw}.  There are
actually two parameters to play with, the hopping parameter of $D_w$,
and the lattice spacing.  When the latter is finite and gauge fields
are present, the location of the critical hopping parameter in $D_w$
is expected to shift from that of the free fermion theory.  As we saw
when discussing the Aoki phase, there is potentially a rather complex
phase structure in the plane of these two parameters, with various
numbers of doublers becoming massless as the hopping is varied.  The
Ginsparg-Wilson relation in and of itself does not in general
determine the number of physical massless fermions.

Although the Wilson operator entering this construction is local and
quite sparse, the resulting overlap action is not.  Because of the
inversion in Eq.~(\ref{projection}), it involves direct couplings
between arbitrarily separated sites
\cite{Hernandez:1998et,Horvath:1998cm,Horvath:1999bk}.  How rapidly
these couplings fall with distance depends on the gauge fields and is
not fully understood.  The five dimensional domain-wall theory is
local in the most naive sense; all terms in the action only couple
nearest neighbor sites.  However, were one to integrate out the heavy
modes, the resulting low energy effective theory would also involve
couplings with arbitrary range.  Despite these non-localities,
encouraging studies
\cite{Neuberger:1998wv,Edwards:1998wx,Borici:1999ws,
  Dong:2000mr,Gattringer:2000js} show that it may indeed be practical
to implement the required inversion in large scale numerical
simulations.  The overlap operator should have memory advantages over
the domain wall approach since a large number of fields corresponding
to the extra dimension do not need to be stored.

The overlap approach hides the infinite sea of heavy fermion states in
the extra dimension of the domain wall approach.  This tends to
obscure the possible presence of singularities in the required
inversion of the Wilson kernel.
Detailed analysis
\cite{Luscher:2000zd,Kikukawa:1998pd} shows that this operator is
particularly well behaved order by order in perturbation theory.  This
has led to hopes that this may eventually lead to a rigorous
formulation of chiral models, such as the standard model.

Despite being the most elegant known way to have an exact remnant of
chiral symmetry on the lattice, the overlap operator raises several
issues.  These complications probably become insignificant as the
continuum limit is approached, but should be kept in mind given the
high computational cost of this approach.  To begin with, the overlap
is highly non-unique.  It explicitly depends on the kernel being
projected onto the unitary circle.  Even after choosing the Wilson
kernel, there is a dependence on the input mass parameter.  One might
want to define topology in terms of the number of exact zero modes of
the overlap operator.  However the non-uniqueness leaves open the
question of whether the winding number of a gauge configuration might
depend on this choice.  Later we will return to the question of
possible ambiguities in defining topological susceptibility in the
continuum limit.

In this connection, it is possible to make a bad choice for the mass
parameter.  In particular, if it is chosen below the continuum kappa
critical value of $1/8$, no low modes will survive.  This is true
despite the fact that the corresponding operator will still satisfy
the Ginsparg-Wilson condition.  This explicitly shows that just
satisfying the Ginsparg-Wilson condition is not a sufficient condition
for a chiral theory.  Conversely, if one chooses the mass parameter
too far in the supercritical region, additional low modes will be
produced from the doublers.  As mentioned earlier, the Ginsparg-Wilson
condition does not immediately determine the number of flavors in the
theory.

Another issue concerns the one flavor case, discussed earlier.
Because of the anomaly, this theory is not supposed to show any chiral
symmetry and has no Goldstone bosons.  Nevertheless, one can construct
the overlap operator and it will satisfy the Ginsparg-Wilson
condition. This shows that the consequences of this condition are
weaker than for the usual continuum chiral symmetry.  With a
conventional chiral symmetry, the spectrum cannot show a gap.  Either
we have the Goldstone bosons of spontaneous chiral breaking or we have
massless fermions \cite{'tHooft:1979bh}.

It should also be noted that the overlap behaves peculiarly for
fermions in higher representations than the fundamental.  As we
discussed earlier, the number of zero modes associated with a
non-trivial topology in the continuum theory depends on the fermion
representation being considered.  It has been observed in numerical
simulations that the appropriate multiplicity is not always seen for
the overlap operator constructed on rough gauge configurations
\cite{Edwards:1998dj}.

As a final comment, note that these actions preserving a chiral
symmetry all involve some amount of non-locality.  With minimal
doubling this has a finite range, but is crucial for allowing the
anomaly to work out properly.  An important consequence is that the
operator product expansion, a standard perturbative tool, must involve
operators with a similar non-locality.  The ambiguities in defining
non-degenerate quark masses lie in these details.

\subsection{Staggered fermions}

Another fermion formulation that has an exact chiral symmetry is the
so called ``staggered'' approach.  To derive this it is convenient to
begin with the ``naive'' discretization of the Dirac equation from
before.  This considers fermions hopping between nearest neighbor
lattice sites while picking up a factor of $\pm i\gamma_\mu$ for a hop
in direction $\pm \mu$.  Going to momentum space, the discretization
replaces powers of momentum with trigonometric functions, for example
\begin{equation}
\gamma_\mu p_\mu\rightarrow \gamma_\mu {\sin(p_\mu)}.
\end{equation}
Here we work in lattice units and thus drop factors of $a$.  As
discussed before, this formulation reveals the famous ``doubling''
issue, arising because the fermion propagator has poles not only for
small momentum, but also whenever any component of the momentum is at
$\pi$.  The theory represents not one fermion, but sixteen.  And the
various doublers have differing chiral properties.  This arises from
the simple observation that
\begin{equation}
{d\over dp}\sin(p)|_{p=\pi}=-{d\over dp}\sin(p)|_{p=0}.
\end{equation}
The consequence is that the helicity projectors $(1\pm\gamma_5)/2$ for
a travelling particle depend on which doubler one is observing.

\begin{figure*}
\centering
\includegraphics[width=1.6in]{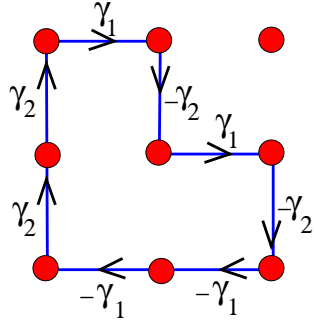}
\caption{When a fermion circumnavigates a loop in the naive
  formulation, it picks up a factor that always
  involves an even power of any particular gamma matrix.
Figure from Ref.~\cite{Creutz:2007rk}.}
\label{loop} 
\end{figure*}

Now consider a fermion traversing a closed loop on the lattice.  As
illustrated in Fig.~\ref{loop}, the corresponding gamma matrix factors
will always involve an even number of any particular $\gamma_\mu$.
Thus the resulting product is proportional to the identity.  If a
fermion starts off with a particular spinor component, it will wind up
in the same component after circumnavigating the loop.  This means
that the fermion determinant exactly factorizes into four equivalent
pieces.  The naive theory has an exact $U(4)$ symmetry, as pointed out
some time ago by Karsten and Smit \cite{Karsten:1980wd}.  Indeed, for
massless fermions this is actually a $U(4)\otimes U(4)$ chiral
symmetry.  This symmetry does not contradict any anomalies since it is
not the full naive $U(16)\otimes U(16)$ of 16 species.  The chiral
symmetry generated by $\gamma_5$ remains exact, but is allowed because
it is actually a flavored symmetry. As mentioned above, the helicity
projectors for the various doubler species use different signs for
$\gamma_5$.

The basic idea of staggered fermions is to divide out this $U(4)$
symmetry \cite{Kogut:1974ag,Susskind:1976jm,Sharatchandra:1981si}  by
projecting out a single component of the fermion spinor on each site.
Taking $\psi\rightarrow P\psi$, one projector that accomplishes this
is
\begin{equation}
P={1\over 4} \bigg(1+i\gamma_1\gamma_2 (-1)^{x_1+x_2}
+i\gamma_3\gamma_4 (-1)^{x_3+x_4}
 +\gamma_5(-1)^{x_1+x_2+x_3+x_4}\bigg)
\end{equation}
where the $x_i$ are the integer coordinates of the respective lattice
sites.  This immediately reduces the doubling from a factor of sixteen
to four.  It is the various oscillating sign factors in this formula
that give staggered fermions their name.

At this stage the naive $U(1)$ axial symmetry remains.  Indeed, the
projector used above commutes with $\gamma_5$.  This symmetry is
allowed since four species, often called ``tastes,'' remain.  Among
them the symmetry is a taste non-singlet; under a chiral rotation, two
rotate one way and two the other.

The next step taken by most of the groups using staggered fermions is
the rooting trick.  In the hope of reducing the multiplicity down from
four, the determinant is replaced with its fourth root,
$|D|\rightarrow |D|^{1/4}$.  With several physical flavors this trick
is applied separately to each.  In simple perturbation theory this is
correct since each fermion loop gets multiplied by one quarter,
cancelling the extra factor from the four ``tastes.''

At this point one should be extremely uneasy: the exact chiral
symmetry is waving a huge red flag.  Symmetries of the determinant
survive rooting, and thus the exact $U(1)$ axial symmetry for the
massless theory remains.  For the unrooted theory this was a flavored
chiral symmetry.  But, having reduced the theory to one flavor, how
can there be a flavored symmetry without multiple flavors?  We will
now show why this rooting trick fails non-perturbatively when applied
to the staggered quark operator.

\subsection{The rooting issue}
In previous sections we have seen that the chiral symmetry with $N_f$
fermion flavors has a rather complicated dependence on $N_f$.  With
only one flavor there is no chiral symmetry at all, while in general
if the fermions are massless, there are $N_f^2-1$ Goldstone bosons.
We have also seen a qualitative difference in the mass dependence
between an even and an odd number of species.  Physics does not behave
smoothly in the number of flavors and this raises issues for fermion
formulations that inherently have multiple flavors, such as staggered
fermions.

Starting with four flavors, the basic question is whether one can
adjust $N_f$ down to one using the formal expression
\begin{equation}
\label{root}
\left| \matrix{
D+m & 0 & 0 & 0 \cr 0 & D+m & 0 & 0 \cr 0 & 0 & D+m & 0 \cr 0 & 0 & 0
& D+m \cr }\right |^{1\over 4} =|D+m|?
\end{equation}
This has been proposed and is widely used as a method for eliminating
the extra species appearing with staggered fermion simulations.

At this point it is important to emphasize that asking about the
viability of Eq.~(\ref{root}) is a vacuous question outside the
context of a regulator.  Field theory has divergences that need to be
controlled, and, as we have seen above, the appearance of anomalies
requires care.  In particular, the regulated theory must explicitly
break all anomalous symmetries in a way that survives in the continuum
limit.

So we must apply Eq.~(\ref{root}) before removing the regulator.  This
is generally expected to be okay as long as the regulator breaks any
anomalous symmetries appropriately on each of the four factors.  For
example, we expect rooting to be valid for four copies of the overlap
operator.  This satisfies a modified chiral symmetry
$D\gamma_5=-\hat\gamma_5 D$ where the gauge winding $\nu$ appears in
the gauge dependent matrix $\hat\gamma$ through ${\rm
Tr}\hat\gamma_5=2\nu$.

Section~\ref{mass} showed that in the continuum with $N_f$ degenerate
flavors there is a $Z_{N_f}$ symmetry in mass parameter space
corresponding to taking $m\rightarrow e^{2\pi i\gamma_5/N_f}m$.
Suppose we try to force the $Z_4$ symmetry in the regulated theory
before we root.  This is easily accomplished by considering the
determinant
\begin{equation}
\left| \matrix{
D+me^{i\pi\gamma_5\over 4} & 0 & 0 & 0 \cr
0 & D+me^{-i\pi\gamma_5\over 4} & 0 & 0 \cr
0 & 0 & D+me^{3i\pi\gamma_5\over 4} & 0 \cr
0 & 0 & 0 & D+me^{-3i\pi\gamma_5\over 4} \cr
}\right |.
\end{equation}
This maintains the above symmetry through a permutation of the four
flavors.  This modification of the determinant still gives a valid
formulation of the four flavor theory at vanishing $\Theta$ because
the imposed phases cancel.  But expressed in this way, we start with
four one-flavor theories each with a different value of $\Theta$.
Were we to root this form, we would be averaging over four
inequivalent theories.  This is not expected to be correct, much as we
would not expect rooting two different masses to give a theory of the
average mass; {\it i.e.}
\begin{equation}
\left(|D+m_1||D+m_2|\right)^{1/2}\ne \left |D+\sqrt{m_1m_2}\right|.
\end{equation}

So we have presented both a correct and an incorrect way to root a
four flavor theory down to one.  What is the situation with staggered
fermions, the primary place where rooting has been applied?  The
problem is that the kinetic term of the staggered action maintains one
exact chiral symmetry even at finite lattice spacing.  Without rooting
this is flavor non-singlet amongst the ``tastes.''  As discussed
earlier, there are two tastes of each chirality.  But, because of this
exact symmetry, which contains a $Z_4$ subgroup, rooting to reduce the
theory to one flavor is similar to the second case above and is not
expected to be valid.  In particular, rooting does not remove the
$Z_4$ discrete symmetry in the mass parameter, a symmetry which must
not be present in the one flavor theory.  Thus, just as in the above
example, the tastes are not equivalent and rooting averages
inequivalent theories.

The conclusion is that rooted staggered fermions are not QCD.  So,
what is expected to go wrong?  The unbroken $Z_4$ symmetry will give
rise to extra minima in the effective potential as a function of
$\sigma$ and $\eta^\prime$.  In particular, for one flavor QCD one
will get an effective potential with minima along the lines of
Fig.~\ref{potential} instead of the desired structure of
Fig.~\ref{potential3}.  Forcing the extra minima would most likely
drive the $\eta^\prime$ mass down from its physical value.  This shift
should be rather large, of order $\Lambda_{QCD}$.  This is testable,
but being dominated by disconnected diagrams, may be rather difficult
to verify in practice.  In addition, if we vary the quark masses, the
extra minima will result in phase transitions occurring whenever any
single quark mass passes through zero.  The previous discussion of the
one flavor theory and the two flavor theory with non-degenerate quarks
both show that this is unphysical; no structure is expected when only
a single mass passes through zero.

This problem is admittedly subtle.  Formula (\ref{root}) seems
intuitively obvious and does work if the individual factors take care
of the possible anomalies, as with four copies of the overlap
operator.  \footnote{
A few still hide behind this wall so frail,\\
\phantom{123}  So blind to chiral twists that made it fail.}
It is also correct perturbatively, since the rooting factor
reduces any fermion loop by the correct amount.  However, the basic
structure built up in earlier sections makes it indisputable that the
dependence of QCD on the parameter Theta is real.  With the staggered
action, the distinct tastes are not equivalent due to their different
behavior under chiral rotations.  It is this inequivalence that is at
the heart of the failure of rooting for this particular action.

Despite these problems, several lattice collaborations continue to
pursue staggered fermions using the rooting trick
\cite{Aoki:2006br,Cheng:2007jq,Bazavov:2009bb}.  The justification is
partly because the simulations are slightly faster than using Wilson
fermions, and partly because the exact chiral symmetry simplifies
operator mixing.  The success of a variety of calculations which are
not strongly dependent on the anomaly shows the approach, while
technically incorrect, is often a good approximation.  On the other
hand, if one's goal is to test QCD as the theory of the strong
interactions or to estimate QCD corrections to standard model
processes \cite{Lunghi:2010gv}, then one must be extremely wary of any
discrepancies found using this method.

\newpage
\Section{Other issues}
\subsection{Quantum fluctuations and topology}

We have seen how zero modes of the Dirac operator are closely tied to
the anomaly.  And we have seen that for smooth classical fields,
configurations that give zero modes for the classical Dirac operator
do indeed exist.  However, when getting into more detail with defining
a lattice Dirac operator, we found subtle issues about which operator
to use.  And way back in Section \ref{pathintegrals} we saw that
typical fields in path integrals are non-differentiable.  This leads
to the question of uniqueness for the winding number of a given gauge
configuration.  Indeed, is something like the topological
susceptibility of the vacuum a true physical observable?

\begin{figure}\centering{
\includegraphics[width=3in, angle=-90]{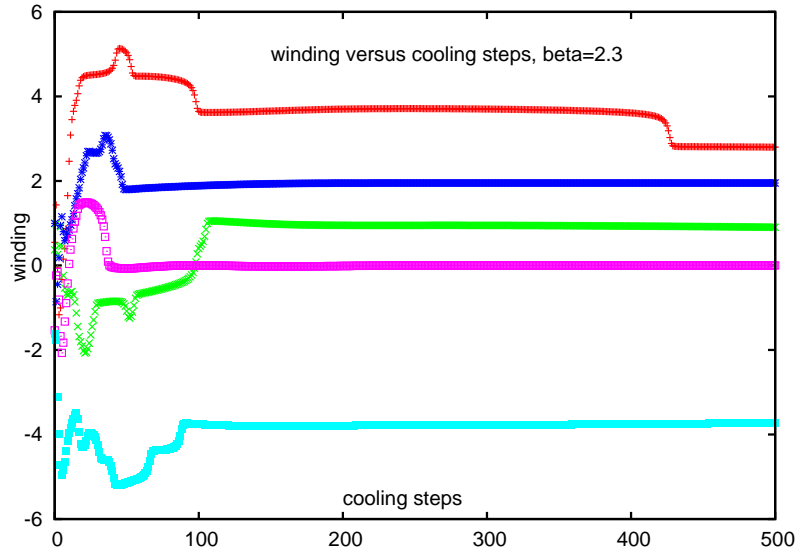}
\caption{ The winding number as a function of cooling steps for a set
  of 5 lattices of size $16^4$ at $\beta=2.3$. Note how it settles
  into approximately integer values with occasional jumps between
  different windings.  Figure from Ref.~\cite{Creutz:2010ec}.}
\label{windingcooling} 
}
\end{figure}

In \cite{Creutz:2010ec} a definition of topological charge was
constructed using the naive fermion operator as a regulator for the
trace of $\gamma_5$ as in the earlier derivation of the index theorem.
This operator does not generally give an integer value for a typical
gauge configuration in simulations.  However, it does reduce to such
after a cooling procedure is used to remove short distance
fluctuations.  The results of such are shown in
Fig.~\ref{windingcooling}.  On the other hand, the gauge field space
in lattice gauge theory is simply connected.  Empirically with enough
cooling, any $SU(2)$ gauge configuration appears to eventually decay to
a state of zero action, gauge equivalent to the vacuum.

Since configurations appear to cool ultimately to trivial winding,
using a cooling algorithm to define topology requires an arbitrary
selection for cooling time.  Modifying the Wilson action can prevent
the winding decay.  For example, forbidding the lattice action on any
given plaquette from becoming larger than a small enough number can
prevent instanton decay \cite{Luscher:1981zq}.  Such an
``admissibility'' condition, however, violates reflection
positivity \cite{Creutz:2004ir} and arbitrarily selects a special
instanton size where the action is minimum.

\begin{figure}\centering{
\includegraphics[width=3in, angle=-90]{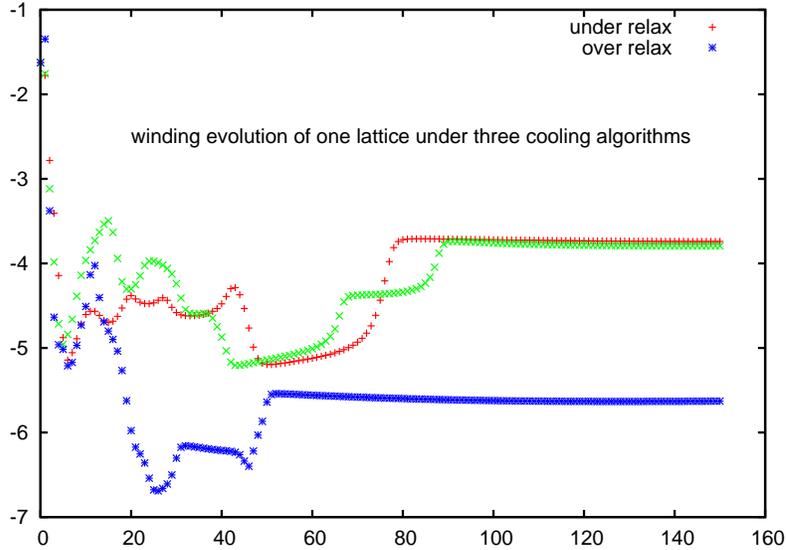}
\caption{ The topological charge evolution for three different cooling
  algorithms on a single $\beta=2.3$ lattice configuration for $SU(2)$
  on a $16^4$ lattice.  Figure from Ref.~\cite{Creutz:2010ec}.  With
  the higher winding numbers, lattice artifacts shift the plateaus
  slightly away from integer values.}
\label{relaxparm} 
}
\end{figure}

Cooling time is not the only issue here.  While attaining an integer
winding requires cooling, note in Fig.~\ref{windingcooling} that the
initial cooling stages seem quite chaotic.  This raises the question
of whether the discrete stages reached after some cooling might depend
rather sensitively on the cooling algorithm.  Figure \ref{relaxparm}
shows the evolution of a single lattice with three different
relaxation algorithms.  One algorithm consists of sweeping over the
lattice using checkerboard ordering and replacing each link with the
group element that minimizes the action associated with the given
link.  This is done by projecting the sum of staples that interact
with the link onto the group.  For the second approach, an
under-relaxed algorithm adds the old link to the sum of the
neighborhood staples before projecting onto the new group element.
Finally, an over-relaxed approach subtracts the old element from the
staple sum.  The resulting windings not only depend on cooling time,
but also on the specific algorithm chosen.

In an extensive analysis, Ref.~\cite {Bruckmann:2006wf} has compared a
variety of filtering methods to expose topological structures in gauge
configurations.  All schemes have some ambiguities, but when the
topological structures are clear, the various approaches when
carefully tuned give similar results.  Nevertheless the question
remains of whether there is a rigorous and unambiguous definition of
topology that applies to all typical configurations arising in a
simulation.  Luscher has discussed using a differential flow
with the Wilson gauge action to accomplish the cooling
\cite{Luscher:2010iy}.  This corresponds to the limit of maximal
under-relaxation.  This approach still allows the above topology
collapse unless prevented by something like the admissibility
condition or the selection of an arbitrary flow time.  In addition, if
one wishes to determine the topological charge of a configuration
obtained in some large scale dynamical simulation, it is unclear why
one should take the particular choice of the Wilson gauge action for
the cooling procedure.

The high sensitivity to the cooling algorithm on rough gauge
configurations suggests that there may be an inherent ambiguity in
defining the topological charge of typical gauge configurations and
consequently a small ambiguity in the definition of topological
susceptibility.  It also raises the question of how smooth is a given
definition of topological charge as the gauge fields vary; how much
correlation is there between nearby gauge configurations?  Although
such issues are quite old \cite{Teper:1985rb}, they continue to be of
considerable interest
\cite{Bruckmann:2009cv,Moran:2010rn,Luscher:2004fu}.

As topological charge is suppressed by light dynamical quarks, this is
connected to the question discussed earlier of whether the concept of
a single massless quark is well defined \cite{Creutz:2003xc}.
Dynamical quarks are expected to suppress topological structures, and
the chiral limit with multiple massless quarks should give zero
topological susceptibility with a chiral fermion operator, such as the
overlap.  However, with only a single light quark, the lack of chiral
symmetry indicates that there is no physical singularity in the
continuum theory as this mass passes through zero.  Any scheme
dependent ambiguity in defining the quark mass would then carry
through to the topological susceptibility.

One might argue that the overlap operator solves this problem by
defining the winding number as the number of zero eigenvalues of this
quantity.  Indeed, it has been shown
\cite{Giusti:2004qd,Luscher:2004fu} that this definition gives a
finite result in the continuum limit.  As one is using the fermion
operator only as a probe of the gluon fields, this definition can be
reformulated directly in terms of the underlying Wilson operator
\cite{Luscher:2010ik}.  While the result may have a finite continuum
limit, the earlier discussion showed that the overlap operator is not
unique.  In particular it depends on the initial Dirac operator being
projected onto the overlap circle.  For the conventional Wilson
kernel, there is a dependence on the parameter commonly referred to as
the domain-wall height.  Whether there is an ambiguity in the index
defined this way depends on the density of real eigenvalues of the
kernel in the vicinity of the point from which the projection is
taken.  Numerical evidence
\cite{Edwards:1998wx} suggests that this density decreases with
lattice spacing, but it is unclear if this decrease is rapid enough to
give a unique susceptibility in the continuum limit.  The
admissibility condition also successfully eliminates this ambiguity;
however, as mentioned earlier, this condition is inconsistent with
reflection positivity.

Whether topological susceptibility is well defined or not seems to
have no particular phenomenological consequences.  Indeed, this is not
a quantity directly measured in any scattering experiment.  It is only
defined in the context of a technical definition in a particular
non-perturbative simulation.  Different valid schemes for regulating
the theory might well come up with different values; it is only
physical quantities such as hadronic masses that must match between
approaches.  The famous Witten-Veneziano relation
\cite{Witten:1979vv,Veneziano:1979ec} does connect topological
susceptibility of the pure gauge theory with the eta prime mass in the
large number of colors limit.  This mass, of course, remains well
defined in the physical case of three colors, but the finite $N_c$
corrections to topology can depend delicately on gauge field
fluctuations, which are the concern here.

\subsection{The standard model}

Throughout the above we have concentrated on the strong interactions.
It is only for this sector of the standard model that perturbation
theory fails so spectacularly.  But the weak and electromagnetic
interactions are crucial parts of the full standard model (gravitation
is ignored here since it has even more serious unsolved problems).
And for these interactions it is also true that the perturbative
expansion does not converge.  Because the underlying couplings are so
small, this does not appear to be of any practical importance;
however, conceptually it is quite desirable to have a lattice
formulation for these interactions as well.  From a purist point of
view, the continuum limit of a lattice theory defines a continuum
field theory.  Thus without a lattice description of the other
interactions it is unclear whether we can say they are well founded
field theories.

In this context we note that the general picture of the standard model
has changed dramatically over the years.  Originally it was the
successes of quantum electrodynamics that made it the model
relativistic field theory.  Before QCD, the strong interactions were a
mystery.  But now we see that because of asymptotic freedom, QCD on
its own is likely to be a well defined and self contained theory.  It
is the electroweak theory, where both the electric charge and the
Higgs couplings are not asymptotically free, for which we lack a
non-perturbative formulation.  Indeed, a speculative topic such as the
possibility of emergent gravity may be intimately tied to these issues
with the weaker forces.

For the electromagnetic interactions, a lattice formulation at first
seems straightforward, involving the introduction of an additional
$U(1)$ gauge field for the photons.  Unlike the strong interaction
case, however, for electrodynamics we do not have asymptotic freedom
to tell us how to take the continuum limit.  And the physical coupling
$\alpha\sim 1/137$ seems to be an unnaturally small number.  Perhaps
electrodynamics on its own does not actually exist as a field theory,
much as believed for the scalar $\phi^4$ theory.  But photons and
electrons are essential components to the world around us.  One
interesting possibility is that electromagnetism is actually only a
part of a higher level theory, perhaps in some unification with the
strong interactions.

With the weak interactions we hit a more serious snag in that they are
known to violate parity.  The $W$ bosons appear to interact only with
left handed fermions.  As such we need to couple the fermions in a
chiral manner, and it is not known how to do this in any
non-perturbative scheme.  The problem here is closely tied to the
anomaly and the fact that not all currents can be simultaneously
conserved.  Indeed, when applied to the weak interactions, the 't
Hooft vertex gives rise to effective interactions that do not conserve
baryon number.  Any complete non-perturbative formulation must allow
for such processes \cite{Eichten:1985ft}.  Some attempts to include
such in a domain wall formulation have been
presented \cite{Creutz:1994ny,Poppitz:2010at}, but these generally
involve heavy additional states such as ``mirror'' fermions
\cite{Montvay:1987ys}.  While potentially viable, such approaches so
far lack the theoretical elegance of the original Wilson lattice gauge
theory.  Indeed, it is the problem of chiral gauge theories that
encourages studies of chiral symmetry from all possible
angles. 

Perhaps a lattice formulation more intimately tied to unification
ideas could help here.  The group $SO(10)$ looks quite interesting in
this context \cite{Georgi:1979dq}.  Here a single generation of
fermions fits nicely into a single 16 dimensional representation of
this group.  And in this picture anomalies are automatically
cancelled.  This would seem to indicate that there should be no
obvious requirement for doublers as an obstacle to a lattice
construction.  However, the usual Wilson approach seems to require a
term that is not a singlet under this group.  This could be overcome
with some added Higgs-like scalar fields, but then we get closer to
the above mentioned models with the doublers playing the role of
mirror fermions.

\subsection{Where is the parity violation?}

The standard model of elementary particle interactions is based on the
product of three gauge groups, $SU(3)\otimes SU(2) \otimes U(1)_{em}$.
Here the $SU(3)$ represents the strong interactions of quarks and
gluons, the $U(1)_{em}$ corresponds to electromagnetism, and the
$SU(2)$ gives rise to the weak interactions.  We ignore here the
technical details of electroweak mixing.  The full model is, of
course, parity violating, as necessary to describe observed helicities
in beta decay.  This violation is normally considered to lie in the
$SU(2)$ of the weak interactions, with both the $SU(3)$ and
$U(1)_{em}$ being parity conserving.  We will show here that this is
actually a convention, adopted primarily because the weak interactions
are small compared to the others.  We show below that reassigning
degrees of freedom allows a reinterpretation where the $SU(2)$ gauge
interaction is vector-like.  Since the full model is parity violating,
this process shifts the parity violation into the strong,
electromagnetic, and Higgs interactions.  The resulting theory pairs
the left handed electron with a right handed anti-quark to form a
Dirac fermion.  With a vector-like weak interaction, the chiral issues
which complicate lattice formulations now move to the other gauge
groups.  Requiring gauge invariance for the re-expressed
electromagnetism then clarifies the mechanism behind one proposal for
a lattice regularization of the standard
model \cite{Creutz:1996xc,Creutz:1997fv}.

For simplicity we consider here only the first generation, which
involves four left handed doublets.  These correspond to the
neutrino/electron lepton pair plus three colors for the up/down quarks
\begin{equation}
\pmatrix{\nu \cr e^-\cr}_L,
\ \pmatrix{{u^r} \cr {d^r}\cr}_L,
\ \pmatrix{{u^g} \cr {d^g}\cr}_L,
\ \pmatrix{{u^b} \cr {d^b}\cr}_L.
\end{equation}
Here the superscripts from the set $\{r,g,b\}$ represent the internal
$SU(3)$ index of the strong interactions, and the subscript $L$ indicates
left-handed helicities. 

If we ignore the strong and electromagnetic interactions, leaving only
the weak $SU(2)$, each of these four doublets is
equivalent and independent.  We now arbitrarily pick two of them
and do a charge conjugation operation, thus switching to their
antiparticles
\begin{equation}\matrix{
\pmatrix{{u^g} \cr {d^g}\cr}_L \longrightarrow 
\pmatrix{\overline{{d^g}} \cr \overline{{u^g}}\cr}_R \cr
\cr
\pmatrix{{u^b} \cr {d^b}\cr}_L \longrightarrow 
\pmatrix{\overline{{d^b}} \cr \overline{{u^b}}\cr}_R .\cr
}
\end{equation}
In four dimensions anti-fermions have the opposite helicity; so, we
label these new doublets with $R$ representing right handedness.

With two left and two right handed doublets, we can combine them into
two Dirac doublets
\begin{equation}
\pmatrix{
\pmatrix{\nu \cr e^-\cr}_L\cr
\pmatrix{\overline{{d^g}} \cr \overline{{u^g}}\cr}_R\cr
}
\qquad
\pmatrix{
\pmatrix{{u^r} \cr {d^r}\cr}_L\cr
\pmatrix{\overline{{d^b}} \cr \overline{{u^b}}\cr}_R \cr
}.
\end{equation}
\def \half { {1\over 2} }
Formally in terms of the underlying fields, the construction takes
\begin{eqnarray}
&&\psi=\half (1-\gamma_5)\psi_{(\nu,e^-)}+\half (1+\gamma_5)
\psi_{({\overline{d^g}},{\overline{u^g}})} \\
&&\chi=\half (1-\gamma_5)\psi_{({u^r}, {d^r})}+\half (1+\gamma_5)
\psi_{({\overline{d^b}},{\overline{u^b}})}.
\end{eqnarray}

From the conventional point of view, these fields have rather peculiar
quantum numbers.  For example, the left and right parts have different
electric charges.  Electromagnetism now violates parity.  The left and
right parts also have different strong quantum numbers; the strong
interactions violate parity as well.  Finally, the components have
different masses; parity is violated in the Higgs mechanism.

The different helicities of these fields also have variant baryon
number.  This is directly related to the known baryon violating
processes through weak ``instantons'' and axial
anomalies\cite{'tHooft:1976fv}.  When a topologically non-trivial weak
field is present, the axial anomaly arises from a level flow out of
the Dirac sea \cite{Ambjorn:1983hp}.  This generates a spin flip in the
fields, {\it i.e.} $e^-_L \rightarrow ({\overline{u^g}})_R$.  Because
of the peculiar particle identification, this process does not
conserve charge, with $\Delta Q= -{2\over 3} +1={1\over 3}$.  This
would be a disaster for electromagnetism were it not for the fact that
simultaneously the other Dirac doublet also flips {${d^r}_L
\rightarrow ({\overline{u^b}})_R$} with a compensating $\Delta Q =
-{1\over 3}$.  This is anomaly cancellation, with the total $\Delta Q
= {1\over 3}-{1\over 3}=0$.  Only when both doublets are considered
together is the $U(1)$ symmetry restored.  In this anomalous process
baryon number is violated, with $L+Q\rightarrow \overline Q +
\overline Q$.  This is the famous `` `t Hooft vertex'' \cite {'tHooft:1976fv}
discussed earlier in the context of the strong interactions.

\subsection {A lattice model}

The above discussion on twisting the gauge groups has been in the
continuum.  Now we turn to the lattice and show how this picture leads
to a possible lattice model for the strong interactions, albeit with
an unusual added coupling that renders the treatment quite difficult
to make rigorous \cite{Creutz:1996xc,Creutz:1997fv}.  Whether this
model is viable remains undecided, but it does incorporate many of the
required features.

For this we use the domain wall approach for the fermions
\cite{Kaplan:1992bt,Shamir:1993zy}.  
As discussed earlier, in this picture, our four dimensional world is a
``4-brane'' embedded in 5-dimensions.  The complete lattice is a five
dimensional box with open boundaries, and the parameters are chosen so
the physical quarks and leptons appear as surface zero modes.  The
elegance of this scheme lies in the natural chirality of these modes
as the size of the extra dimension grows.  With a finite fifth
dimension one doubling remains, coming from interfaces appearing as
surface/anti-surface pairs.  It is natural to couple a four
dimensional gauge field equally to both surfaces, giving rise to a
vector-like theory.

\begin{figure}
\centering
\includegraphics[width=2.5in]{{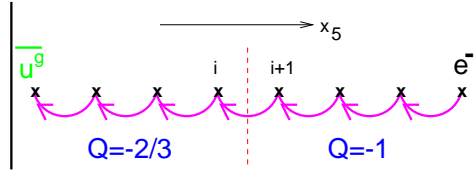}}
\caption{Pairing the electron with the anti-green-up-quark.
Figure taken from \cite{Creutz:1997fv}.}
\label{fig:1}
\end{figure}

We now insert the above pairing into this five dimensional scheme.  In
particular, consider the left handed electron as a zero mode on one
wall and the right handed anti-green-up-quark as the partner zero mode
on the other wall, as sketched in Fig.~\ref{fig:1}.  This provides a
lattice regularization for the $SU(2)$ of the weak interactions.

However, since these two particles have different electric charge,
$U(1)_{EM}$ must be broken in the interior of the extra dimension.  We
now proceed in analogy to the ``waveguide''
picture\cite{Golterman:1993th} and restrict this charge violation to
$\Delta Q$ to one layer at some interior position $x_5=i$.  Using
Wilson fermions, the hopping term from $x_5=i$ to $i+1$
\begin{equation}
\overline\psi_{i}P\psi_{i+1}\qquad{(P=(\gamma_5+r)/2)}
\end{equation}
is a $Q=1/3$ operator.  At this layer, electric charge is not
conserved.  This is unacceptable and needs to be fixed.

\begin{figure}
\centering
\includegraphics[width=2.5in]{{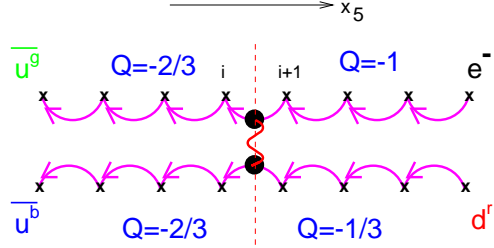}}
\caption {Transferring charge between the doublets. 
Figure taken from \cite{Creutz:1997fv}.}
\label{fig:2}
\end{figure}

To restore the $U(1)$ symmetry one must transfer the charge from
$\psi$ to the compensating doublet $\chi$.  For this we replace
the sum of hoppings with a product on the offending layer
\begin{equation}
\overline\psi_{i}P\psi_{i+1}
{+}\overline\chi_{i}P\chi_{i+1}
{\longrightarrow}
\overline\psi_{i}P\psi_{i+1}
{\times}\overline\chi_{i}P\chi_{i+1}.
\end{equation}
This introduces an electrically neutral four-fermi operator.  Note
that it is baryon violating, involving a ``lepto-quark/diquark''
exchange, as sketched in Fig.~\ref{fig:2}.  One might think of the
operator as representing a ``filter'' at $x_5=i$ through which only
charge compensating pairs of fermions can pass.

In five dimensions there is no chiral symmetry.  Even for the free
theory, combinations like $\overline\psi_{i}P\psi_{i+1} $ have
non-vanishing vacuum expectation values.  We use such as a
``tadpole,'' with $\chi$ generating an effective hopping for $\psi$
and {\it vice versa}.

Actually the above four fermion operator is not quite sufficient for
all chiral anomalies, which can also involve right handed
singlet fermions.  To correct this we need explicitly include the right
handed sector, adding similar four fermion couplings
(also electrically neutral).  The main difference is that this sector
does not couple to the weak bosons.

Having fixed the $U(1)$ of electromagnetism, we restore the strong
$SU(3)$ with an anti-symmetrization of the quark color indices in the
new operator, $ {Q^r}{Q^g}{Q^b}{\longrightarrow
\epsilon^{\alpha\beta\gamma}Q^\alpha Q^\beta Q^\gamma}$.  Note that
similar left-right inter-sector couplings are needed to correctly
obtain the effects of topologically non-trivial strong gauge fields.

An alternative view of this picture folds the lattice about the
interior of the fifth dimension, placing all light modes on one wall
and having the multi-fermion operator on the other.  This is the model
of Ref.~\cite {Creutz:1996xc}, with the additional inter-sector couplings
correcting a technical error \cite{Neuberger:1997cz}.

Unfortunately the scheme is still not completely rigorous.  In
particular, the non-trivial four-fermion coupling represents a new
defect and we need to show that this does not give rise to unwanted
extra zero modes.  Note, however, that the five dimensional mass is
the same on both sides of defect; thus there are no topological
reasons for such.

A second worry is that the four fermion coupling might induce an
unwanted spontaneous symmetry breaking of one of the gauge symmetries.
We need to remain in a strongly coupled paramagnetic phase without
spontaneous symmetry breaking.  Ref.~\cite{Creutz:1996xc} showed that
strongly coupled zero modes do preserve the desired symmetries, but
the analysis ignored contributions from heavy modes in the fifth
dimension.

Assuming all works as desired, the model raises several other
interesting questions.  As formulated, we needed a right handed
neutrino to provide all quarks with partners.  Is there some variation
that avoids this particle, which decouples from all gauge fields in
the continuum limit?  Another question concerns possible numerical
simulations; is the effective action positive?  Finally, we have used
the details of the usual standard model, leaving open the question of
whether this model is somehow special.  Can we always find an
appropriate multi-fermion coupling to eliminate undesired modes in
other chiral theories where anomalies are properly canceled?

\newpage
\Section{Final remarks}

We have seen how many features of QCD are influenced by
non-perturbative physics.  This is particularly important to various
aspects of chiral symmetry breaking.  Taken as a whole, these fit
together into a rather elegant and coherent picture.  In particular,
chiral symmetry is broken in three rather different ways.  We have
concentrated on the interplay of these mechanisms.

The primary and most important effect is the dynamical symmetry
breaking that leads to the pions being light pseudo-Goldstone bosons.
Their dynamics represents the most important physics for QCD at low
energies.  The popular and useful chiral expansion is a natural
expansion in the momenta and masses of these particles.

In addition to the basic dynamical breaking is the anomaly, which
eliminates the flavor-singlet axial $U(1)$ symmetry of the classical
theory.  Thus the $\eta^\prime$ meson is not a Goldstone boson and
acquires a mass of order $\Lambda_{qcd}$.  Understanding this breaking
requires non-perturbative physics associated with the zero modes of
the Dirac operator.

Finally, we have the explicit symmetry breaking from the quark masses.
This is responsible for the pseudo-scalar mesons not being exactly
massless.  Using the freedom to redefine fields using chiral rotations,
the number of independent mass parameters is $N_f+1$ where $N_f$ is
the number fermion species under consideration.  This includes the
possibility of CP violation coming from the interplay of the mass term
with the anomaly.

Throughout we have used only a few widely accepted assumptions, such
as the existence of QCD as a field theory and standard ideas about
chiral symmetry.  Thus it is perhaps somewhat surprising that several
of the conclusions remain controversial.  The first of these is that
chiral symmetry is lost in a theory with only one light quark.  The
resulting additive non-perturbative renormalization of the mass
precludes using a massless up quark to solve the strong CP problem.
Tied to this is the issue of whether topological susceptibility is
well defined when non-differentiable fields dominate the path
integral.  Finally, probably the most bitter controversies revolve
about the symmetries inherent in the staggered formulation and how
these invalidate the use of rooting to remove unwanted degeneracies.

As simple as the overall picture is, it requires understanding effects
that go well beyond perturbation theory.  We need aspects of the Dirac
spectrum that rely on gauge fields of non-trivial topology.  Such
appear already in the classical theory, although their true importance
only appears in the context of the anomaly.  Including this physics
properly in a lattice formulation is a rich and sometimes
controversial topic of active research.

\begin{ack}
The author is grateful to Ivan Horvath for suggesting this article.
He is also indebted to Ivan as well as to Stefano Capitani and
Nobuyoshi Ohta for finding numerous typos in the original version.
The Alexander von Humboldt Foundation provided valuable support for
visits to the University of Mainz where a portion of this research was
carried out.  This manuscript has been authored by employees of
Brookhaven Science Associates, LLC under Contract
No. DE-AC02-98CH10886 with the U.S. Department of Energy. The
publisher by accepting the manuscript for publication acknowledges
that the United States Government retains a non-exclusive, paid-up,
irrevocable, world-wide license to publish or reproduce the published
form of this manuscript, or allow others to do so, for United States
Government purposes.
\end{ack}


\end{document}